\documentclass{article}

\usepackage{microtype}
\usepackage{graphicx}
\usepackage{subcaption}
\usepackage{booktabs} 
\usepackage{amsmath}
\usepackage{siunitx}
\usepackage{enumitem}
\usepackage{tabularx}
\usepackage{array}
\usepackage{dblfloatfix}
\newcolumntype{L}[1]{>{\raggedright\arraybackslash}p{#1}}
\newcolumntype{Y}{>{\raggedright\arraybackslash}X}

\newif\ifincludeappendix
\includeappendixtrue

\usepackage{hyperref}

\usepackage[accepted]{icml2026}

\usepackage{amsmath}
\usepackage{amssymb}
\usepackage{mathtools}
\usepackage{amsthm}

\usepackage[capitalize,noabbrev]{cleveref}

\theoremstyle{plain}

\theoremstyle{definition}

\theoremstyle{remark}

\usepackage[textsize=tiny]{todonotes}

\icmltitlerunning{IDE-Bench: Evaluating Large Language Models as IDE Agents on Real-World Software Engineering Tasks}

\begin{document}

\twocolumn[
  \icmltitle{IDE-Bench: Evaluating Large Language Models as IDE Agents on Real-World Software Engineering Tasks}



  \icmlsetsymbol{equal}{*}

  \begin{icmlauthorlist}
    \icmlauthor{Spencer Mateega}{comp}
    \icmlauthor{Jeff Yang}{comp}
    \icmlauthor{Tiana Costello}{comp}
    \icmlauthor{Shaurya Jadhav}{comp}
    \icmlauthor{Nicole Tian}{comp}
    \icmlauthor{Agustin Garcinuño}{comp}
  \end{icmlauthorlist}

  \icmlaffiliation{comp}{AfterQuery, San Francisco, CA, US}

  \icmlcorrespondingauthor{Spencer Mateega}{spencer@afterquery.com}

  \icmlkeywords{Machine Learning, ICML}

  \vskip 0.3in
]



\printAffiliationsAndNotice{}  

\begin{abstract}
  IDE-Bench is a comprehensive framework for evaluating AI IDE agents on real-world software engineering tasks through an IDE-native tool interface. We present a Dockerized test harness that goes beyond raw terminal execution, granting models a structured tool ecosystem that represents  AI-native IDEs like Cursor and Windsurf. By providing high-level abstractions for codebase search, structured file editing, and tools for testing full-stack applications, IDE-Bench evaluates an agent's ability to act as a true engineering collaborator. For evaluation and to prevent training data contamination, we created 80 tasks across eight never-published repositories spanning C/C++, Java, and MERN stacks, representing modern tech stack production scenarios, including feature implementation, bug fixing, refactoring, and performance optimization that mirror daily developer workflows in private codebases. Our benchmark is the first to systematically correlate agent-reported intent with successful project-level modifications in a multi-language, full-stack environment on completely uncontaminated code. We release IDE-Bench and a public leaderboard at: \url{ide-bench.com}.
\end{abstract}

\section{Introduction}
Since the rise of Cursor, GitHub Copilot, and other agentic-enabled IDE, software engineering workflows have greatly transformed. Engineers have integrated agentic tools into everyday tasks, reducing code repetition and boosting productivity. Thus, agent-enabled IDEs have seen rapid adoption, transforming how developers write and review code \cite{coutinho2024rolegenerativeaisoftware,peng2023impactaideveloperproductivity}. GitHub Copilot reported over 20 million users in July 2025, and Cursor surpassed 360,000 paying customers by November 2025. Despite this shift in software engineering practices, there is no existing benchmark that uses tool calling to rigorously test how well models perform as IDE agents on the types of debugging, refactoring, feature development, and full-stack workflows real developers engage in \cite{kwa2025measuringaiabilitycomplete,SWTBench,BigCodeBench,LiveCodeBench}.

In this work, we evaluate autonomous agents that operate with the full capacities of an IDE agent, including both tool calling and code generation capabilities. We define IDE agents as AI models operating in a chat-based IDE environment with access to the same tools available in agent-enabled IDEs like Cursor. We propose IDE-Bench, a comprehensive framework for evaluating AI IDE agents on real-world software engineering tasks across diverse technology stacks. IDE-Bench offers benefits over existing LLM benchmarks, most notably the tool-calling ability. Our key contributions are:

\begin{itemize}
    \item \textbf{IDE-native structured tool interface}: Unlike SWE-Bench's static context retrieval or Terminal-Bench's raw shell commands, IDE-Bench is designed specifically for chat-based IDE agents, the interaction model used by tools like Cursor and Windsurf. It evaluates agents using the same tool abstractions that define the IDE-agent paradigm (e.g., \texttt{read\_file}, \texttt{edit\_file}, \texttt{codebase\_search}). The benchmark includes conditional tool access for API endpoint testing and MongoDB query verification when tasks require MERN stack components, capabilities absent from current benchmarks. Users can gather evaluation metrics such as full conversation trajectories, tool-call sequences, and behavioral metrics that capture how agents explore, debug, and recover from errors within an IDE environment.
    \item \textbf{Novel Repositories Representing Real Developer Workflows}: We provide eight novel repositories that have never been published to the internet, GitHub, or any public platform, ensuring protection from contamination from training data, a large concern as benchmark tasks published online can potentially be memorized by future model releases. The tasks were created to represent the real-life scenarios that developers may encounter in private code bases, such as implementing user-requested features from specifications, debugging difficult multi-file issues, refactoring legacy systems, and optimizing performance under genuine, real-world constraints. Each repository simulates a realistic production environment scenario such as full dependency management, integration testing requirements, multi-language stacks, and under-specified edge cases that developers may often encounter in actual software projects.
\end{itemize}

\section{Related Work}

\textbf{SWE-Bench} evaluates LLMs' ability to navigate complex repositories, drawing from GitHub issues and pull requests (PRs), and it tests debugging and small-scale corrections rather than large-scale project development \cite{jimenez2024swebenchlanguagemodelsresolve}. Notably, SWE-Bench does not provide model access to the IDE environment and instead evaluates model performance in a single-shot, static paradigm where context retrieval remains the main performance bottleneck; SWE-Bench Verified addresses challenges in the original paper by removing invalid task instances and supplementing static retrieval with system call tools \cite{openai2024swebench}. \textbf{SWE-Bench Pro} extends SWE-Bench with professional-grade tasks from enterprise codebases \cite{deng2025swebenchproaiagents} and shares the goal of more realistic evaluation and minimizing contamination, but it still maintains the single-shot, context-retrieval paradigm. In contrast, IDE-Bench grants agents access to the full IDE tool interface—including search, structured file editing, and tools for testing full-stack applications (API endpoints, databases, WebSockets)—and supports multi-language stacks (C/C++, Java, TypeScript, Python). Lastly, \textbf{Terminal-Bench} evaluates realistic, long-horizon computer tasks in a terminal-centric workflow \cite{merrill2026terminalbenchbenchmarkingagentshard}; it excels at testing lower-level command execution and environment manipulation, but it does not measure higher-level IDE-native behaviors such as lexical search, codebase indexing, and IDE-level navigation. These observations motivate IDE-Bench, designed to evaluate the interactive, tool-enabled, multi-step software development in an IDE environment.

\section{IDE-Bench Benchmark Design}
\subsection{Task Domains}
The IDE-Bench evaluation suite consists of eight repositories with ten tasks each, with each carefully designed to represent some real-world software engineering work. Unlike synthetic benchmarks or isolated coding challenges, our tasks are drawn from real-life development scenarios that engineers encounter in production environments whether it be interpreting underspecified feature requests, debugging state management issues across multiple files, refactoring code while maintaining backward compatibility, or optimizing algorithms under performance constraints. The suite spans Python, C, C++, JavaScript/TypeScript, and Java. Many tasks enforce strict output format requirements and handle edge cases that emerge only in real deployments, created to test an agent's ability to meet production specifications rather than pass artificial unit tests. All repositories remain unpublished to prevent training data contamination, ensuring models are evaluated on truly novel code.

Each task directory contains a \texttt{task\_description.txt} file (with the problem statement, instructions, and goal), a \texttt{task\_diff.txt} reference patch, and a \texttt{task\_tests.py} evaluation file. To maintain evaluation integrity, we remove the \texttt{task\_diff.txt} from the environment before model deployment. Evaluation is conducted via a unified \texttt{run\_tests.sh} script located in the repository root, which executes the task-specific tests and returns a success status based on the script's exit code.

The eight repositories span diverse domains including systems programming (ESIM Management System, Memory Profiling App, Game Engine Service), enterprise applications (SmartHub Operations Center), web services (Event Callback System, Cross-Lingual Document Translator), code analysis (Code Quality Analyzer), and network programming (Network Traffic Analyzer). Detailed descriptions of each repository are provided in Appendix~\ref{app:repository_descriptions}.

\subsection{Harness Environment}
We use the LiteLLM interface to deploy an agent harness, an agent evaluation framework designed to benchmark agents against complex coding tasks in containerized environments, implementing a tool ecosystem that mirrors real agent-enabled IDEs such as Cursor and Windsurf.  The harness architecture utilizes Docker containerization for reproducible test environments, git-based change tracking for precise code-diff analysis, and a multi-phase execution pipeline that includes task-specific requirement verification against golden diffs and comprehensive test suites.  

\subsection{Agent Tool Interface}

The harness equips models with 17 tools in 5 categories following OpenAI's function calling specification: (1) File System \& Code Navigation (e.g., \texttt{read\_file}, \texttt{list\_dir}, \texttt{codebase\_search}), (2) Code Editing (\texttt{edit\_file}), (3) Execution \& Testing (\texttt{run\_terminal\_cmd}), (4) Full-Stack Testing for MERN tasks (e.g., \texttt{api\_call}, \texttt{database\_query}), and (5) Specialized tools. Each tool call requires an \texttt{explanation} parameter to encourage explicit reasoning and enable trajectory analysis. Complete tool specifications with parameters and usage guidelines are provided in Appendix~\ref{app:tool_interface}.

\subsection{Execution Workflow}

\begin{figure*}[t]
    \vspace{-6pt}
    \centering
    \includegraphics[width=0.94\textwidth,trim=0 8 0 8,clip]{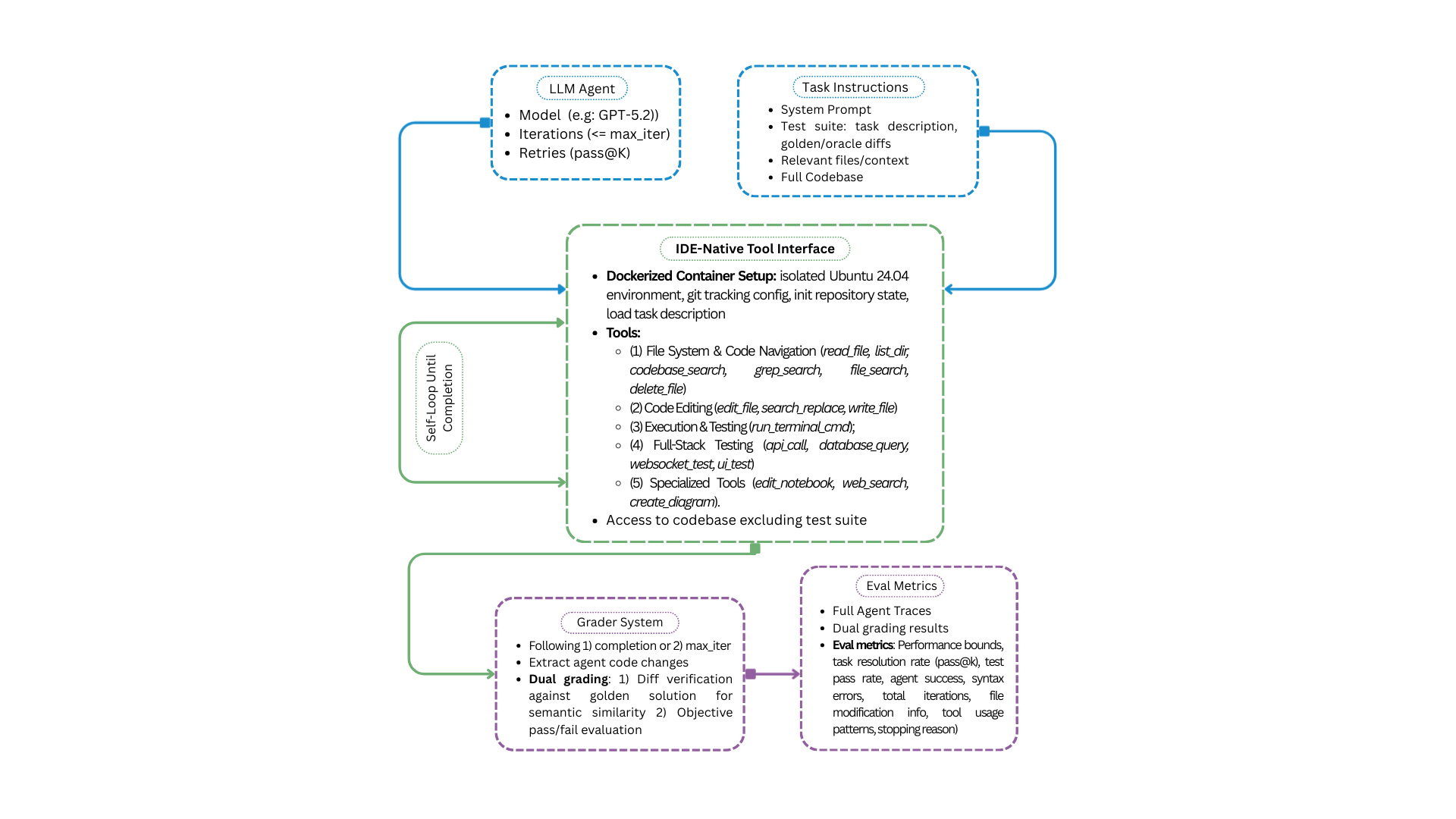}
    \vspace{-10pt}
    \refstepcounter{figure}\label{workflow}
    \caption*{\textbf{IDE-Bench Execution Pipeline:} Each task runs in an isolated Docker container. The harness loads the task description, launches the agent (LLM or Oracle baseline) with access to IDE tools, and captures all interactions. After execution, the grader runs the test suite and extracts code changes via git diff for comparison against the golden solution.}
\end{figure*}

The evaluation pipeline consists of three main stages, illustrated in Figure \ref{workflow}:

\textbf{1. Container Setup:} Each task begins with a fresh Docker container built from the repository's Dockerfile, providing an isolated Ubuntu 24.04 environment. The harness configures git for change tracking, initializes the repository state, and loads the task description.

\textbf{2. Agent Execution:} The harness launches either a Gladiator agent (the model under evaluation using LiteLLM) or an Oracle baseline (which directly applies the reference patch). For Gladiator agents, the system provides a comprehensive system prompt (see Appendix~\ref{app:system_prompt}) with tool descriptions and task requirements. The agent iteratively uses the 17 available tools to explore the codebase, modify files, and test implementations (tool specs: Appendix~\ref{app:tool_interface}). All tool calls, file edits, and responses are logged for trajectory analysis. Security restrictions prevent agents from reading test files or golden solutions, ensuring that they must reason from task requirements alone.

\textbf{3. Evaluation and Grading:} After the run (or the 100-iteration limit), we execute the repository test suite (\texttt{./run\_tests.sh}) and compare the agent's code changes against the reference patch. Appendix~\ref{app:harness_grading} provides the full grading pipeline and diff-extraction details.

\subsubsection{Implementation and Metrics (Summary)}

We use a LiteLLM-based harness with standardized tool calling and safety restrictions; details are provided in Appendix~\ref{app:harness_impl}. The benchmark reports grounded performance bounds (Null vs.\ Oracle baselines), task resolution rate (pass@k), and per-test pass rate; the full metric suite (including iteration counts and tool-usage signals) is provided in Appendix~\ref{app:eval_metrics_full}.

\section{Experiments}

\subsection{Overall Performance}
    \begin{table}[htbp]
        \centering
        \caption{\textbf{Task Resolution Rate by Model}}
        \label{tab:model_performance_condensed}
        \small
        \begin{tabular}{lcc} 
        \toprule
        \textbf{Model} & \textbf{pass@1 (\%)} & \textbf{pass@5 (\%)} \\
        \midrule
        GPT 5.2 & 85.00 $\pm$ 7.81 & 95.00 $\pm$ 5.10 \\
        Claude Sonnet 4.5 & 87.50 $\pm$ 7.28 & 88.75 $\pm$ 6.99 \\
        Claude Haiku 4.5 & 78.75 $\pm$ 8.86 & 87.50 $\pm$ 7.28 \\
        Claude Opus 4.5 & 83.75 $\pm$ 8.05 & 86.25 $\pm$ 7.56 \\
        GPT 5.1 Codex Max & 73.75 $\pm$ 9.48 & 85.00 $\pm$ 7.81 \\
        Gemini 3 Pro Preview & 55.00 $\pm$ 10.65 & 80.00 $\pm$ 8.67 \\
        Qwen3 Max & 65.00 $\pm$ 10.23 & 76.25 $\pm$ 9.19 \\
        Qwen3 Coder & 57.50 $\pm$ 10.59 & 75.00 $\pm$ 9.34 \\
        DeepSeek V3.2 & 31.25 $\pm$ 9.96 & 71.25 $\pm$ 9.74 \\
        Grok 4.1 Fast & 35.00 $\pm$ 10.23 & 67.50 $\pm$ 10.06 \\
        DeepSeek R1 0528 & 20.00 $\pm$ 8.67 & 46.25 $\pm$ 10.67 \\
        Grok Code Fast 1 & 11.25 $\pm$ 6.99 & 32.50 $\pm$ 10.06 \\
        Llama 4 Maverick & 2.50 $\pm$ 3.99 & 8.75 $\pm$ 6.34 \\
        Command-R+ 08 2024 & 0.00 $\pm$ 2.29 & 7.50 $\pm$ 5.96 \\
        Llama 4 Scout & 2.50 $\pm$ 3.99 & 6.25 $\pm$ 5.56 \\
        \bottomrule
        \end{tabular}
        \label{task_resolution_rate}
    \end{table}

    Table \ref{task_resolution_rate} reports pass@1 and pass@5 (the fraction of tasks a model resolves at least once across one and five independent attempts, respectively). We evaluate 15 models across 6,000 runs (80 tasks  times  5 attempts  times  15 models). When the models are sorted by task resolution rate, two patterns stand out.

    First, there is a clear separation between the top frontier models and the open-weight models in our evaluations. GPT 5.2 achieves the highest performance at 95\% pass@5, while Claude Sonnet 4.5, Claude Haiku 4.5, Claude Opus 4.5, and GPT 5.1 Codex Max cluster tightly in the 85.0-88.75\% range, all with relatively tight confidence intervals, suggesting stable performance across the tested domains (\cite{anthropic2025claudesonnet4.5,anthropic2025claudehaiku4.5,anthropic2025claudeopus4.5,openai2025gpt5.1codexmaxsystemcard,openai2025gpt5.2systemcard}). On the other hand, the next tier includes Gemini 3 Pro Preview at 80.00\%, followed by the Qwen3 models (Qwen3 Max at 76.25\%, Qwen3 Coder at 75.00\%) and DeepSeek V3.2 at 71.25\% \cite{geminiteam2025gemini3pro,qwen3coder,qwen3max,deepseekv3.2}. The performance then declines more steeply, with most lower-tier models failing to exceed a 50\% resolution rate. Thus, there is a visible, sizable gap between models that can truly function as containerized IDE agents and those that cannot.

    Second, we see that the model's identity still matters even when aggregate rates appear similar. Codex Max performs as expected, given it is tuned specifically for agentic coding, towards the ceiling of this benchmark. However, more surprisingly, Claude Haiku (a speed-oriented model) performs nearly as well and, for some tasks, better than Sonnet. This suggests that for a substantial fraction of the tasks in IDE-Bench, the limiting factor is often whether the agent can execute a clean tool loop and converge without drifting.

    The lower-performing models show a different pattern in how they benefit from retries. When looking at the gap between pass@1 and pass@5, we see that models below 85\% pass@5 improve by much larger amounts compared to top-tier models (e.g., DeepSeek V3.2 gains 40 points, 31.25\% $\rightarrow$ 71.25\%), while models above 85\% pass@5 show minimal gains (e.g., Claude Sonnet gains only 1.25 points, 87.50\% $\rightarrow$ 88.75\%). This 85\% threshold appears to mark a transition where models shift from inconsistent, iteration-dependent behavior to stable, first-attempt success.

\textbf{First-Attempt Performance (pass@1).} As shown in Table~\ref{task_resolution_rate}, Claude Sonnet 4.5 achieves the highest first-attempt pass rate at 87.50\%, followed closely by GPT 5.2 at 85.00\% and Claude Opus 4.5 at 83.75\%. For deployments where API costs and latency constraints limit retries, this first-attempt reliability acts as a practical selection criteria beyond aggregate success rates.
    \subsection{Binary Metrics Hide Partial Success}

Standard benchmarking often treats task resolution as binary where a run could either be fully correct or fully a failure. However, our per-test analysis shows that this binary framing hides a common outcome mode in IDE-Bench: ``near misses'' where the agent implements the core fix but fails on a small number of tests (often due to output formatting or an underspecified edge case). We include representative examples in Appendix Table~\ref{tab:near_miss}.

One case to note is Event Callback System task-4: Claude Sonnet 4.5, Gemini 3 Pro Preview, and Claude Opus 4.5 all repeatedly achieve 91.7\% test pass rates across all five attempts yet score 0\% under pass@5. Similarly, Cross-Lingual Document Translator task-7 shows Claude Opus 4.5 achieving 92.3\% test pass rate with 0\% pass@5. These tasks require thorough implementation of algorithmic logic and strict output formatting. In these runs, the models implement the core functionality correctly but miss a small number of tests, most likely due to output specification mismatches or edge cases in formatting compliance.

We conclude that this illustrates a recurring pattern in agentic coding that specification precision can be more challenging than algorithmic correctness. An 8.3\% gap in tests passed can correspond to a disproportionate amount of engineering effort, because the remaining work is often ``small'' but brittle (such as formatting, ordering, off-by-one behavior, or requirements hidden in the tests). For developers in the real world, the actual implication is that many benchmark ``failures'' are not complete failures. They may be closer to ``almost there,'' and the optimal response might be manual context correction rather than full regeneration from scratch.

\subsection{Computational Efficiency and Resource Utilization}
Table~\ref{token_consumption} (Appendix) reports average token usage per successful task and an efficiency score (pass@5 divided by tokens consumed in thousands). We find that success rate and computational cost do not necessarily correlate: Grok 4.1 Fast is the most token-efficient model (67.50\% pass@5 at 182k tokens per success; efficiency 0.37), while Claude Opus achieves strong coverage (86.25\%) but at much higher cost (1,354k tokens per success) \cite{xai2025grok4.1Fast}.

Table \ref{token_consumption} demonstrates the average token usage per successful task completion, with an efficiency score (pass@5 divided by tokens consumed in thousands). Here we are able to see that success rate and computational cost do not necessarily correlate. Higher pass@5 does not mean that more tokens were consumed, and, likewise, high token budgets do not necessarily guarantee higher success.

Grok 4.1 Fast is the most token-efficient model in this set (67.50\% pass@5 at 182k tokens per success with an efficiency of 0.37). GPT 5.1 Codex Max is also notable because it is able to reach great metrics of success (85.00\%) while using much fewer tokens per success than other top-tier models (282k tokens; efficiency 0.30). On the other hand, we are able to see that some models spend tokens rather inefficiently. For example, DeepSeek V3.2 requires 1159k tokens per successful task while achieving 71.25\% pass@5, and Gemini 3 Pro Preview requires 983.1k tokens per success with 80.00\% pass@5. Even Claude Haiku 4.5, despite achieving 87.50\% pass@5, requires 726k tokens per successful task. We find this pattern to be consistent with iterative refinement loops that continue to generate and test patches without converging.

These results also help us distinguish between two different refinement styles: "fast" vs. "thorough." Models like Grok 4.1 Fast and DeepSeek R1 tend to be cheap when they succeed; however, they succeed much more rarely \cite{deepseekr1}. On the other hand, models such as Claude Haiku and GPT 5.2 succeed more often, but their successful runs are more expensive due to longer iterative trajectories. We should note that the low token counts of the weakest models (e.g., Llama 4 series) should not be read as a strength; instead, it  reflects early termination without resolution \cite{meta2025llama4maverick}.

From these results, we propose a \textbf{two-tier routing architecture} where a faster, efficient model attempts first pass, and a more thorough (but more expensive) model acts as fallback. In our framework, such a scheme could preserve high task coverage while reducing cost relative the most expensive model.

\subsection{Task-Level Specialization}

Our analysis reveals that aggregate pass@5 scores can be misleading: models with similar overall performance often exhibit dramatically different strengths at the task level. We identify multiple discriminative tasks where adjacently-ranked models show 100-point performance gaps, suggesting that specific capabilities (TypeScript async handling, Java web framework navigation, C/C++ memory management) matter more than global rankings for deployment decisions. For example, Event Callback System task-4 demonstrates task-level specialization: GPT-5.2 and Codex Max achieve 100\% pass@5 task success (all tests passing) while Claude Sonnet, Haiku, and Gemini have 0\% pass@5 task success, despite similar aggregate scores. These patterns demonstrate that task domain and framework-specific competence are critical factors for model selection. Detailed analysis including task-level heatmaps and discriminative performance tables is provided in Appendix~\ref{app:task_level_specialization}.

\subsection{Language and Domain-Specific Performance Patterns}

Our task-level analysis reveals distinct capabilities across models, with performance varying significantly by programming language and application domain. To avoid over-interpreting individual tasks, we aggregate results by repository tech stack (from dataset metadata) and report mean pass@5 per model within each repository; ranges indicate the minimum--maximum mean across repositories in that bucket.

Table~\ref{tab:model_specialization} (Appendix) summarizes these stack-specific patterns using mean pass@5 aggregated within each repository; ranges denote min-max of repository means within each bucket.

\textbf{C/C++ systems and low-level tooling.} Across the C/C++-heavy repositories (ESIM Management, Game Engine, Memory Profiling, and C++-based code analysis), Claude Sonnet and Claude Opus are consistently strong, with GPT-5.2 close behind (Table~\ref{tab:model_specialization}). These tasks stress pointer-level reasoning, stateful invariants, and build-system navigation (Make/CMake), where small mistakes often cause hard test failures.

\textbf{TypeScript/Node.js services.} GPT-5.2 leads on the TypeScript/Node.js repositories on average, but the specialization is clearest on discriminative tasks. For example, on Event Callback System task-4, only GPT-5.2 and GPT-5.1 Codex Max achieve 100\% success while most other models score 0\%, highlighting sensitivity to asynchronous control-flow and event-driven edge cases.

\textbf{Java web backends (lightweight frameworks).} SmartHub Operations Center uses Java with Javalin and exhibits lower overall success rates than the other repositories. However, we still observe sharp task-level divergence: on task-8, GPT-5.2, Gemini 3 Pro, Qwen3 Max, and Claude Opus reach 100\% while Codex Max is 0\% and Sonnet is 60\%, illustrating that framework-specific conventions and Java ecosystem navigation remain challenging and non-uniform across models.

\textbf{Python analysis pipelines.} On Network Traffic Analyzer, top models nearly saturate performance (Opus at 100\%, GPT-5.2 and Sonnet at 98\%), suggesting that data-parsing and analytics-style scripting tasks are comparatively well-covered by current frontier models in our setting.

These specialization patterns show that task domain and stack-specific competence are critical factors beyond aggregate pass@5 rankings, and they motivate routing strategies that condition on repository language and framework.

\subsection{Model Consistency and Variance Analysis}

Beyond mean success rates, predictability matters: the same task can go either way depending on the attempt. Appendix Table~\ref{tab:consistency} summarizes this directly. Claude Opus 4.5 is the most consistent model in our evaluation ($\sigma=0.027$), followed by Claude Sonnet 4.5 ($\sigma=0.045$), while DeepSeek V3.2 is the most variable ($\sigma=0.323$).

We also decompose variance into between-task versus within-task components (Appendix Table~\ref{tab:variance_decomp}). Opus exhibits the highest reliability (ICC=0.804, R=4.09), GPT 5.2 has moderate reliability (ICC=0.493, R=0.97) despite the highest pass@5, and DeepSeek V3.2 is unstable (ICC=0.331, R=0.50). In practice, predictability can matter as much as peak performance for IDE integration, since inconsistent assistance reduces trust even when average performance is competitive.

\subsection{Failure Modes}

Appendix~\ref{app:failure_taxonomy} provides the full failure taxonomy (definitions, detection criteria, and per-model breakdown). Here we summarize the dominant failure dynamics that explain why models with similar pass@5 can still feel very different in an IDE.

\begin{itemize}[noitemsep, topsep=0pt]
    \item \textbf{Early action dominates failure.} Among failed runs, the most common failure modes are Premature Editing (63.0\%), Thrashing/Backtracking (28.2\%), and Context Loss (27.6\%). Runs may exhibit multiple modes simultaneously, so percentages sum above 100\%.
    \item \textbf{Open-weight failures skew toward ``act too early.''} Open-weight and lightweight agents exhibit extremely high Premature Editing rates (80--95\% of their failed runs), suggesting they begin patching before they have a correct map of the codebase. This is consistent with failure trajectories where early edits trigger downstream instability rather than convergence.
    \item \textbf{Frontier failures skew toward non-convergence.} Several frontier and mid-tier models show disproportionate Context Loss and Thrashing when they fail (e.g., Claude Sonnet has 74.6\% Context Loss among its failed runs; Grok 4.1 Fast has 69.7\% Thrashing), indicating that failure often comes from unstable convergence under longer tool loops and not from total inability to implement the core fix.
    \item \textbf{Failure modes concentrate by stack.} Tool Call Failures are disproportionately concentrated in the Java web repository and the full-stack translator (together accounting for roughly 52\% of Tool Call Failures), while Syntax Error Loops are concentrated in the Python-heavy repositories (network-traffic-analyzer and code-quality-analyzer together account for roughly 82\% of Syntax Error Loops). This suggests that workflow brittleness and ``paper cuts'' (tool misuse, syntax churn) are not uniformly distributed across domains.
\end{itemize}

\subsection{Tool-Sequence Behavior and Workflow Patterns}

Appendix~\ref{app:tool_sequences} analyzes tool-call sequences and transition probabilities to characterize agent workflow inside the IDE. We find that tool behavior is not random; it follows a small number of repeatable patterns that help explain both success and failure.

\begin{itemize}[noitemsep, topsep=0pt]
    \item \textbf{Read--edit alternation is the core loop.} After \texttt{read\_file}, agents transition to \texttt{edit\_file} 37.0\% of the time, and after \texttt{edit\_file}, they return to \texttt{read\_file} 55.9\% of the time. This suggests iterative local reasoning rather than one-shot patching.
    \item \textbf{Tool usage is self-chaining.} Search and execution tools self-chain at high rates, reflecting refinement loops: \texttt{codebase\_search}$\rightarrow$\texttt{codebase\_search} 81.5\%, \texttt{run\_terminal\_cmd}$\rightarrow$\texttt{run\_terminal\_cmd} 66.2\%, \texttt{list\_dir}$\rightarrow$\texttt{list\_dir} 63.1\%, and \texttt{grep\_search}$\rightarrow$\texttt{grep\_search} 59.3\%.
    \item \textbf{Edits are rarely followed immediately by tests.} Only 8.0\% of edits transition directly to \texttt{run\_terminal\_cmd}, implying that many agents re-check context before testing (often by reading or searching first).
    \item \textbf{These signatures line up with the failure taxonomy.} Short-circuiting the read phase is consistent with Premature Editing, while repeated self-chaining without a stabilizing read/test cycle aligns with Thrashing/Backtracking and longer-horizon Context Loss. For full plots and additional breakdowns (transition matrix, top bigrams/trigrams, read-to-edit ratios, and tools-before-first-edit), see Appendix~\ref{app:tool_sequences}.
\end{itemize}

\subsection{Iteration Efficiency and Debugging Dynamics}

Figure~\ref{fig:cum_iter} (Appendix) shows how task success accumulates as agents spend more iterations inside the conversational/tool loop. To understand why some models gain quickly while others stall, we complement pass@k with an iteration-level efficiency analysis (Appendix~\ref{app:iteration_efficiency}) that decomposes each run into: (i) time to first successful edit, and (ii) how much of the trajectory is spent on productive iterations (successful edits), exploration iterations (tool use without edit attempts), and non-productive iterations (failed edit attempts) (Table~\ref{tab:iteration_efficiency}; Figures~\ref{fig:iter_cdf}--\ref{fig:iter_stacked}).

Across all runs, agents use a median of 21 iterations per run and reach their first successful edit at median iteration 7, indicating that when progress happens it typically happens early. However, models differ sharply in how they allocate iterations:

\begin{itemize}[noitemsep, topsep=0pt]
    \item \textbf{Early traction vs.\ delayed traction.} Fast starters (xAI Grok 4.1 Fast at 4.4 iterations; Claude Opus 4.5 at 5.2) secure an initial foothold quickly, while slower starters (DeepSeek V3.2 at 13.1; GPT 5.1 Codex Max at 11.8) spend substantially longer before their first successful edit (Figure~\ref{fig:iter_cdf}).
    \item \textbf{Exploration dominates, but strategy varies.} The median model spends 82.3\% of iterations in exploration, yet this ranges from highly deliberate behavior (Opus: 99.1\% exploration) to more action-biased behavior (Grok 4.1 Fast: 58.4\%) (Figure~\ref{fig:iter_stacked}).
    \item \textbf{Wasted iterations expose tool-loop brittleness.} Several strong models maintain near-zero non-productive rates (e.g., GPT 5.2: 1.0\%), whereas Cohere Command R+ wastes 18.0\% of iterations on failed edits, consistent with brittle edit attempts and insufficient pre-edit validation (Figure~\ref{fig:iter_stacked}) \cite{commandr+082024}.
    \item \textbf{Iteration budget interaction.} Some models frequently run to the 100-iteration cap (e.g., Opus shows extreme spread), while others terminate far earlier (e.g., Llama 4 and Grok Code Fast) (Figure~\ref{fig:iter_boxplot}) \cite{xai2025grokcodefastl}. This suggests that ``more iterations'' is not uniformly beneficial; efficiency and convergence behavior differ by model.
\end{itemize}

These iteration-level signatures complement pass@k: two models can achieve similar aggregate success while exhibiting different efficiency profiles (time-to-first-edit, exploration fraction, and edit failure rate), which directly impacts developer experience (latency, cost, and predictability) in IDE-agent deployments.

\subsection{Deployment Implications}

The results from our benchmark provide concrete guidance for deploying IDE agents in production environments. We present recommendations organized by deployment objective: maximizing success rate, minimizing cost, ensuring predictable behavior, and implementing multi-model strategies.

\textbf{Production-Ready Threshold (85\% pass@5).} Our retry benefit analysis reveals a natural threshold at 85\% pass@5 that separates production-ready models from those requiring multiple attempts. Models above this threshold (GPT 5.2 at 95\%, Claude Sonnet at 88.75\%, Claude Haiku at 87.50\%, Claude Opus at 86.25\%, GPT 5.1 Codex Max at 85\%) show minimal gains between pass@1 and pass@5 (1.25--11.25 points), indicating stable, first-attempt success. Models below 85\% exhibit dramatically higher retry benefits—DeepSeek V3.2 gains 40 points (31.25\% $\rightarrow$ 71.25\%), Grok 4.1 Fast gains 32.5 points (35\% $\rightarrow$ 67.5\%)—reflecting inconsistent, iteration-dependent behavior unsuitable for production where developers expect deterministic results.

\textbf{Single-Model Deployments by Objective.} For maximum task resolution regardless of cost, GPT 5.2 (95.00\% pass@5) represents the optimal choice. For cost-sensitive deployments, Grok 4.1 Fast achieves the highest efficiency score (0.37, computed as pass@5 / tokens\_per\_success\_in\_thousands) while maintaining 67.50\% coverage. Production environments requiring consistent, predictable behavior favor Claude Sonnet ($\sigma=0.045$, ICC=0.723) or Claude Opus ($\sigma=0.027$, ICC=0.804) over higher-variance alternatives like Gemini 3 Pro ($\sigma=0.191$) or DeepSeek V3.2 ($\sigma=0.323$). For first-attempt reliability (pass@1), Claude Sonnet leads at 87.50\%, followed by GPT 5.2 at 85.00\%.

\textbf{Language-Specific Routing.} Task-level specialization suggests stack-aware routing strategies: TypeScript/Node.js async tasks to GPT 5.2 or GPT 5.1 Codex Max (both achieve 100\% on Event Callback System task-4 where Claude models and Gemini score 0\%), Java web backends (Javalin) to GPT 5.2 or Gemini 3 Pro (both reach 100\% on SmartHub Operations Center task-8), Python data/analysis tasks to Claude Opus, GPT 5.2, or Claude Sonnet (near-saturating performance on Network Traffic Analyzer), and C/C++ systems programming to GPT 5.2, GPT 5.1 Codex Max, or Claude Sonnet (all achieve 100\% on Memory Profiling task-8).

\textbf{Two-Tier Architecture Strategies.} Model similarity analysis (Appendix Table~\ref{tab:model_similarity}) informs multi-model deployments:

\begin{enumerate}
    \item \textbf{Fast-then-thorough:} Deploy Grok 4.1 Fast for initial attempts (covers 67.5\% of tasks at high efficiency 0.37), falling back to Claude Haiku for unresolved cases. Empirically, the union solves 71/80 tasks (88.75\%), while reducing computational costs for the 67.5\% of tasks that Grok resolves without requiring the more expensive fallback.
    
    \item \textbf{High-coverage pairing:} GPT-5.2 for initial attempts (95\% coverage), falling back to Claude Sonnet for failures. The Jaccard index of 0.909 indicates high overlap (90.9\% of tasks solved by either are solved by both); empirically, the union solves 77/80 tasks (96.25\%), with incremental gains concentrated in the Java web repository where success is less uniform across models.
    
    \item \textbf{Reliability-aware pairing:} Claude Opus 4.5 (ICC=0.804, highest reliability) for production tasks requiring deterministic behavior, with GPT 5.2 (ICC=0.493, moderate reliability but 95\% pass@5) as fallback for tasks where Opus reaches iteration limits. This leverages Opus's predictability while achieving high overall coverage.
\end{enumerate}

\textbf{Avoiding Redundancy.} Claude Haiku and Claude Sonnet exhibit 93.2\% Jaccard overlap, the highest among all pairs, indicating heavy redundancy; their union solves 73/80 tasks (91.25\%). Similarly, GPT-5.2 shows 89.6\% overlap with Claude Haiku and 89.5\% with Codex Max. For cost-effective coverage, pair models with lower Jaccard indices (e.g., Grok 4.1 Fast + Claude Haiku), or accept high-overlap pairings only when optimizing for peak coverage rather than diversity.

\textbf{Consistency vs. Exploratory Contexts.} Production IDE assistance requiring predictable behavior favors Claude Opus or Claude Sonnet (low $\sigma$, high ICC). GPT 5.2 offers the highest coverage (95\% pass@5) but exhibits only moderate reliability (ICC=0.493), making it better suited for settings that can tolerate more attempt-to-attempt variability. Research or experimental contexts can leverage Gemini 3 Pro's higher variance ($\sigma=0.191$, ICC=0.567) to explore diverse solution approaches, accepting occasional inconsistency for broader exploration.

\section{Conclusion and Future Work}
\textbf{Performance stratification and ceilings.} There is a substantial gap between the frontier tier, led by GPT 5.2 at 95\% pass@5, and the next cluster at 85--89\% (Claude Sonnet, Claude Haiku, Claude Opus, GPT 5.1 Codex Max). Open-weight alternatives span a wide range: Qwen3 Max and Qwen3 Coder are competitive at 76.25\% and 75.00\% pass@5, DeepSeek V3.2 reaches 71.25\%, while the lowest tier remains below 50\%. At the same time, the top four models are tightly clustered, suggesting that on this benchmark many state-of-the-art models are operating near a shared ceiling. Importantly, we identify several tasks that none of the top models consistently solve, providing concrete examples of current limitations under iterative tool-based debugging.

\textbf{Near misses and the limits of binary evaluation.} We observe repeated cases where models pass the vast majority of tests (e.g., 11/12 on Event Callback System task-4) yet receive zero credit under pass@5. This is not an edge case; it is a common failure mode where the agent gets the ``substance'' right but misses strict specification details. For deployment, this matters because these outputs may be closer to ``minimal correction'' than ``start over.''

\textbf{Task-level specialization dominates aggregate rankings.} We find multiple tasks with 100-point gaps between models that are adjacent (or tied) in overall pass@5. Some tasks filter strongly for framework-specific competence (e.g., Java web framework navigation in SmartHub Operations Center, TypeScript async patterns in Event Callback System, or C/C++ memory safety in Memory Profiling). This implies that global leaderboards are a weak proxy for performance on any given real workload.

\textbf{Efficiency and reliability vary independently.} Token usage per success varies by over an order of magnitude across models (131k to 1,354k tokens), and outcome consistency varies by 12.0 times  between the most stable (Claude Opus 4.5, $\sigma=0.027$) and least stable (DeepSeek V3.2, $\sigma=0.323$) evaluated models. First-attempt success rates (pass@1) reveal that Claude Sonnet 4.5 maintains 87.50\% pass@1 performance (with only a 1.25-point gap to its 88.75\% pass@5), while DeepSeek V3.2 shows only 31.25\% pass@1 despite achieving 71.25\% pass@5, indicating high variability. This suggests that the ``best'' model depends heavily on the deployment objective: minimizing cost, maximizing coverage, or maximizing predictability. In practice, these profiles motivate portfolio or routing-based deployments rather than a single-model strategy.

IDE-Bench provides a full-stack benchmark for evaluating LLMs as containerized IDE agents and shows that frontier models can solve a large fraction of real-world engineering tasks. We evaluate whether an agent can reason, navigate, and use tools inside a containerized environment resembling real software engineering practice. The benchmark consists of 80 multi-file tasks spanning eight domains (systems programming in C/C++, enterprise Java web applications (Javalin/Thymeleaf), web services in TypeScript/Node.js, and data processing in Python, among others), and it measures both task-level success (pass@k) and finer-grained signals such as per-test pass rate, iteration trajectories, token usage, and outcome variance. Our evaluation revealed clear performance ceilings in specification compliance, reliability, and domain coverage. We find that our benchmark analysis indicates that single-number rankings are not enough, as LLM deployment evaluation must account for task specialization, cost, and consistency. We hope IDE-Bench serves both as a practical guide for current IDE integrations and as a set of concrete targets for improving the next generation of software engineering agents.

\clearpage
\section*{Impact Statement}

``This paper presents work whose goal is to advance the field of Machine
Learning. There are many potential societal consequences of our work, none
which we feel must be specifically highlighted here.''



\bibliography{example_paper}
\bibliographystyle{icml2026}

\ifincludeappendix
\newpage
\appendix
\onecolumn
\raggedbottom

\noindent\textbf{Appendix roadmap.} For convenience, we group supplementary material into four blocks: (i) supplementary tables referenced in the main text; (ii) behavioral analyses derived from run logs; (iii) task-level breakdowns and additional figures; and (iv) reproducibility and access details.\section{Supplementary Tables}

\subsection{Near-Miss Examples}
    \begin{table}[htbp]
        \centering
        \caption{\textbf{Near-Miss Examples}: Tasks with high test pass rates yet 0\% pass@5. Event Callback System (ECS), Cross-Lingual Document Translator (CLDT)}
        \label{tab:near_miss}
        \small
        \begin{tabular}{lccc}
        \toprule
        \textbf{Model} & \textbf{Dataset} & \textbf{Task} & \textbf{Test Pass} \\
        \midrule
        Claude Opus 4.5 & CLDT & task-7 & 92.3\% \\
        Claude Sonnet 4.5 & ECS & task-4 & 91.7\% \\
        Gemini 3 Pro & ECS & task-4 & 91.7\% \\
        Claude Opus 4.5 & ECS & task-4 & 91.7\% \\
        Claude Sonnet 4.5 & ECS & task-10 & 90.0\% \\
        \bottomrule
        \end{tabular}
    \end{table}

\subsection{Token Consumption}
    \begin{table}[htbp]
    \centering
    \caption{\textbf{Token Consumption for Successful Tasks by Model} Values represent average tokens consumed per successfully completed task, in thousands. Models are sorted by descending efficiency score. pass@k reported with k=5.}
    \label{tab:token_consumption_condensed}
    \small
    \begin{tabular}{lccc}
    \toprule
    \textbf{Model} & \textbf{pass@5} & \textbf{Avg Tokens} & \textbf{Efficiency} \\
     & \textbf{(\%)} & \textbf{(k)} & \textbf{Score\textsuperscript{†}} \\
    \midrule
    Grok 4.1 Fast & 67.50 & 181.7 & 0.37 \\
    GPT 5.1 Codex Max & 85.00 & 282.2 & 0.30 \\
    DeepSeek R1 0528 & 46.25 & 167.7 & 0.28 \\
    Grok Code Fast 1 & 32.50 & 166.1 & 0.20 \\
    GPT 5.2 & 95.00 & 648.2 & 0.15 \\
    Qwen3 Max & 76.25 & 519.8 & 0.15 \\
    Claude Sonnet 4.5 & 88.75 & 663.3 & 0.13 \\
    Claude Haiku 4.5 & 87.50 & 726.7 & 0.12 \\
    Qwen3 Coder & 75.00 & 694.1 & 0.11 \\
    Gemini 3 Pro Preview & 80.00 & 983.1 & 0.08 \\
    Llama 4 Maverick & 8.75 & 131.1 & 0.07 \\
    Claude Opus 4.5 & 86.25 & 1354.1 & 0.06 \\
    DeepSeek V3.2 & 71.25 & 1159.0 & 0.06 \\
    Command-R+ 08 2024 & 7.50 & 244.5 & 0.03 \\
    Llama 4 Scout & 6.25 & 243.5 & 0.03 \\
    \bottomrule
    \multicolumn{4}{l}{\textsuperscript{†}Efficiency Score = pass@5 / (Tokens/1000)} \\
    \end{tabular}
    \label{token_consumption}
\end{table}

\subsection{Language/Stack Specialization (Table~\ref{tab:model_specialization})}
    \begin{table}[htbp]
        \centering
    \caption{\textbf{Model Specialization Patterns}: Stack-specific strengths using mean pass@5 aggregated within each repository. Ranges denote min--max of repository means within the bucket.}
        \label{tab:model_specialization}
    \footnotesize
    \setlength{\tabcolsep}{4pt}
    \renewcommand{\arraystretch}{1.15}
    \begin{tabularx}{\textwidth}{L{3.1cm} Y Y}
        \toprule
    \textbf{Stack / Domain} & \textbf{Repositories} & \textbf{Top performers (mean pass@5; range)} \\
        \midrule
    C/C++ systems \& tooling &
    \begin{tabular}[t]{@{}l@{}}
    ESIM Management System (C)\\
    Game Engine Service (C++)\\
    Memory Profiling App (C++)\\
    Code Quality Analyzer (C++/Python)
    \end{tabular} &
    \begin{tabular}[t]{@{}l@{}}
    Sonnet 90.5\% (76--96)\\
    Opus 89.5\% (84--94)\\
    GPT-5.2 88.5\% (76--96)
    \end{tabular} \\
    TypeScript/Node.js services &
    \begin{tabular}[t]{@{}l@{}}
    Event Callback System (TypeScript)\\
    Cross-Lingual Document Translator (Node/Express)
    \end{tabular} &
    \begin{tabular}[t]{@{}l@{}}
    GPT-5.2 79.0\% (74--84)\\
    Opus 77.0\% (76--78)\\
    Sonnet 76.0\% (68--84)
    \end{tabular} \\
    Python data/analysis &
    Network Traffic Analyzer (Python) &
    \begin{tabular}[t]{@{}l@{}}
    Opus 100\%\\
    GPT-5.2 98\%\\
    Sonnet 98\%
    \end{tabular} \\
    Java web (Javalin) &
    SmartHub Operations Center (Java + Thymeleaf) &
    \begin{tabular}[t]{@{}l@{}}
    GPT-5.2 64\%\\
    Gemini 58\%\\
    Sonnet 54\%
    \end{tabular} \\
        \bottomrule
    \end{tabularx}
    \end{table}

\subsection{Consistency and Variance Decomposition}
We report two complementary views of consistency across repeated attempts on the same task. First, we compute a simple consistency score: for each model and each task, we compute the standard deviation of the task's test pass rate across $k=5$ attempts, then average this standard deviation across tasks. Lower $\sigma$ means that rerunning the same task yields more similar outcomes.

Second, we decompose variance into \textbf{between-task} and \textbf{within-task} components. Between-task variance captures differences driven by task difficulty; within-task variance captures attempt-to-attempt drift on the same task. We summarize this decomposition using two metrics: \textbf{ICC} (Intraclass Correlation Coefficient), \( \mathrm{ICC}=\sigma^2_{\text{between}} / (\sigma^2_{\text{between}}+\sigma^2_{\text{within}}) \), and \textbf{Reliability Ratio}, \( R=\sigma^2_{\text{between}} / \sigma^2_{\text{within}} \). ICC is bounded [0,1]; higher values mean outcomes are more task-determined rather than sample-dependent. The ratio is unbounded and provides a direct task-to-attempt variance comparison.

    \begin{table}[htbp]
        \centering
        \caption{\textbf{Model Consistency}: Lower standard deviation indicates more predictable outcomes.}
        \label{tab:consistency}
        \small
        \begin{tabular}{lcc}
        \toprule
        \textbf{Model} & \textbf{Std Dev ($\sigma$)} & \textbf{Variance level} \\
        \midrule
        Claude Opus 4.5 & 0.027 & Lowest \\
        Claude Sonnet 4.5 & 0.045 & Very low \\
        Claude Haiku 4.5 & 0.091 & Low \\
        GPT 5.2 & 0.092 & Low \\
        GPT 5.1 Codex Max & 0.101 & Low-moderate \\
        Qwen3 Max & 0.124 & Moderate \\
        Qwen3 Coder & 0.157 & Moderate-high \\
        Gemini 3 Pro Preview & 0.191 & High (7.1 times Claude Opus) \\
        Grok 4.1 Fast & 0.257 & Very high (9.5 times Claude Opus) \\
        DeepSeek V3.2 & 0.323 & Highest (12.0 times Claude Opus) \\
        \bottomrule
        \end{tabular}
    \end{table}

    \begin{table}[htbp]
        \centering
        \caption{\textbf{Variance Decomposition \& Reliability}: ICC (Intraclass Correlation) is bounded [0,1] and standard in ML; Reliability Ratio is unbounded and shows task-to-attempt variance ratio.}
        \label{tab:variance_decomp}
        \small
        \begin{tabular}{@{}lcccc@{}}
        \toprule
        \textbf{Model} & \textbf{Between-Task} & \textbf{Within-Task} & \textbf{ICC} & \textbf{Reliability} \\
                       & \textbf{Variance} & \textbf{Variance} & & \textbf{Ratio} \\
        \midrule
        Claude Opus 4.5 & 0.0439 & 0.0107 & 0.804 & 4.09 \\
        Claude Sonnet 4.5 & 0.0373 & 0.0143 & 0.723 & 2.61 \\
        Qwen3 Max & 0.1221 & 0.0473 & 0.721 & 2.58 \\
        Qwen3 Coder & 0.1173 & 0.0591 & 0.665 & 1.99 \\
        Claude Haiku 4.5 & 0.0547 & 0.0288 & 0.655 & 1.90 \\
        GPT 5.1 Codex Max & 0.0764 & 0.0405 & 0.654 & 1.89 \\
        Gemini 3 Pro Preview & 0.1046 & 0.0800 & 0.567 & 1.31 \\
        Command-R+ 08 2024 & 0.0304 & 0.0304 & 0.500 & 1.00 \\
        GPT 5.2 & 0.0372 & 0.0383 & 0.493 & 0.97 \\
        DeepSeek V3.2 & 0.0662 & 0.1335 & 0.331 & 0.50 \\
        \bottomrule
        \end{tabular}
    \end{table}

\subsection{Model Similarity}
    \begin{table}[htbp]
        \centering
        \caption{\textbf{Model Similarity Matrix (Jaccard Index)}: Selected pairs showing high and low similarity.}
        \label{tab:model_similarity}
        \small
        \begin{tabular}{@{}lcc@{}}
        \toprule
        \textbf{Model Pair} & \textbf{Jaccard} & \textbf{Interpretation} \\
        \midrule
        \multicolumn{3}{@{}l}{High Similarity (Redundant Pairs)} \\
        Claude Haiku $\leftrightarrow$ Claude Sonnet & 0.932 & 93.2\% overlap \\
        GPT-5.2 $\leftrightarrow$ Claude Haiku & 0.896 & 89.6\% overlap \\
        Codex Max $\leftrightarrow$ GPT-5.2 & 0.895 & 89.5\% overlap \\
        \midrule
        \multicolumn{3}{@{}l}{Low Similarity (Complementary Pairs)} \\
        Llama 4 Scout $\leftrightarrow$ Claude Sonnet & 0.070 & 7.0\% overlap \\
        Llama 4 Scout $\leftrightarrow$ Claude Haiku & 0.071 & 7.1\% overlap \\
        \bottomrule
        \end{tabular}
    \end{table}

We quantify model similarity using the Jaccard index, which measures overlap in the sets of tasks solved under pass@5. Similarity is a deployment diagnostic: it tells you whether adding a second model increases coverage or mostly adds cost. High Jaccard overlap indicates redundancy (two models tend to solve the same tasks), so a two-model portfolio yields limited incremental coverage and is best justified by latency/cost tradeoffs or different reliability profiles. Lower overlap suggests complementary coverage, which is valuable for fallback strategies: when the primary model fails a task, the fallback has a better chance of solving a different subset.

In our setting, the strongest redundancy is between Claude Haiku and Claude Sonnet (Jaccard 0.932). Even GPT-5.2 and Claude Sonnet exhibit high overlap (Jaccard 0.909), with union coverage of 77/80 tasks (96.25\%). This implies that most two-tier gains come from targeted routing (e.g., by repository/language) rather than naive pairing.

\clearpage
\section{Behavioral Analyses: Iteration Efficiency}
\label{app:iteration_efficiency}

In practice, two models can tie (or nearly tie) on pass@5 and still feel totally different in an IDE, as one may converge quickly while the other continues to go through iterations without getting closer to a passing test run. Here we break each run into three iteration types: productive (successful edits), exploration (tool use without edit attempts), and non-productive (failed edit attempts). These help us see where iteration budget actually goes.

\subsection{Methodology}

We analyze iteration efficiency by combining data from \texttt{trajectory.jsonl} (agent actions per iteration) and \texttt{edits.jsonl} (edit outcomes) files. For each run, we extract:

\begin{itemize}[noitemsep, topsep=0pt]
    \item \textbf{Total iterations}: Number of reasoning cycles executed
    \item \textbf{Time to first success}: Iteration number of first successful edit
    \item \textbf{Productive iterations}: Iterations containing $\geq$1 successful edit
    \item \textbf{Exploration iterations}: Iterations with tool calls but no edit attempts (information gathering)
    \item \textbf{Non-productive iterations}: Iterations with edit attempts where all edits failed (e.g., rejected for syntax errors)
\end{itemize}

\textbf{Categorization Logic}: We assign each iteration to exactly one category, based on what the agent attempted in that step. If the iteration contains at least one successful edit, it is labeled \textbf{productive}. If the iteration contains edit attempts but none of them succeed, it is labeled \textbf{non-productive}. If the iteration contains no edit attempts (only reading/searching/testing), it is labeled \textbf{exploration}. This makes the categories mutually exclusive and collectively exhaustive, so the counts of productive, non-productive, and exploration iterations sum to the total iterations in the run.

\subsection{Key Findings}

Across our runs, agents averaged 30.7 iterations per run (median: 21) and achieved their first successful edit at iteration 8.5 (median: 7). However, models allocate iterations very differently. Table~\ref{tab:iteration_efficiency} summarizes per-model metrics.

\begin{table}[h]
    \centering
        \small
\caption{Iteration efficiency metrics by model. Percentages show proportion of total iterations.}
\label{tab:iteration_efficiency}
\begin{tabular}{@{}lrrrrr@{}}
        \toprule
\textbf{Model} & \textbf{Total} & \textbf{First} & \textbf{Prod.} & \textbf{Expl.} & \textbf{Non-P.} \\
 & \textbf{Iters} & \textbf{Succ.} & \textbf{(\%)} & \textbf{(\%)} & \textbf{(\%)} \\
        \midrule
xAI Grok 4.1 Fast     & 16.3 & 4.4 & 32.7 & 58.4 & 8.9 \\
xAI Grok Code Fast 1  & 10.9 & 7.6 & 11.7 & 83.5 & 4.9 \\
Llama 4 Maverick      & 11.4 & 8.2 & 11.4 & 83.5 & 5.1 \\
DeepSeek R1 0528      & 12.1 & 5.8 & 21.0 & 70.0 & 9.0 \\
Llama 4 Scout         & 15.9 & 8.3 & 13.1 & 80.3 & 6.6 \\
\midrule
OpenAI GPT 5.1 Codex  & 25.7 & 11.8 & 10.9 & 87.4 & 1.7 \\
Claude Sonnet 4.5     & 33.5 & 8.0 & 18.2 & 80.2 & 1.6 \\
Claude Haiku 4.5      & 32.8 & 8.1 & 11.0 & 87.0 & 2.0 \\
DeepSeek V3.2         & 33.3 & 13.1 & 11.5 & 85.0 & 3.5 \\
Cohere Command R+     & 36.1 & 5.4 & 2.7 & 79.3 & 18.0 \\
\midrule
Qwen3 Coder           & 37.3 & 11.5 & 14.9 & 79.9 & 5.2 \\
OpenAI GPT 5.2        & 38.8 & 8.1 & 16.6 & 82.3 & 1.0 \\
Qwen3 Max             & 42.8 & 8.9 & 13.4 & 84.0 & 2.7 \\
Gemini 3 Pro Preview  & 44.4 & 9.9 & 14.9 & 83.6 & 1.6 \\
Claude Opus 4.5       & 65.5 & 5.2 & 0.9 & 99.1 & 0.1 \\
        \bottomrule
        \end{tabular}
    \end{table}

\textbf{Key Insights}:
\begin{itemize}[noitemsep, topsep=0pt]
    \item \textbf{Exploration dominates.} Models spend 58--99\% of iterations in exploration (median: 82.3\%), with Claude Opus 4.5 the most deliberate at 99.1\%.
    \item \textbf{Fast starters exist, but they are not always the best overall.} xAI Grok 4.1 Fast reaches first success fastest (4.4 iterations; minimum across all models) and has the highest productive share (32.7\%).
    \item \textbf{Wasted iterations are where workflows break.} Cohere Command R+ spends 18.0\% of iterations on non-productive failed edits, versus 1.0\% for GPT 5.2 and 2.7\% for Qwen3 Max.
\end{itemize}

\subsubsection{Time to First Success}

Figure~\ref{fig:iter_cdf} shows the cumulative distribution of iterations needed to achieve the first successful edit. Models with steeper curves make progress faster.

\begin{figure}[h]
        \centering
\includegraphics[width=0.95\columnwidth]{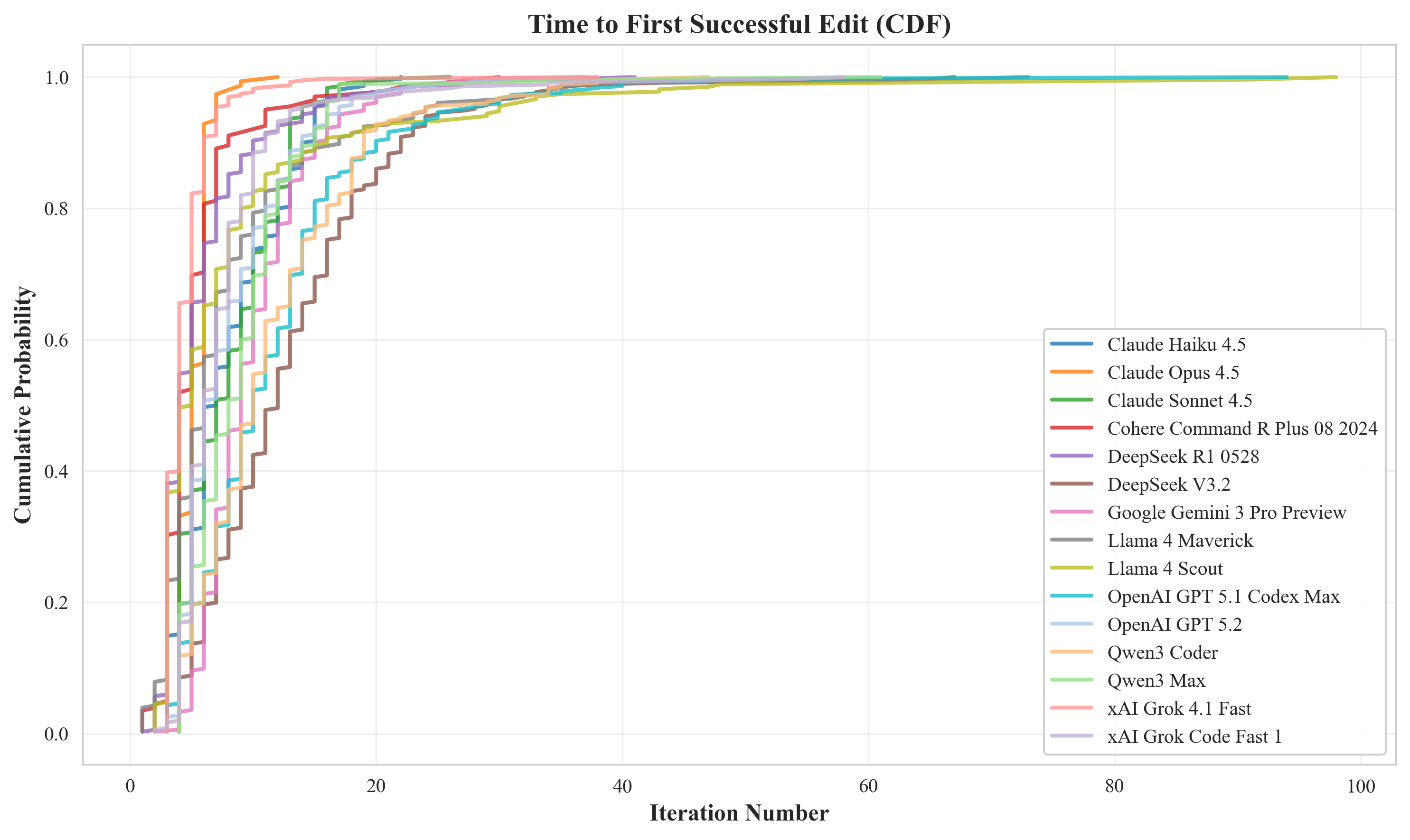}
\caption{Cumulative distribution function (CDF) of iterations to first successful edit by model. Steeper curves indicate faster initial progress. Only includes runs with at least one successful edit.}
\label{fig:iter_cdf}
    \end{figure} 

\textbf{Key Observations}:
\begin{itemize}[noitemsep, topsep=0pt]
    \item \textbf{Fast starters}: xAI Grok 4.1 Fast (4.4 iterations) and Claude Opus 4.5 (5.2 iterations) achieve first success 37\% and 26\% faster than the overall median (7 iterations), suggesting superior initial problem understanding.
    \item \textbf{Median performance}: 50\% of all runs achieve first success within 7 iterations, indicating that most tasks permit rapid initial progress. The mean (8.5) exceeds the median due to long-tail outliers.
    \item \textbf{Slow starters}: DeepSeek V3.2 (13.1 iterations) and GPT 5.1 Codex (11.8 iterations) require 1.7-1.9 times  the median before first success, potentially indicating more conservative exploration strategies or initial plan formulation overhead.
\end{itemize}

\subsubsection{Iteration Count Distribution}

Figure~\ref{fig:iter_boxplot} displays the distribution of total iterations across models using box plots.

\begin{figure}[h]
        \centering
\includegraphics[width=0.95\columnwidth]{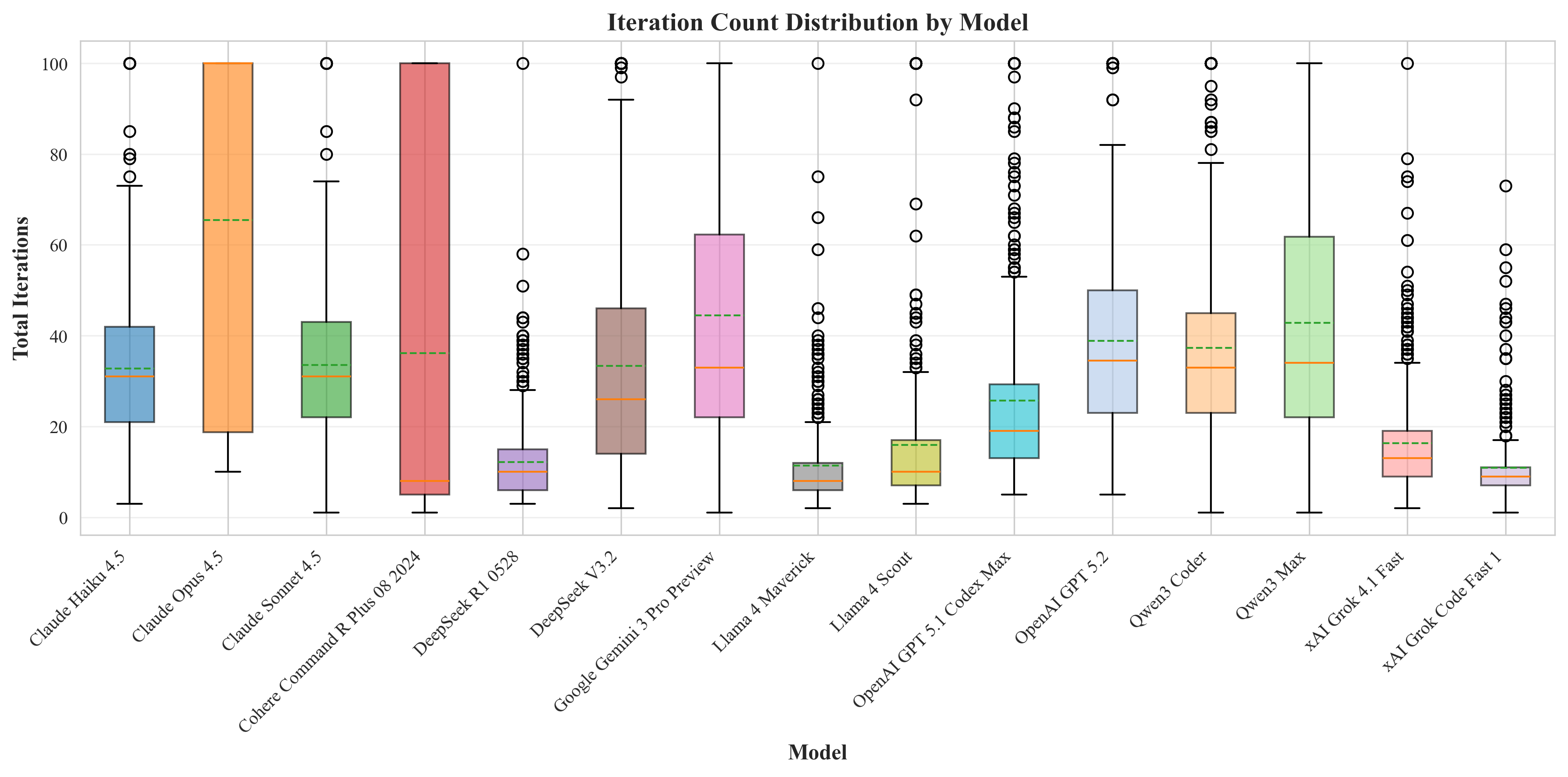}
\caption{Distribution of total iterations by model. Box shows quartiles (Q1, median, Q3), whiskers extend to 1.5 times IQR, and points show outliers. Lower boxes indicate more efficient completion.}
\label{fig:iter_boxplot}
    \end{figure}

\textbf{Notable Patterns}:
\begin{itemize}[noitemsep, topsep=0pt]
    \item \textbf{Iteration minimalists}: Llama 4 models (Maverick: 11.4, Scout: 15.9) and xAI Grok Code (10.9) complete tasks with 48-62\% fewer iterations than the dataset median (21), demonstrating great efficiency or earlier convergence/termination, for the tasks that they did pass.
    \item \textbf{High variance models}: Claude Opus 4.5 exhibits extreme spread (median: 100, mean: 65.5), suggesting bimodal behavior likely hitting max-iteration limits on difficult tasks while completing simple ones quickly.
    \item \textbf{Outlier analysis}: Runs exceeding 100 iterations (typically max-limit timeouts) comprise roughly 15\% of Opus 4.5 runs, but $<$5\% for Grok/Llama models, indicating different stopping criteria or persistence strategies.
\end{itemize}

\subsubsection{Iterations vs. Success}

Figure~\ref{fig:iter_scatter} examines the relationship between total iterations and task success.

\begin{figure}[h]
    \centering
\includegraphics[width=0.95\columnwidth]{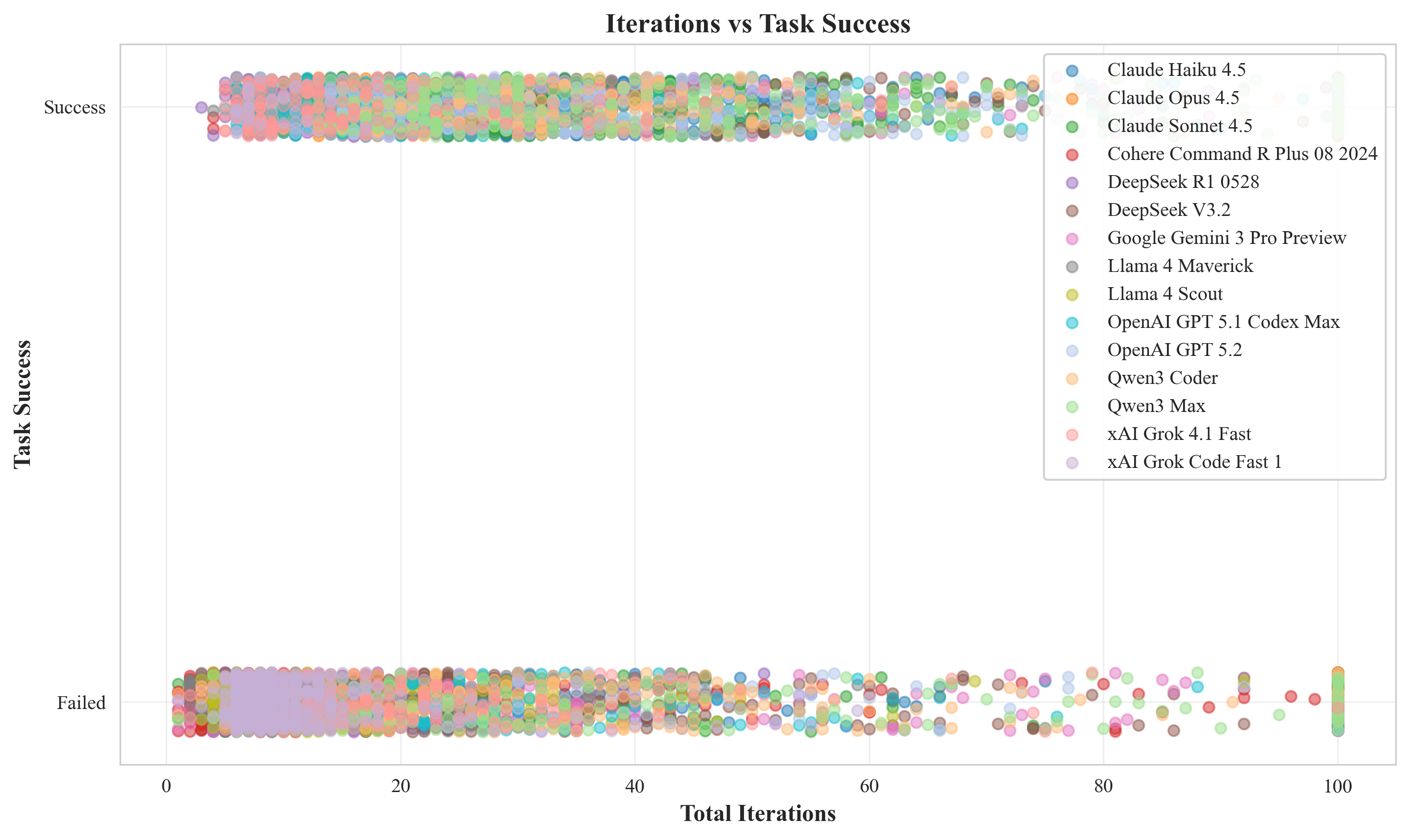}
\caption{Scatter plot of total iterations vs. task success (jittered for visibility). Each point represents one run. Successful runs (top) and failed runs (bottom) show distinct iteration patterns.}
\label{fig:iter_scatter}
\end{figure}

\textbf{Analysis}:
\begin{itemize}[noitemsep, topsep=0pt]
    \item \textbf{Success concentration}: Successful runs concentrate in the 5-35 iteration range, with peak density around 10-20 iterations. This suggests an ``optimal window'' where agents have gathered sufficient context without exhausting iteration budgets.
    \item \textbf{Bimodal failure}: Failed runs exhibit two distinct modes: (1) early failures ($<$10 iterations), likely immediate recognition of task infeasibility or critical errors, and (2) timeout failures ($>$80 iterations), agents unable to converge despite extensive attempts.
    \item \textbf{Efficiency champions}: xAI Grok and Llama 4 models achieve success at significantly lower iteration counts (median 8-9) compared to Claude Opus 4.5 (median 100), suggesting fundamentally different approaches; aggressive convergence vs. exhaustive exploration.
\end{itemize}

\subsubsection{Productivity Breakdown}

Figure~\ref{fig:iter_stacked} breaks down mean iterations into productive, exploratory, and non-productive categories.

\begin{figure}[h]
    \centering
\includegraphics[width=0.95\columnwidth]{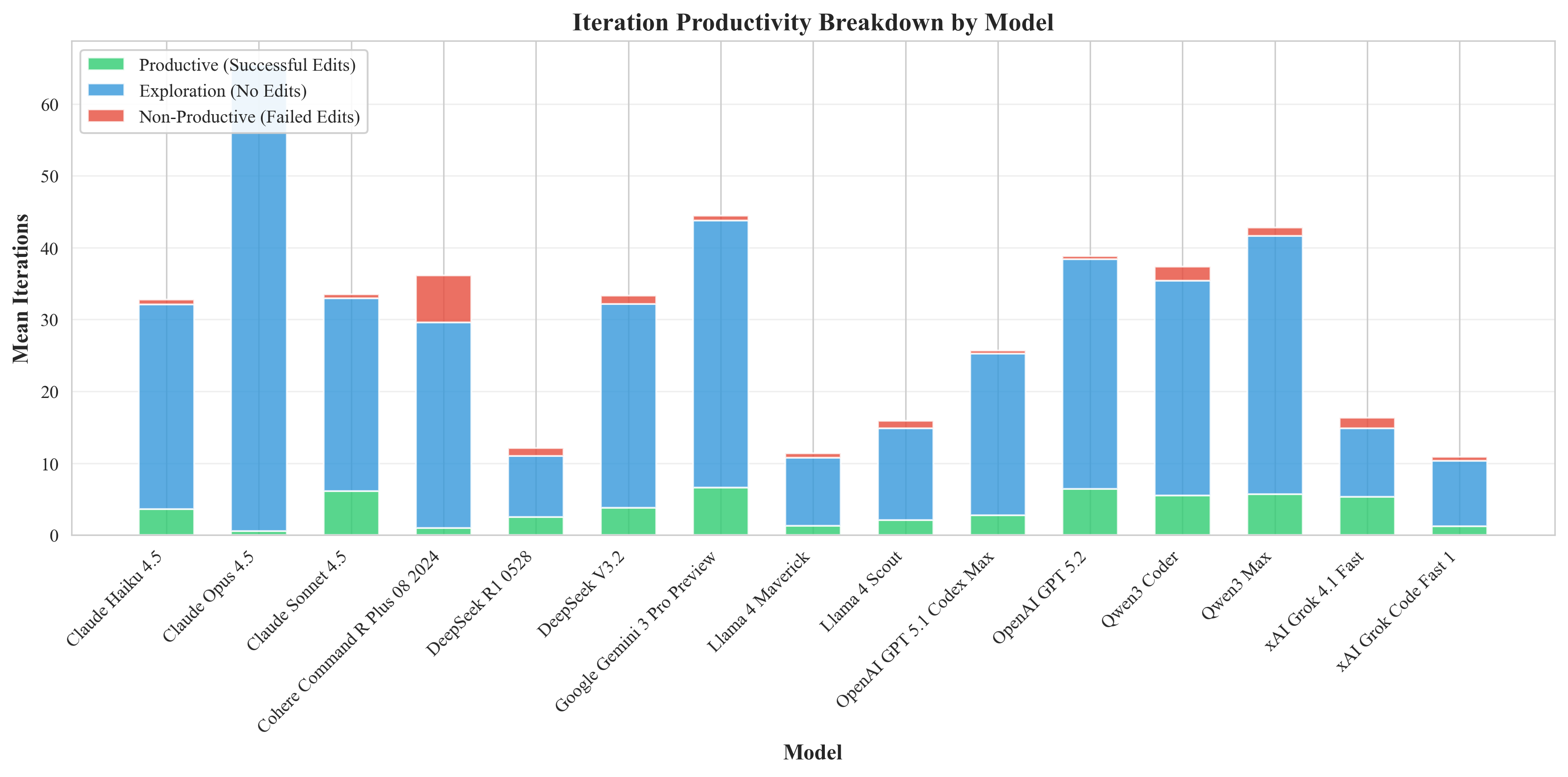}
\caption{Stacked bar chart showing iteration productivity breakdown by model. Green: productive iterations (successful edits), Blue: exploration iterations (no edit attempts), Red: non-productive iterations (failed edits).}
\label{fig:iter_stacked}
\end{figure}

\textbf{Key Insights}:
\begin{itemize}[noitemsep, topsep=0pt]
    \item \textbf{Exploration spectrum}: Claude Opus 4.5 dedicates 99.1\% of iterations to exploration (extreme deliberation), while xAI Grok 4.1 Fast uses only 58.4\% (aggressive action bias), a 41 percentage point difference reflecting fundamentally different planning strategies.
    \item \textbf{Error prevention effectiveness}: Top models (Claude Opus: 0.1\%, GPT 5.2: 1.0\%, Gemini 3: 1.6\%, Sonnet 4.5: 1.6\%) maintain minimal non-productive rates, while Cohere Command R+ suffers 18.0\% waste, 180 times  worse than Claude Opus and 18 times  worse than GPT 5.2, suggesting inadequate pre-edit validation.
    \item \textbf{Productivity leaders}: xAI Grok 4.1 Fast achieves 32.7\% productive ratio (highest), followed by DeepSeek R1 (21.0\%) and Claude Sonnet 4.5 (18.2\%), indicating these models convert iterations to progress most efficiently.
    \item \textbf{Exploration-error relationship}: Models with $>$85\% exploration (GPT 5.1 Codex: 87.4\%, Haiku: 87.0\%) exhibit $<$2\% non-productive rates, suggesting (though not statistically confirmed, $p=0.061$) that extensive information gathering may reduce edit errors, potentially at the cost of lower absolute productivity.
\end{itemize}

\subsection{Discussion}

\subsubsection{Behavioral Clustering Analysis}

Unsupervised clustering (k-means, $k=3$) on the three-dimensional space (exploration \%, productive \%, non-productive \%) reveals natural model groupings:

\begin{itemize}[noitemsep, topsep=0pt]
    \item \textbf{Cluster 1 - High Exploration Strategy} (91\% exploration avg, 1\% waste avg): \textbf{Claude Opus 4.5}, \textbf{Claude Haiku 4.5}, \textbf{GPT 5.1 Codex Max}. These models prioritize extensive information gathering ($>$85\% exploration) before attempting edits, achieving near-zero error rates ($<$2\% non-productive) but low productivity (8\% avg).
    
    \item \textbf{Cluster 2 - Balanced Strategy} (76\% exploration avg, 4\% waste avg): \textbf{Claude Sonnet 4.5}, \textbf{GPT 5.2}, \textbf{Qwen3 models}, \textbf{Gemini 3}, \textbf{xAI Grok 4.1 Fast}, \textbf{DeepSeek R1}. These models balance exploration with action, achieving highest productivity (20\% avg) while maintaining low error rates.
    
    \item \textbf{Cluster 3 - Mixed Performance} (83\% exploration avg, 8\% waste avg): \textbf{Cohere Command R+}, \textbf{DeepSeek V3.2}, \textbf{Llama 4 models}, \textbf{Grok Code Fast 1}. This heterogeneous cluster includes both efficient models (Llama 4) and the statistical outlier (Cohere, 18\% waste).
\end{itemize}

\textbf{Note}: Cluster assignments derived from data-driven k-means, not predetermined thresholds.

\subsubsection{Implications for Model Development}

These results suggest several areas for improvement:

\begin{enumerate}[noitemsep, topsep=0pt]
    \item \textbf{Adaptive Exploration}: The 41-point exploration range (58-99\%) suggests models lack task-difficulty calibration. Dynamic adjustment, such as aggressive on simple tasks, deliberate on complex ones, could optimize iteration usage.
    
    \item \textbf{Error Prevention}: Cohere's 18\% non-productive rate (detected as statistical outlier via IQR method, 18-180 times  worse than best performers) indicates critical need for pre-edit validation. Syntax checking, type verification, or edit simulation could reduce waste.
    
    \item \textbf{Early Success Momentum}: Models achieving first success within 5 iterations (Grok 4.1, Opus 4.5, Cohere, DeepSeek R1) demonstrate varied final success rates, suggesting early progress alone is insufficient without sustained productive iteration patterns.
    
    \item \textbf{Iteration Budget Recalibration}: 75\% of runs complete within 30 iterations (median 21), yet the 100-iteration budget appears set for worst-case scenarios. Adaptive budgets (e.g., 30 baseline, extending to 100 for promising trajectories) could improve throughput without sacrificing coverage.
\end{enumerate}

\subsubsection{Comparison to Terminal-Bench}

Terminal-Bench reports similar iteration analysis for command-line agents but focuses on command-level granularity \cite{merrill2026terminalbenchbenchmarkingagentshard}. Our analysis adds:

\begin{itemize}[noitemsep, topsep=0pt]
    \item \textbf{Edit-level success tracking}: Distinguishing between successful and failed edit attempts within iterations
    \item \textbf{Productivity categorization}: Three-way split (productive/exploration/non-productive) vs. binary success/failure
    \item \textbf{IDE-specific patterns}: File editing iterations have different characteristics than shell command iterations
\end{itemize}

\subsection{Limitations}

Several caveats apply to this analysis:

\begin{enumerate}[noitemsep, topsep=0pt]
    \item \textbf{Categorization Granularity}: An iteration with both successful and failed edits is categorized as ``productive'', which may oversimplify mixed-outcome iterations.
    
    \item \textbf{Exploration Quality}: We count exploration iterations but don't assess whether the exploration was productive (reading relevant files) vs. aimless (reading unrelated files).
    
    \item \textbf{Non-Productive Nuance}: Failed edits due to syntax errors (fixable) vs. semantic errors (may require reapproach) are not distinguished.
    
    \item \textbf{Temporal Dynamics}: We analyze aggregate statistics but don't capture learning within a run (e.g., improving edit success rate over time).
    
    \item \textbf{Task Complexity Confounds}: Some tasks naturally require more iterations. Future work should normalize by task difficulty.
\end{enumerate}

\clearpage
\newpage
\section{Behavioral Analyses: Failure Mode Taxonomy}
\label{app:failure_taxonomy}

When an IDE agent fails, it rarely fails in a random way. Through our analysis, we are able to see repeatable failure patterns across models and repositories, and those patterns explain much of what developers actually experience (early wrong edits, getting stuck in a loop, losing track of requirements). This appendix defines our failure taxonomy and shows how these failure modes distribute across models and task domains.

\subsection{Methodology}

We analyzed agent trajectories by examining four data streams for each run:
\begin{itemize}[noitemsep, topsep=0pt]
    \item \textbf{run\_summary.json}: High-level metrics (success, test results, iterations, edits)
    \item \textbf{tool\_calls.jsonl}: Timestamped sequence of all tool invocations
    \item \textbf{trajectory.jsonl}: Agent reasoning steps and decisions
    \item \textbf{edits.jsonl}: All code modifications with success/failure status
\end{itemize}

\subsubsection{Failure Mode Definitions}

We define 12 distinct failure modes based on observable patterns in agent behavior. Each mode has explicit detection criteria.

\begin{enumerate}[noitemsep, topsep=0pt]
    \item \textbf{Premature Editing}: Agent makes code changes before sufficiently exploring the codebase. Detected when fewer than 3 unique files were read before the first edit.
    
    \item \textbf{Syntax Error Loop}: Agent repeatedly introduces syntax errors without recovery. Detected when $\geq 3$ consecutive failed edits occur with syntax errors present.
    
    \item \textbf{Wrong File Targeting}: Agent edits files unrelated to the task requirements. Detected when $>50\%$ of edits fail and $<30\%$ of available files were explored.
    
    \item \textbf{Infinite Exploration}: Agent reads files excessively without making progress. Detected when more than 15 consecutive iterations occur without an edit attempt.
    
    \item \textbf{Large Risky Edits}: Agent makes overly broad changes that break existing functionality. Detected when any single edit exceeds 50 lines.
    
    \item \textbf{Timeout/Iteration Exhaustion}: Agent exhausts maximum iterations without completion. Detected when iteration count $\geq 100$ (max limit).
    
    \item \textbf{Infrastructure Errors}: Failure due to external factors (container issues, rate limits, etc.). Detected from explicit error categorization in run\_summary.
    
    \item \textbf{No Successful Edits}: Agent makes no successful code modifications. Detected when \texttt{successful\_edits} $= 0$ despite \texttt{made\_code\_changes} $= \text{true}$.
    
    \item \textbf{Tool Call Failures}: High count of failed tool invocations. Detected when $\geq 5$ tool calls fail due to invalid parameters or execution errors.
    
    \item \textbf{Thrashing/Backtracking}: Agent repeatedly modifies the same file without progress. Detected when agent makes $> 4$ edits to the same file.
    
    \item \textbf{Context Loss}: Agent fails to maintain workspace state. Detected when agent re-reads the same file $> 5$ times.
    
    \item \textbf{Test Misinterpretation}: Agent misunderstands test requirements despite correct implementation. Detected when agent reports completion but tests fail with assertion errors on edge cases.
\end{enumerate}

\subsubsection{Plain-Language Descriptions (Developer View)}

Below we restate the same failure modes in practical terms, or what a developer would actually observe while using an IDE agent:

\begin{description}[leftmargin=1.2em, labelsep=0.6em, itemsep=0.3em, topsep=0.3em]
    \item[\textbf{Premature Editing}] The agent starts changing code before it has found the right files or understood the existing design. You see quick, confident edits followed by confusion or lots of follow-up thrashing.
    \item[\textbf{Syntax Error Loop}] The agent repeatedly produces code that does not even parse/compile (e.g., missing imports, broken indentation, unmatched braces) and fails to recover cleanly.
    \item[\textbf{Wrong File Targeting}] The agent is editing the wrong layer or module (e.g., tweaking a helper or UI file when the bug is in a backend service). The changes look plausible but cannot affect the failing tests.
    \item[\textbf{Infinite Exploration}] The agent keeps opening/reading files and summarizing them but never commits to a concrete fix, often circling the same areas without converging.
    \item[\textbf{Large Risky Edits}] The agent makes a sweeping refactor or broad rewrite when the task calls for a small, precise change. This often introduces new breakages and increases review burden.
    \item[\textbf{Timeout/Iteration Exhaustion}] The agent runs out of steps/iterations without reaching a working solution. In practice this looks like a long back-and-forth where each attempt gets stuck on a different detail.
    \item[\textbf{Infrastructure Errors}] The run fails due to the environment rather than the code (e.g., container/test runner issues, transient tool failures, rate limits). The fix is not in the repo but instead, it is operational.
    \item[\textbf{No Successful Edits}] The agent attempted edits, but none were actually applied correctly (or they were reverted/invalid). From the user's perspective, the agent “did work” but the repo never meaningfully changed.
    \item[\textbf{Tool Call Failures}] The agent misuses tools (bad paths, wrong commands, invalid arguments) or repeatedly hits execution errors. It is blocked by workflow friction rather than code reasoning.
    \item[\textbf{Thrashing/Backtracking}] The agent ping-pongs between alternative fixes (or repeatedly rewrites the same area) without making measurable progress. Tests oscillate or fail in the same way across attempts.
    \item[\textbf{Context Loss}] The agent loses track of key requirements or earlier discoveries (e.g., forgets what the failing test expected, re-discovers the same facts, contradicts earlier conclusions).
    \item[\textbf{Test Misinterpretation}] The agent implements something reasonable but misunderstands the specification the tests enforce (often an edge case or exact output format). It gets “almost right” but cannot close the last gap.
\end{description}

\textbf{Note}: A single run may exhibit multiple failure modes. We observe co-occurrence patterns between thrashing and context loss across several open-weight models. 

\subsection{Results}

Across our runs runs (53.8\% failures and 46.2\% successes), we identify 12 distinct failure modes, the most common failure modes being Premature Editing (63.0\% of failures), Thrashing/Backtracking (28.2\%), and Context Loss (27.6\%). Runs may exhibit multiple failure modes simultaneously, so percentages can sum above 100\%.

\subsubsection{Overall Failure Distribution}

Figure~\ref{fig:failure_distribution} shows the overall distribution of failure modes across all models and tasks. The top three failure modes are:

\begin{itemize}[noitemsep, topsep=0pt]
    \item \textbf{Premature Editing}: 1,998 occurrences (63.0\% of failed runs): agents make code changes before adequately exploring the codebase
    \item \textbf{Thrashing/Backtracking}: 894 occurrences (28.2\%): agents repeatedly undo and redo similar changes without converging
    \item \textbf{Context Loss}: 875 occurrences (27.6\%): agents lose track of task requirements mid-execution
\end{itemize}

These top three failure modes account for the majority of agent failures, suggesting that exploration strategy, convergence behavior, and context management are critical bottlenecks for current IDE agents.

\begin{figure}[h]
\centering
\includegraphics[width=0.8\columnwidth]{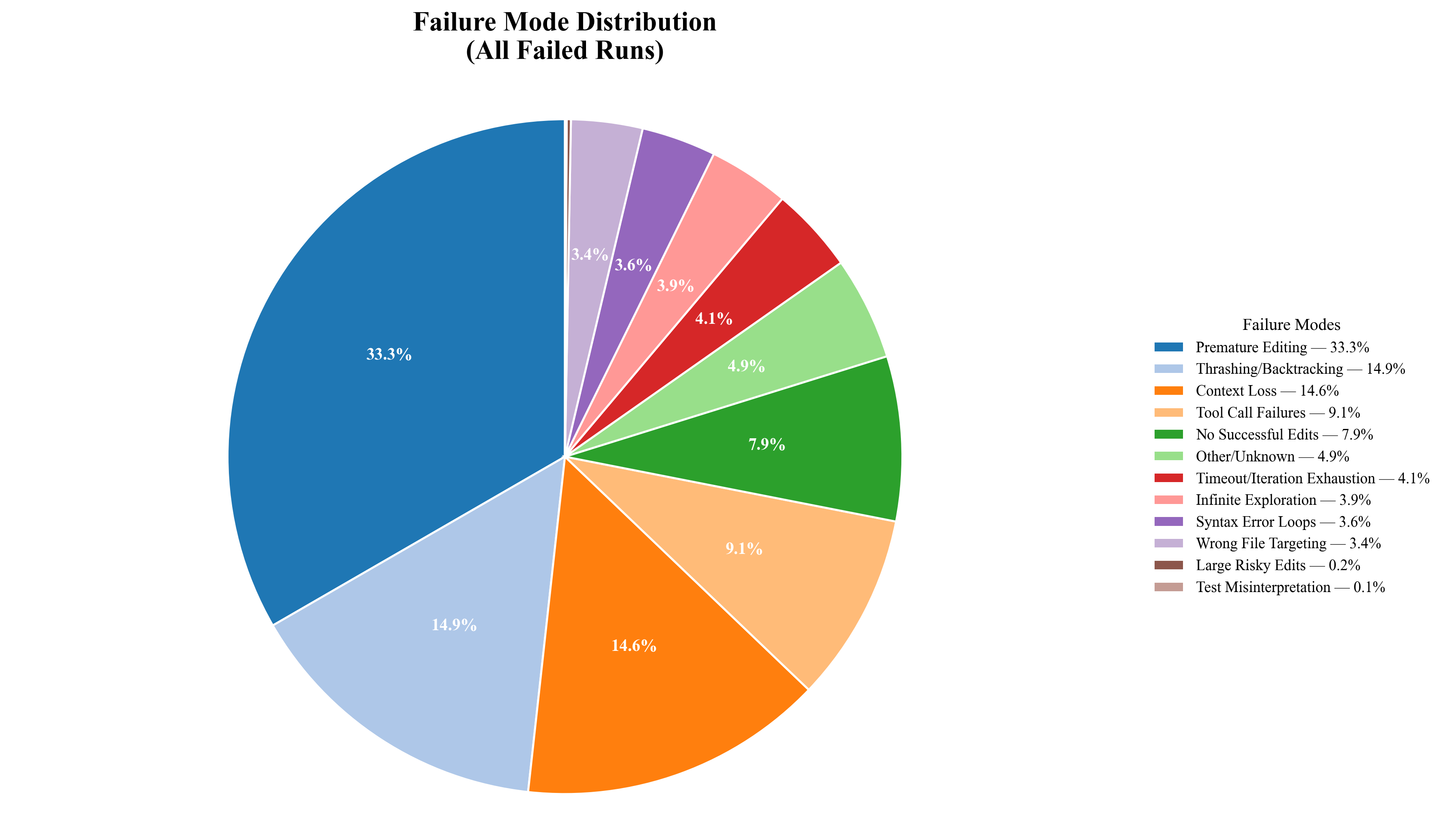}
\caption{Distribution of failure modes across all failed runs. Note that runs may exhibit multiple failure modes; percentages sum to $>100\%$.}
\label{fig:failure_distribution}
\end{figure}

\subsubsection{Model-Specific Failure Patterns}

Figure~\ref{fig:failure_by_model} reveals that different models exhibit distinct failure signatures:

\textbf{Frontier Models (GPT 5.2, Claude Sonnet/Haiku/Opus, GPT 5.1 Codex Max):} These models exhibit low Premature Editing counts (8--30 occurrences) and comparatively high success rates (72--84\%). When they fail, the dominant failure patterns shift from insufficient exploration to convergence failures: Thrashing/Backtracking and Context Loss account for much of their remaining errors, suggesting they explore adequately but sometimes fail to stabilize an edit-test loop or maintain task focus in longer runs.

\textbf{Open-Weight Models (Llama 4 Scout/Maverick, Grok Code Fast 1):} Show extremely high Premature Editing rates (290--352 occurrences, 73--92\% of their runs), indicating a systematic failure to explore before acting. Llama 4 Scout also exhibits elevated No Successful Edits (52 cases), suggesting code modifications frequently fail to apply correctly.

\textbf{Mid-Tier Models (DeepSeek R1, Qwen3 models):} Display moderate Premature Editing (144-298 occurrences) with balanced distributions across multiple failure modes. DeepSeek R1 shows 298 Premature Editing cases (74.5\% of runs), while Qwen3 Coder exhibits Context Loss (101 cases) and Infinite Exploration (35 cases), suggesting inconsistent exploration strategies.

\textbf{Specialized Failures:} Command-R+ uniquely exhibits high Timeout/Iteration Exhaustion (103 cases, 25.8\% of its runs) and No Successful Edits (187 cases, 46.8\%), indicating it explores extensively but struggles to produce working code modifications. Grok 4.1 Fast shows high Thrashing (166 cases, 70\% of failed runs) and Context Loss (152 cases, 64\% of failed runs), suggesting unstable editing behavior despite adequate exploration.

\begin{figure}[h]
\centering
\includegraphics[width=\columnwidth]{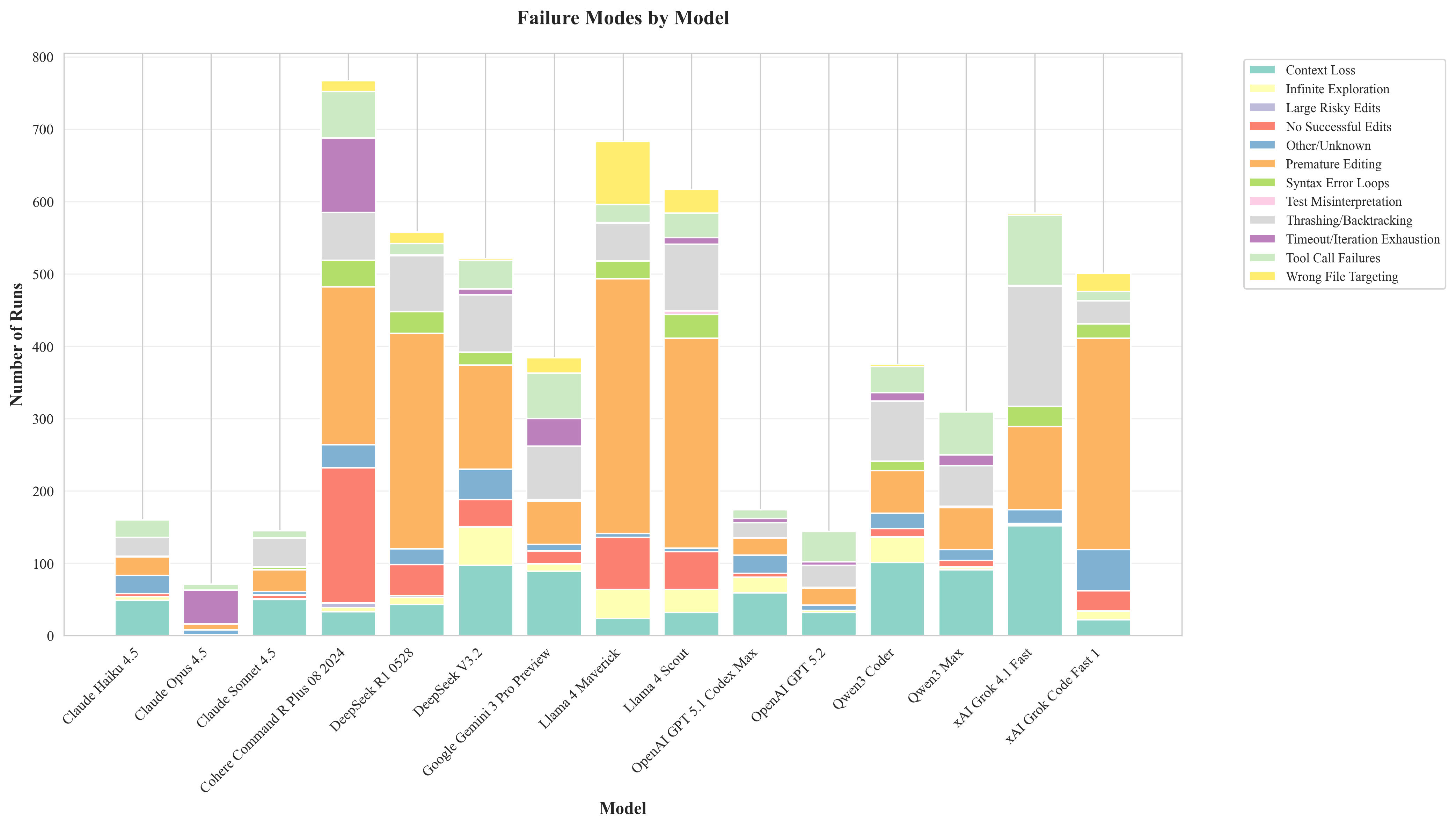}
\caption{Failure mode distribution by model (stacked bar chart). Each bar shows the proportion of different failure modes for failed runs from that model.}
\label{fig:failure_by_model}
\end{figure}

\subsubsection{Failure Mode Heatmap (Model $\times$ Failure Type)}

Figure~\ref{fig:failure_heatmap} breaks down failure modes by model. Each cell shows how many failed runs for that model were tagged with that failure mode (runs may have multiple modes, so counts can exceed the number of failed runs).

\subsubsection{Co-occurrence Patterns}

Failure modes are not independent: a single failed run can exhibit multiple modes at once. We compute co-occurrence by counting how often pairs of failure modes appear together within the same failed run. Key co-occurrences include:

\textbf{Thrashing/Backtracking and Context Loss (most common pair)}: Co-occur in 599 failed runs (18.9\%), making it the strongest co-occurring failure signature. This indicates a common failure trajectory where agents lose track of requirements and enter unstable edit cycles.

\textbf{Premature Editing with convergence failures}: Premature Editing co-occurs frequently with Thrashing/Backtracking (584 runs, 18.4\%) and Context Loss (435 runs, 13.7\%), suggesting that ``act too early'' often cascades into iterative non-convergence.

\textbf{Premature Editing and No Successful Edits}: Co-occur in 240 failed runs (7.6\%), capturing cases where agents both explore insufficiently and fail to land any working patch.

\textbf{Timeout/Iteration Exhaustion and No Successful Edits}: Co-occur in 106 failed runs (3.3\%), consistent with long-running episodes that exhaust the iteration budget while producing no effective progress.

Overall, these patterns demonstrate that failure modes are not independent: early mistakes (e.g., premature edits) often trigger downstream instability (thrashing/context loss), while repeated unproductive behavior is associated with timeouts.

\clearpage
\begingroup
\captionsetup{skip=2pt}
\begin{figure}[H]
\centering
    \includegraphics[width=0.85\linewidth]{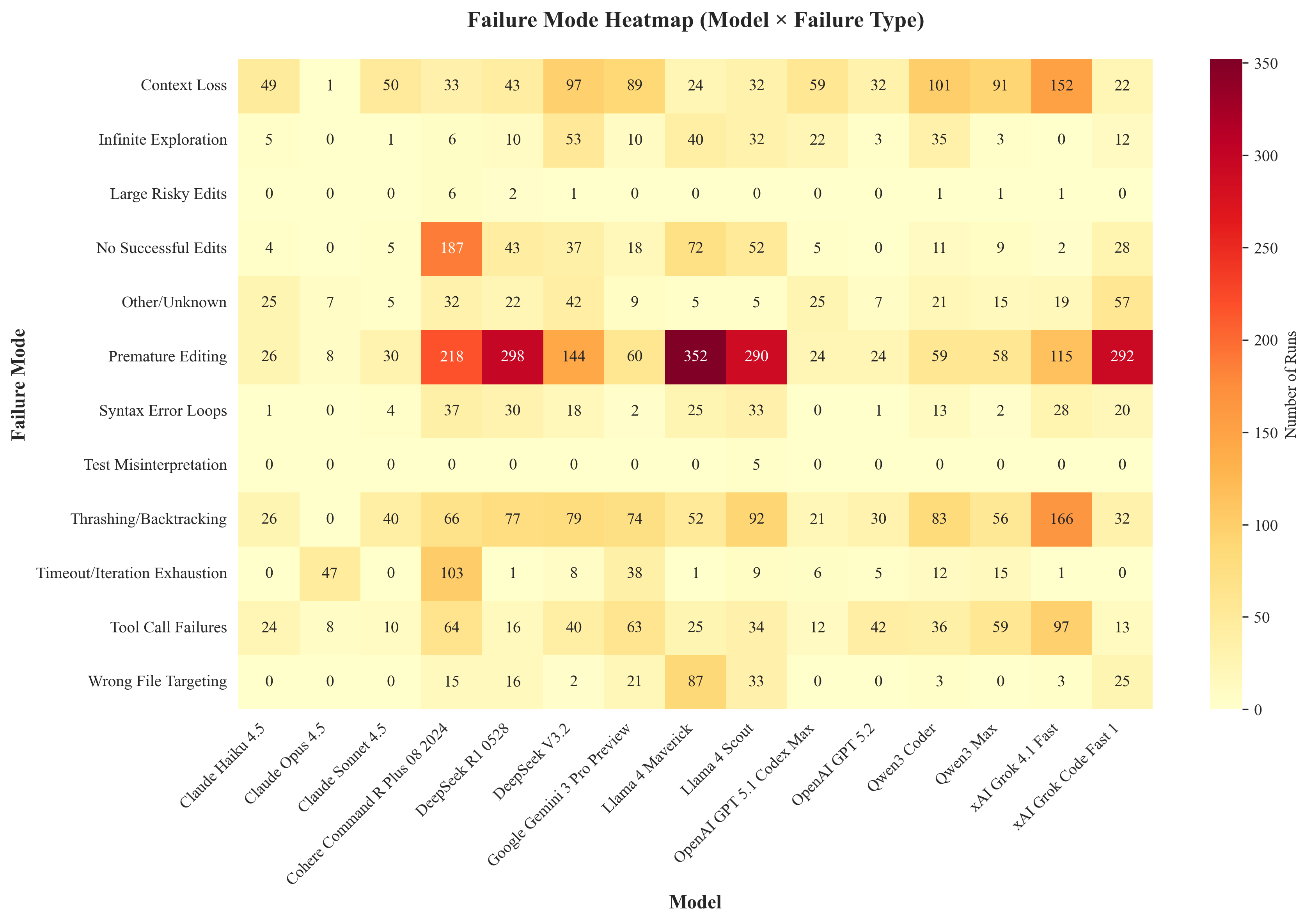}
\caption{Failure mode heatmap (model $\times$ failure type). Color intensity indicates how often each failure mode appears among failed runs for each model.}
\label{fig:failure_heatmap}
\end{figure}
\vspace{-6pt}

\subsubsection{Prevalence Across Task Types}

Figure~\ref{fig:failure_prevalence} breaks down failure mode prevalence by task domain, revealing language and framework-specific failure patterns:

\textbf{Python Tasks (Network Traffic Analyzer, Code Quality Analyzer):} Have the highest rates of Syntax Error Loops (94 and 81 occurrences respectively, 81.8\% of all syntax loop failures). This demonstrates that agents struggle with Python-specific syntax requirements (indentation, type hints, generator expressions) more than other languages.

\textbf{C/C++ Systems Tasks (Memory Profiling, ESIM Management):} Show high levels of premature editing (302 and 227 cases), indicating agents underestimate the complexity of memory management and pointer arithmetic, attempting modifications before fully understanding data structures.

\textbf{TypeScript/JavaScript Tasks (Event Callback System):} Display the highest premature editing rate (334 cases, 16.7\% of all premature editing). Asynchronous patterns, callback chains, and event-driven architecture require thorough exploration, which agents frequently skip.

\textbf{Java web tasks (SmartHub Operations Center):} Uniquely exhibit high tool call failures (142 cases, 26.2\% of all tool failures) and Infinite Exploration (73 cases, 31.5\%). The multi-module Java codebase, framework conventions (Javalin routing + Thymeleaf templates), and MVC-style layering cause agents to explore extensively without finding correct modification points.

\textbf{Full-Stack MERN Tasks (Cross-Lingual Document Translator):} Show the highest Context Loss (200 cases, 22.9\% of all context loss failures), suggesting that multi-layer architectures (MongoDB, Express, React, and Node.js) overwhelm agent context windows, potentially causing them to lose track of which layer they are modifying.

\begin{figure}[H]
\centering
    \includegraphics[width=0.85\linewidth]{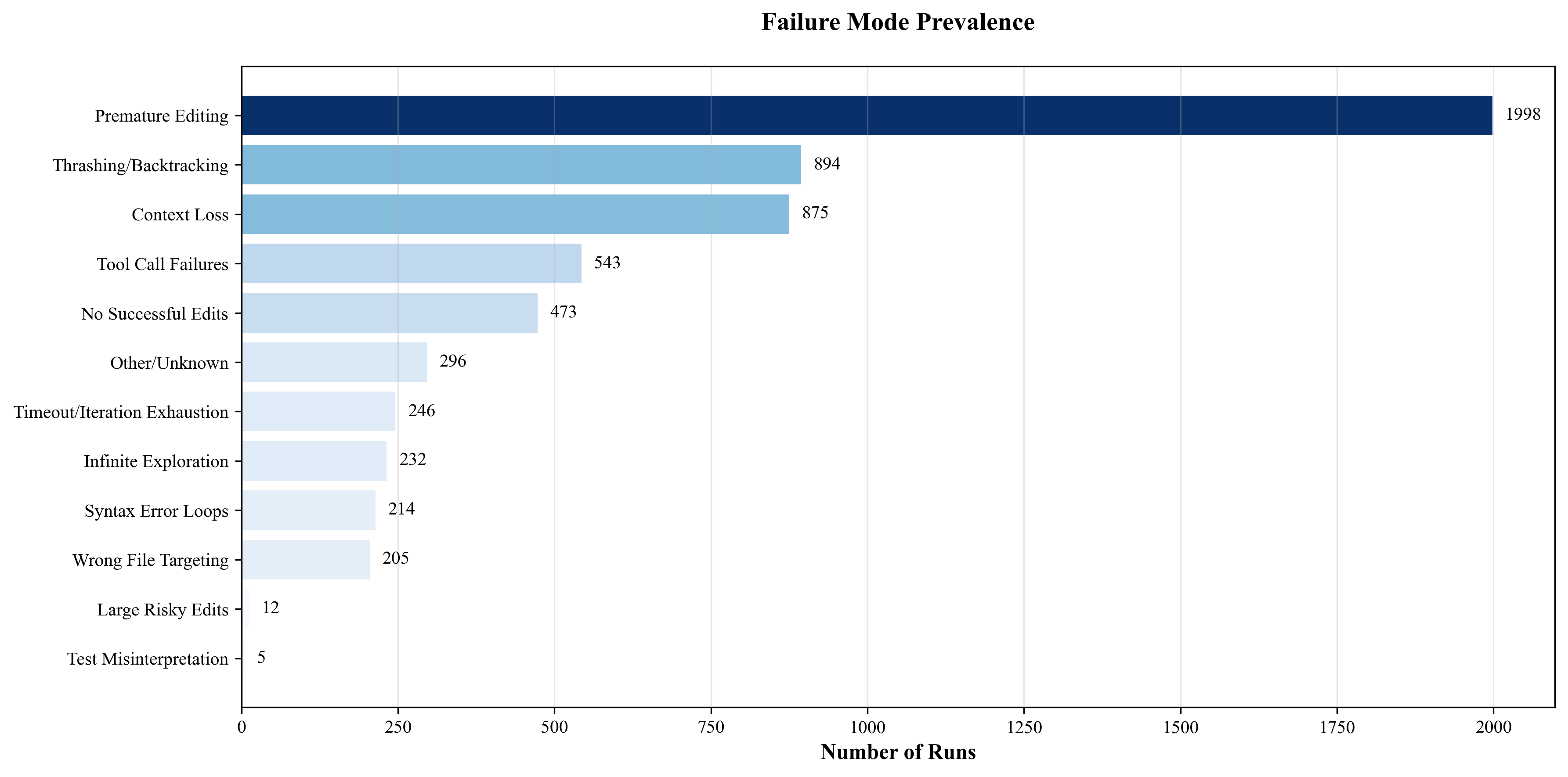}
\caption{Failure mode prevalence by task domain. Shows which failure types are most common in different types of programming tasks.}
\label{fig:failure_prevalence}
\end{figure}
\endgroup

\subsection{Discussion}

\subsubsection{Implications for Model Development}

The failure taxonomy reveals several areas for targeted improvement:

\begin{itemize}[noitemsep, topsep=0pt]
    \item \textbf{Exploration Strategy}: High rates of Premature Editing and Infinite Exploration suggest current models lack effective heuristics for balancing information gathering vs. action taking.
    
    \item \textbf{Error Recovery}: Syntax Error Loops and Thrashing indicate that many models struggle with debugging feedback loops, repeatedly making similar mistakes.
    
    \item \textbf{Specification Understanding}: Test Misinterpretation failures highlight the need for better instruction following and edge case reasoning.
    
    \item \textbf{Risk Assessment}: Large Risky Edits suggest models underestimate the potential impact of broad changes, lacking a ``measure twice, cut once'' principle.
\end{itemize}

\clearpage
\section{Behavioral Analyses: Tool Call Sequences}
\label{app:tool_sequences}

If an agent is unable to run a clean loop (of commands including read, edit, verify, and test), it will not behave like a real IDE collaborator, even if it can write code that seems and looks reasonable. This appendix quantifies tool-call transitions and common sequences so we can see the ``default behaviors'' models fall into (and where they get stuck).
\subsection{Methodology}

We analyze tool call sequences extracted from \texttt{trajectory.jsonl} files, which record every tool invocation (with arguments and results) for each agent run. We include all available runs for the 15 selected models after de-duplicating to the most recent run per (model, dataset, task, attempt) and preferring non-infrastructure-error runs when duplicates exist. For each run, we extract:

\begin{itemize}[noitemsep, topsep=0pt]
    \item \textbf{Tool sequences}: Ordered list of tool names (e.g., \texttt{read\_file}, \texttt{edit\_file})
    \item \textbf{Tool transitions}: Which tools typically follow which (first-order Markov analysis)
    \item \textbf{Behavioral features}: Read-to-edit ratios, tool diversity, exploration patterns
    \item \textbf{Temporal patterns}: Tools used before first edit, iteration efficiency
\end{itemize}

We categorize tools into three primary types:
\begin{itemize}[noitemsep, topsep=0pt]
    \item \textbf{Read tools}: \texttt{read\_file}, \texttt{grep}, \texttt{codebase\_search}, \texttt{list\_dir}, etc.
    \item \textbf{Edit tools}: \texttt{edit\_file}, \texttt{write\_file}, \texttt{search\_replace}
    \item \textbf{Execution tools}: \texttt{run\_terminal\_cmd}
\end{itemize}

\subsection{Tool Transition Patterns}

Figure~\ref{fig:tool_transition_matrix} presents a transition probability matrix showing which tools typically follow which across all runs.

\begin{figure}[h]
\centering
\includegraphics[width=0.95\columnwidth]{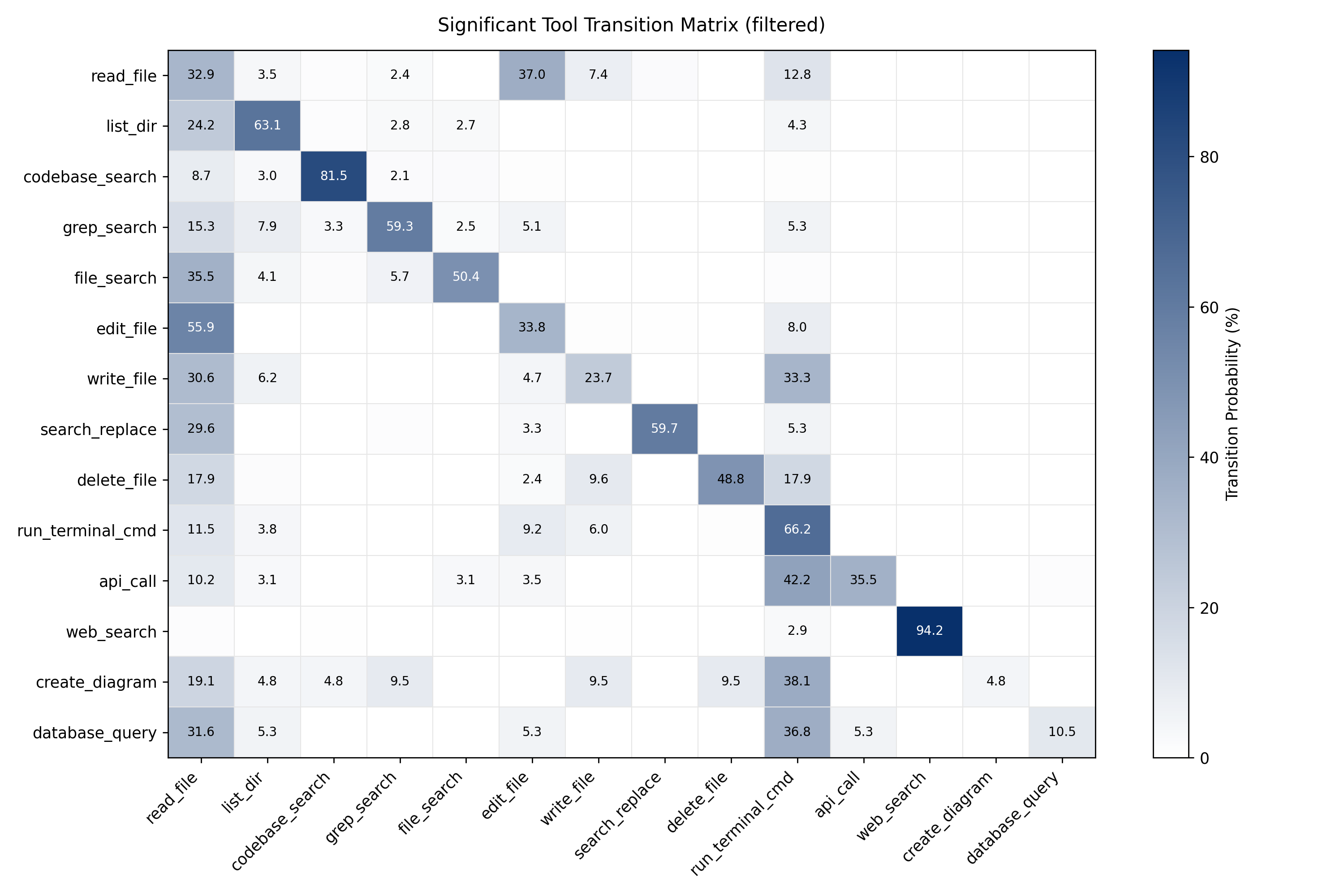}
\caption{Tool transition probability matrix. Color intensity indicates the probability P(tool\_B | tool\_A) that tool B is used immediately after tool A. Darker colors indicate higher transition probabilities.}
\label{fig:tool_transition_matrix}
\end{figure}

\textbf{Significance filtering.} For readability, we visualize only transitions that are individually \(\geq 1\%\) of a source tool's outgoing probability, plus the top-5 transitions per source tool and any additional transitions needed to cover 95\% of outgoing mass. The remaining low-frequency transitions are omitted; these are often due to rare tool usage (e.g., \texttt{web\_search}, \texttt{create\_diagram}) or provider/tooling failures that produce malformed tool-call entries.

\textbf{Key Observations}:
\begin{itemize}[noitemsep, topsep=0pt]
    \item \textbf{Strong read-edit coupling}: After reading a file, agents edit it 37.0\% of the time (which is the most common cross-category transition). However, we see that after editing, agents return to reading 55.9\% of the time, suggesting an iterative cycle of reading and editing.
    \item \textbf{Tool chaining dominance}: Tools exhibit high self-transition rates, indicating sequential batching: \texttt{codebase\_search} chains 81.5\% of the time (search refinement), \texttt{run\_terminal\_cmd} 66.2\% (multi-step testing), \texttt{list\_dir} 63.1\% (directory exploration), and \texttt{grep\_search} 59.3\% (iterative searching).
    \item \textbf{Edit-test feedback loops}: After editing, only 8.0\% transition directly to \texttt{run\_terminal\_cmd}; this suggests that most models verify edits by reading (55.9\%) before testing, indicating a possible deliberate validation strategy.
    \item \textbf{Read multiplicity}: Read tools self-chain at 32.9\%, indicating agents read multiple related files before taking action. We see this to be consistent with the observed high read-to-edit ratios (1.70--5.29).
\end{itemize}

\subsection{Most Common Tool Sequences}

Figure~\ref{fig:tool_sequence_frequency} shows the 15 most frequent tool bigrams (two-tool sequences) across all runs.

\begin{figure}[h]
\centering
\includegraphics[width=\columnwidth]{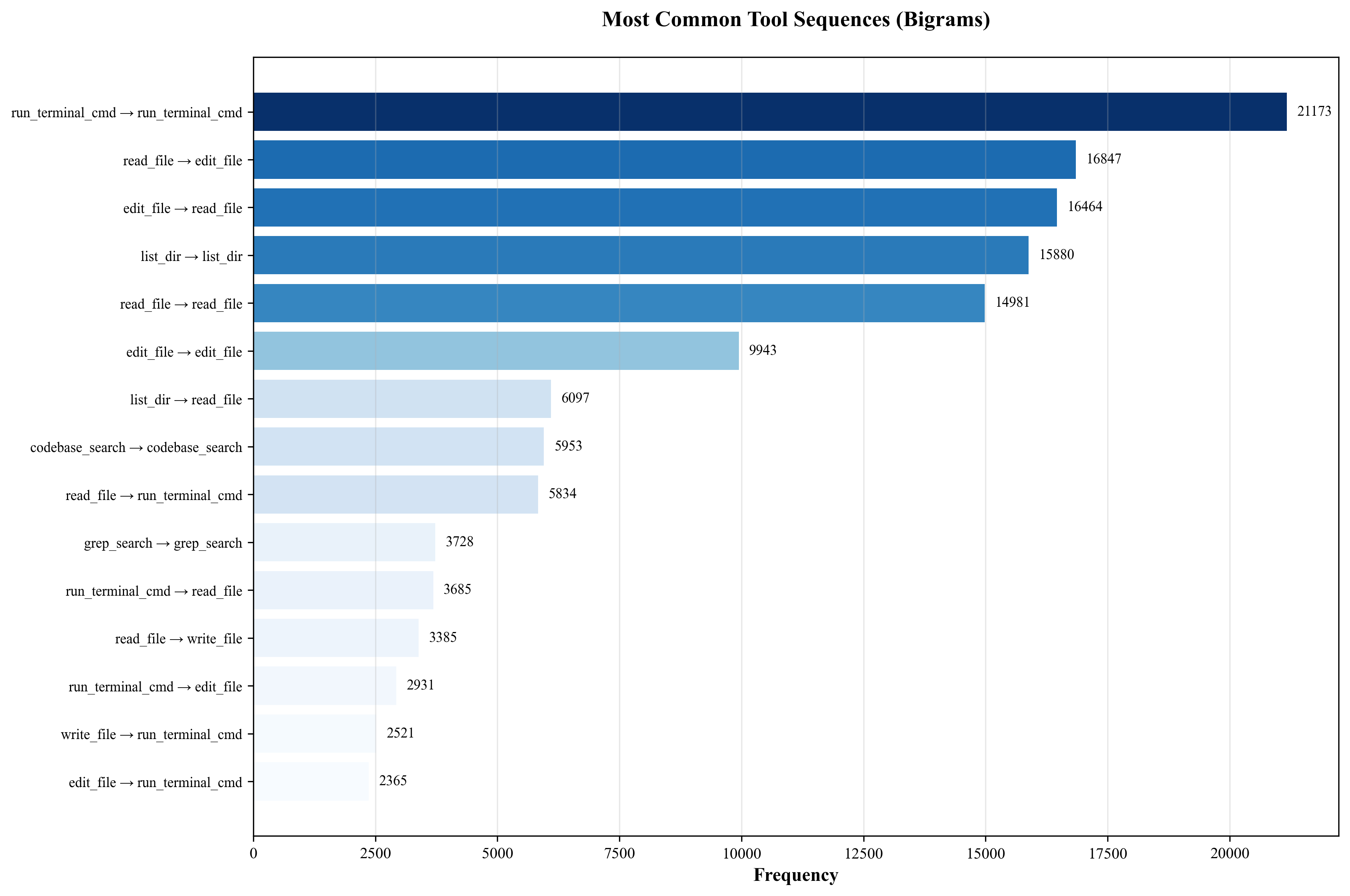}
\caption{Top 15 most common tool sequences (bigrams). The frequency indicates how many times each two-tool pattern appears across all runs.}
\label{fig:tool_sequence_frequency}
\end{figure}

Across all 15 models, the top 5 most frequent bigrams are:

\begin{enumerate}[noitemsep, topsep=0pt]
    \item \texttt{run\_terminal\_cmd → run\_terminal\_cmd} (21,173 occurrences):testing and verification sequences
    \item \texttt{read\_file → edit\_file} (16,847 occurrences): the exploration to action
    \item \texttt{edit\_file → read\_file} (16,464 occurrences): verifying and gathering context
    \item \texttt{list\_dir → list\_dir} (15,880 occurrences): exploring directories
    \item \texttt{read\_file → read\_file} (14,981 occurrences): gathering information from multiple files
\end{enumerate}

\subsection{Patterns of Exploration vs. Exploitation}

\subsubsection{Read-to-Edit Ratios}

Figure~\ref{fig:read_to_edit_ratio} compares our models' exploration-exploitation balance by measuring the ratio of read-type tool calls to edit-type tool calls.

\begin{figure}[h]
\centering
\includegraphics[width=\columnwidth]{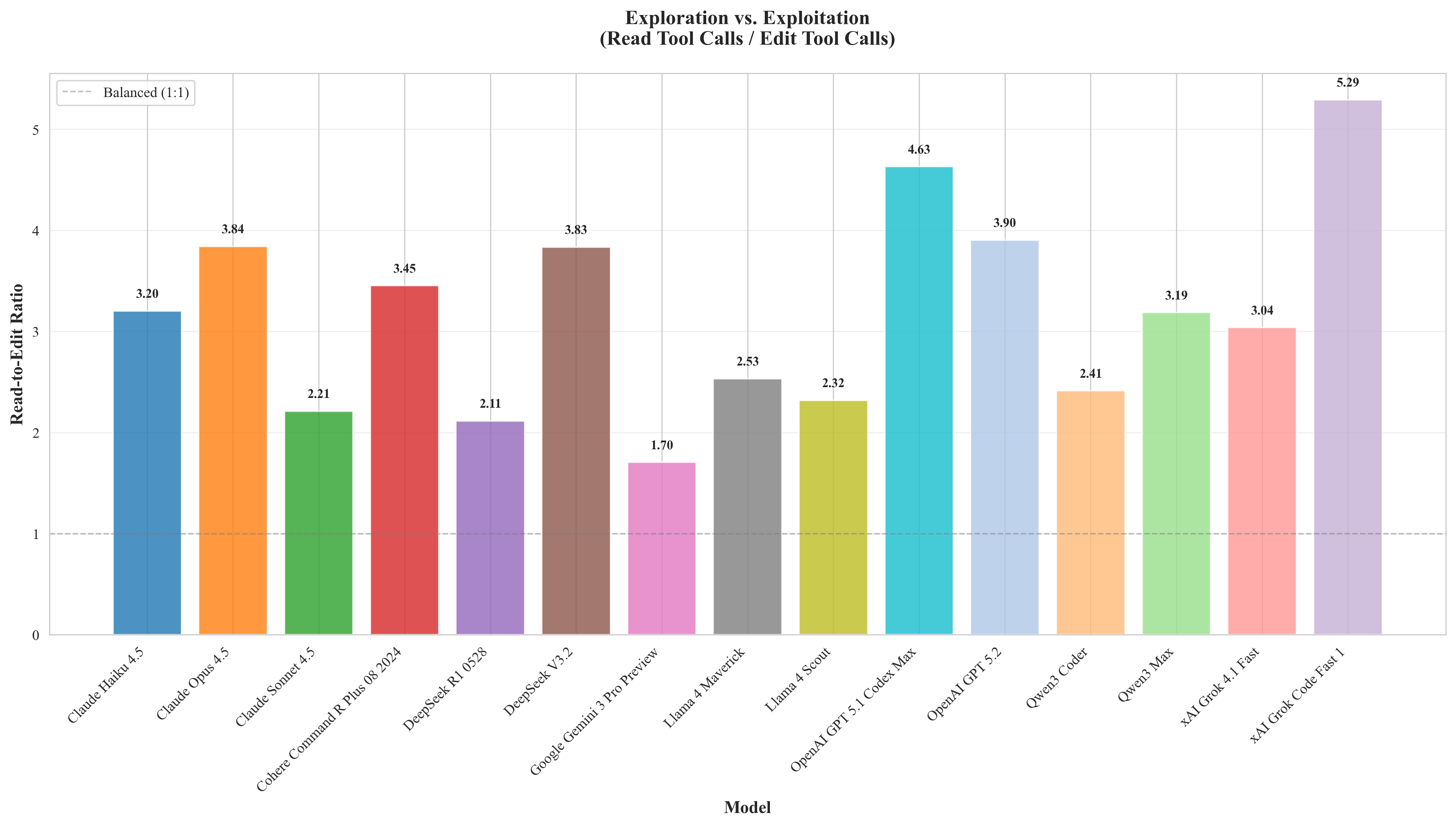}
\caption{Read-to-edit ratio by model. Values $>1$ indicate more exploration (reading) than exploitation (editing). Reference line at 1.0 shows balanced behavior.}
\label{fig:read_to_edit_ratio}
\end{figure}

\textbf{Findings}:
\begin{itemize}[noitemsep, topsep=0pt]
    \item \textbf{Wide strategy spectrum}: Models exhibit read-to-edit ratios from 1.70 to 5.29, roughly a 3.1 times difference reflecting different exploration philosophies. Google Gemini 3 Pro Preview is most action-oriented (1.70), while xAI Grok Code Fast 1 is most deliberative (5.29).
    \item \textbf{Success correlation}: Higher performing models show diverse ratios. GPT 5.2 (84.2\% success, ratio=3.90), Sonnet 4.5 (83.2\% success, ratio=2.21), and GPT 5.1 Codex (72.0\% success, ratio=4.63) all succeed even though they have their own different strategies, demonstrating that there may be multiple viable approaches.
    \item \textbf{Low ratios correlating with risk}: The three lowest-ratio models (Gemini 3: 1.70, DeepSeek R1: 2.11, Sonnet 4.5: 2.21) show highly variable success (61.4\%, 16.8\%, 83.2\%), with the model DeepSeek R1's poor performance suggesting insufficient exploration can be detrimental.
    \item \textbf{Very high ratios showing mixed results}: xAI Grok Code Fast 1's extreme ratio (5.29) yields only 9.0\% success, suggesting that over-exploration without effective action can also fail. However, GPT 5.1 Codex's high ratio (4.63) with 72\% success shows this strategy can work for capable models.
\end{itemize}

\subsubsection{Tools Used Before First Edit}

\textbf{Statistical Summary}:
\begin{itemize}[noitemsep, topsep=0pt]
    \item \textbf{Average across all 15 models}: 6.82 tools before first edit (median: 6.92)
    \item \textbf{Range}: 3.12 to 11.00 tools (which is a 3.5 times difference in initial exploration thoroughness)
    \item \textbf{Most deliberate models}: GPT 5.2 (11.00 tools, 84.2\% success), DeepSeek V3.2 (9.28 tools, 32.8\% success), and Claude Haiku 4.5 (8.69 tools, 75.5\% success) explore most before acting
    \item \textbf{Most aggressive models}: Llama 4 Maverick (3.12 tools, 2.4\% success), DeepSeek R1 (3.18 tools, 16.8\% success), and Llama 4 Scout (3.45 tools, 1.5\% success) act quickly but with they tend to have relatively poor results
    \item \textbf{Success correlation}: Higher performing models (GPT 5.2, Sonnet 4.5, Opus 4.5, Haiku 4.5) average 8.1 tools before first edit, compared to 4.9 for bottom performers (Llama 4 models, xAI Grok Code); this then suggesting enough initial exploration predicts success
\end{itemize}

\subsection{Implications}

\subsubsection{For Prompt Engineering}

From these tool-sequence patterns, we find the following actionable prompt-design insights:

\begin{enumerate}[noitemsep, topsep=0pt]
    \item \textbf{Encourage Exploration Before Editing}: Models that average $\geq$8 tool calls before the first edit have mean run-level success of 60.2\%, compared to 6.9\% for models averaging $<4$ tools before first edit (a +53.3 percentage point difference; 8.7 times  relative lift). Prompts should explicitly require reading/searching before making changes. 
    \item \textbf{Tool Diversity}: Encourage using multiple complementary tools (grep + read\_file + codebase\_search)
    \item \textbf{Delay Editing}: Explicitly instruct agents to explore thoroughly before making changes
    \item \textbf{Model-Specific Strategies}: Different models benefit from different exploration strategies (e.g., Gemini 3 is more action-oriented at 1.70 read-to-edit ratio, while Codex Max is more deliberative at 4.63).
\end{enumerate}

\subsubsection{For Model Development}

Tool sequence analysis identifies areas for model improvement:

\begin{itemize}[noitemsep, topsep=0pt]
    \item \textbf{Planning Capability}: Models that explore more before editing are more successful, suggesting that improving planning/reasoning before action would be beneficial.
    
    \item \textbf{Tool Selection Strategy}: High predictive power of early tool choices suggests models could benefit from explicit training on optimal tool usage patterns.
    
    \item \textbf{Exploration-Exploitation Balance}: The correlation between read-to-edit ratio and success suggests models need better heuristics for when to transition from exploration to action.
    
    \item \textbf{Early Stopping}: The ability to predict failure from first 5-10 tools suggests potential for early termination strategies to save compute.
\end{itemize}

\subsubsection{For Routing Strategies}

Behavioral signatures could inform intelligent model routing:
\begin{itemize}[noitemsep, topsep=0pt]
    \item Tasks requiring heavy exploration → Route to models with higher read-to-edit ratios
    \item Tasks requiring rapid prototyping → Route to models comfortable with lower exploration
    \item Early failure prediction could trigger automatic fallback to stronger (but more expensive) models
\end{itemize}


\clearpage
\section{Task-Level Breakdown and Additional Figures}

\subsection{Task-Level Specialization and Discriminative Performance}
\label{app:task_level_specialization}

Please see individual repository heatmaps in Figures \ref{fig:heatmap_esim}-\ref{fig:heatmap_smarthub}

Aggregate pass@5 is useful, but it can also be misleading: two models may tie in overall resolution rate while behaving very differently at the task level. To make this visible, we isolate the most discriminative tasks (high across-model variability; sample standard deviation $\sigma \ge 0.35$ on per-task attempt success rate, i.e., fraction of the five attempts that pass all tests). Table \ref{tab:discriminative} lists examples where adjacently-ranked models exhibit large swings on specific tasks.

    \begin{table}[htbp]
        \centering
        \caption{\textbf{Most Discriminative Tasks}: Tasks revealing high performance variance across models. SmartHub Operations Center (SH-O), Event Callback System (ECS), Memory Profiling (MP). Performance is the fraction of attempts (out of 5) that pass all tests, scaled to 0--1.}
        \label{tab:discriminative}
        \small
        \begin{tabular}{lcccccc}
        \toprule
        \textbf{Task} & \textbf{$\sigma$} & \textbf{GPT-5.2} & \textbf{Codex} & \textbf{Sonnet} & \textbf{Haiku} & \textbf{Gemini} \\
        \midrule
        SH-O:3 & 0.477 & 1 & 1 & 1 & 1 & 1 \\
        ECS:4 & 0.350 & 1 & 1 & 0 & 0 & 0 \\
        SH-O:2 & 0.444 & 1 & 1 & 1 & 0 & 0.8 \\
        MP:8 & 0.475 & 1 & 1 & 1 & 0 & 0.2 \\
        \bottomrule
        \end{tabular}
    \end{table}

SmartHub Operations Center task-3 is among the most discriminative tasks in the benchmark ($\sigma=0.477$). Many strong models achieve 100\% success (5/5 attempts), while several others collapse to 0--20\%, producing a large spread in outcomes. Event Callback System task-4 ($\sigma=0.350$) also demonstrates clear task-level specialization: GPT 5.2 and Codex Max achieve 100\% success, while Sonnet, Haiku, Gemini, and Opus score 0\%. Given that these models are otherwise close in aggregate pass@5, this is not a small difference; instead, it suggests that the task is selecting for a specific capability (here, TypeScript async handling combined with brittle output specification compliance) that some models have and others do not (at least under our prompting and tool loop).

We see similar specialization effects elsewhere. SmartHub Operations Center task-2 appears to filter for Java web framework navigation and MVC/service-layer reasoning: GPT 5.2, Codex Max, Sonnet, and Qwen3 Max solve it reliably (100\%), Gemini succeeds frequently (80\%), while Claude Haiku fails entirely (0\%). Memory Profiling task-8 filters for C/C++ systems precision, where Haiku and Gemini struggle. Therefore, we are able to see that aggregate rankings do not predict task-level capability; if one is to deploy an IDE agent, the relevant unit of analysis is often the task domain, not the global leaderboard. One caveat to note is that these discriminative results come from specific tasks in this benchmark (and our benchmark has a particular domain and language distribution). For example, Event Callback System task-4 is one TypeScript async scenario (2.5\% of tasks), not a universal statement about TypeScript. These patterns are still useful for model selection, but they should not be over-generalized without broader coverage.

The systems-oriented repositories (e.g., Memory Profiling, ESIM) emphasize memory safety, manual resource management, and pointer arithmetic. These domains act as a hard filter for models that lack robust low-level code priors. In contrast, Event Callback System and Cross-Lingual Document Translator shift the bottleneck toward stateful orchestration and specification adherence (retry logic, auth flows, strict formatting). From our results, Cross-Lingual Document Translator is the most challenging overall in our set, with SmartHub Operations Center close behind; both compound statefulness of web systems with framework-specific navigation costs.

Finally, we identify three tasks that none of the 15 models solve in our evaluation: Event Callback System task-10 and SmartHub Operations Center tasks 9 and 10. These provide concrete examples of capability ceilings for current agents and targets for future model and tooling improvements.

\subsection{Additional Figures}
\begin{figure} [H]
\centering
        \includegraphics[width=0.9\linewidth]{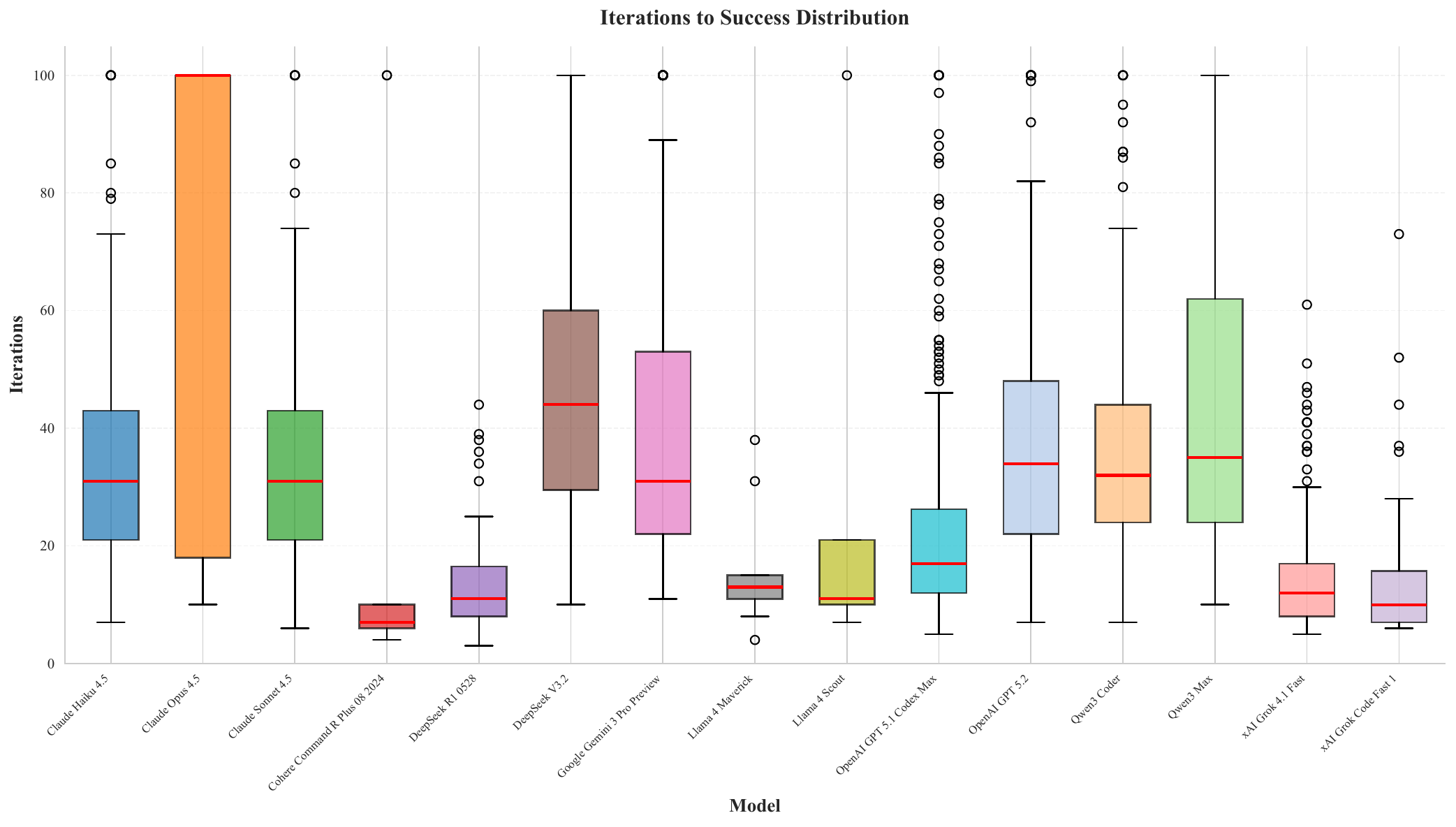}
        \caption{\textbf{Iterations Needed for Success by Model} Distribution of iteration counts at which successful task completions occurred. Models are sorted alphabetically. The box plots show median, quartiles, and outliers, revealing variation in solution efficiency across models.}
        \label{fig:iter_needed}
    \end{figure} 

\begin{figure} [H]
        \centering
        \includegraphics[width=0.9\linewidth]{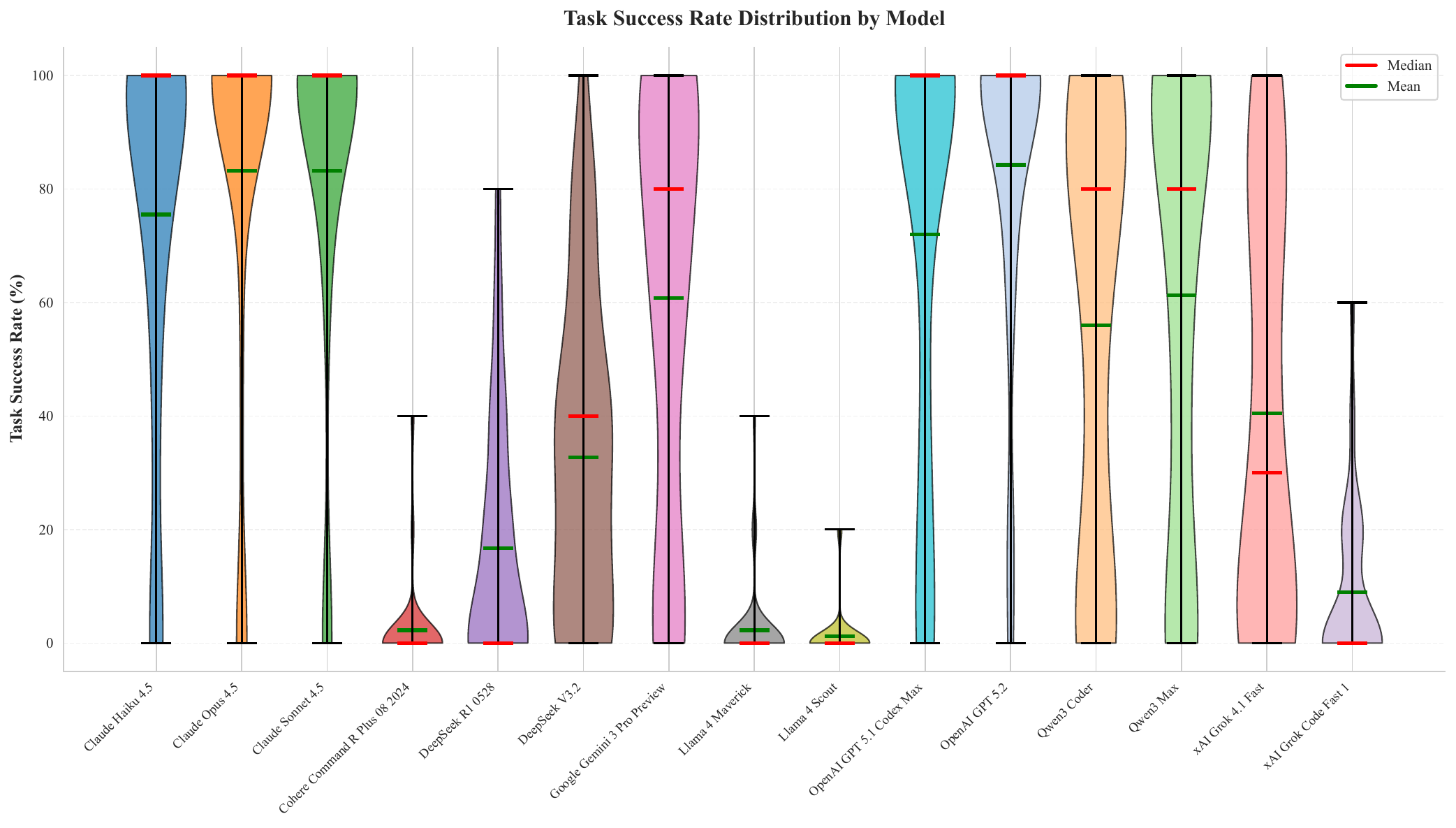}
        \caption{\textbf{Task Success Distribution} Visualization showing task completion patterns across the benchmark repositories. Each data point represents a task's success rate across models, revealing which tasks are universally challenging versus those that distinguish top performers.}
        \label{fig:task_dist}
    \end{figure} 

\begin{table}[H]
    \centering
    \caption{\textbf{Average Task Duration (minutes)} by Dataset in IDE-Bench. Datasets sorted by average duration descending.}
    \label{tab:dataset_durations}
\small
    \begin{tabular}{lccc}
\toprule
    Dataset & Avg Duration & Min & Max \\
\midrule
    SmartHub Operations Center & 12.43 & 0.54 & 325.96 \\
    Game Engine Service & 9.63 & 0.39 & 146.09 \\
    Network Traffic Analyzer & 8.65 & 0.23 & 81.61 \\
    Cross-Lingual Document Translator & 8.42 & 0.58 & 41.38 \\
    Code Quality Analyzer & 7.58 & 0.15 & 45.49 \\
    E-SIM Management System & 6.50 & 0.20 & 40.77 \\
    Event Callback System & 6.14 & 0.18 & 88.08 \\
    Memory Profiling App & 4.39 & 0.20 & 29.52 \\
    \bottomrule
    \end{tabular}
\end{table}

\vspace{-6pt}
\begin{table}[H]
\centering
\caption{\textbf{Token Consumption Distribution.} Statistics showing the mean, median, and interquartile range (P25–P75) in thousands of tokens. Models sorted by mean token consumption descending.}
\label{tab:token_consumption_distribution}
\small
\begin{tabular}{lS[table-format=4.1]S[table-format=4.1]S[table-format=4.1]S[table-format=4.1]}
\toprule
Model & {Mean} & {Median} & {P25} & {P75} \\
 & {(k tokens)} & {(k tokens)} & {(k tokens)} & {(k tokens)} \\
\midrule
Claude Opus 4.5 & 1404.5 & 1752.3 & 319.2 & 2134.1 \\
Gemini 3 Pro Preview & 1116.9 & 984.9 & 367.9 & 1557.2 \\
DeepSeek V3.2 & 835.6 & 589.4 & 291.0 & 1230.0 \\
Qwen3 Coder & 780.3 & 638.0 & 433.2 & 877.4 \\
Claude Haiku 4.5 & 705.5 & 544.5 & 327.9 & 903.4 \\
GPT 5.2 & 682.2 & 536.5 & 334.0 & 901.7 \\
Claude Sonnet 4.5 & 652.7 & 567.1 & 310.0 & 823.6 \\
Qwen3 Max & 514.3 & 485.9 & 317.8 & 648.7 \\
Command-R+ 08 2024 & 317.6 & 303.8 & 137.1 & 464.5 \\
GPT 5.1 Codex Max & 303.0 & 210.5 & 140.3 & 362.1 \\
Grok 4.1 Fast & 209.2 & 187.2 & 132.2 & 265.0 \\
DeepSeek R1 0528 & 143.5 & 130.6 & 97.3 & 173.9 \\
Grok Code Fast 1 & 125.4 & 100.1 & 80.1 & 131.8 \\
Llama 4 Scout & 111.3 & 85.1 & 49.1 & 129.9 \\
Llama 4 Maverick & 104.8 & 78.9 & 57.1 & 112.1 \\
\bottomrule
\end{tabular}
\end{table}

\vspace{-6pt}
\begin{figure} [H]
        \centering
        \includegraphics[width=0.9\linewidth]{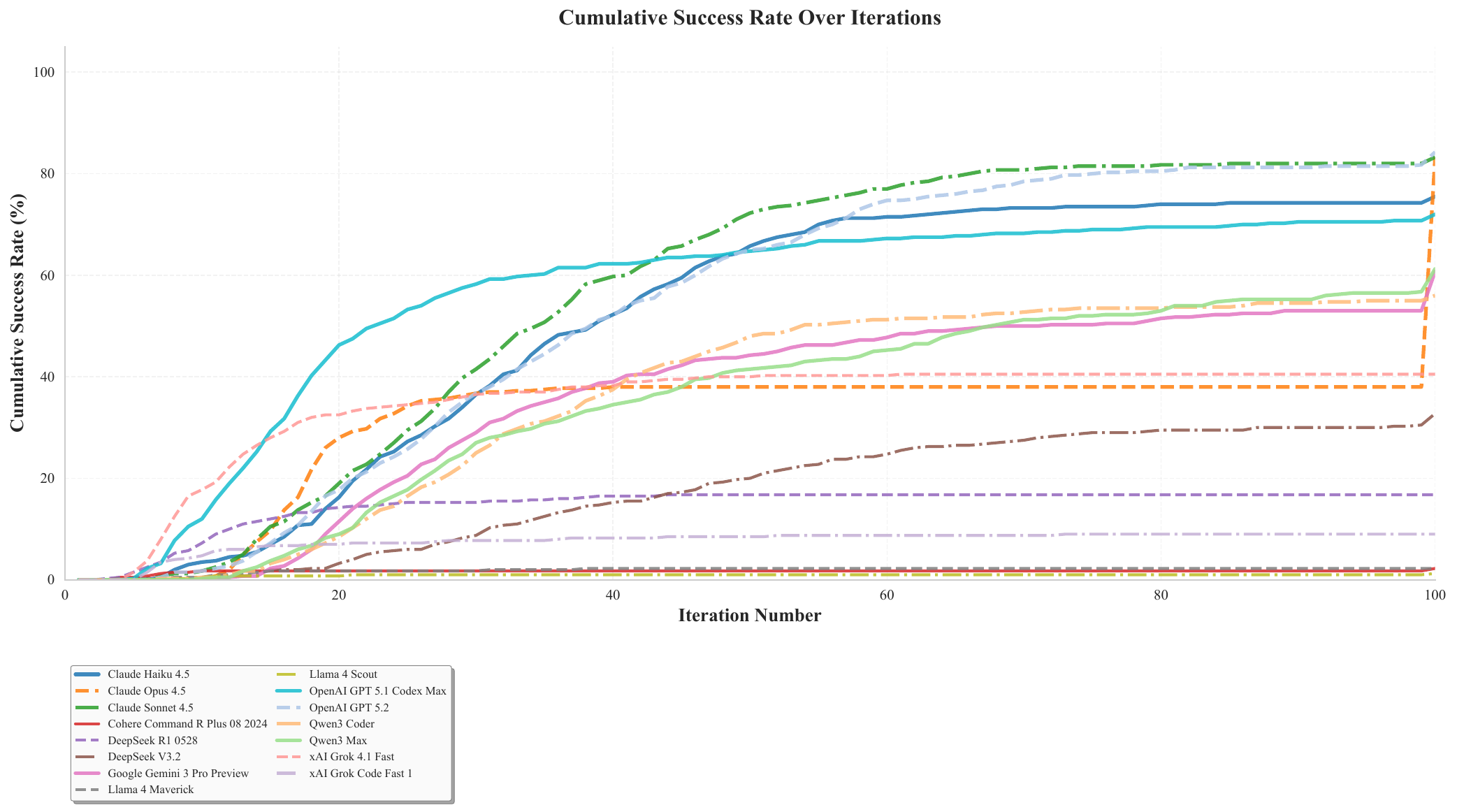}
        \caption{\textbf{Cumulative Success Rate Over Iterations} Shows how models accumulate successful task completions as iterations progress. Steeper initial slopes indicate faster convergence, while plateaus reveal practical iteration limits. This figure includes its legend showing all 15 evaluated models.}
        \label{fig:cum_iter}
    \end{figure}

\vspace{-6pt}
\begin{table}[H]
\centering
\caption{\textbf{First-Attempt Success Rate (pass@1)} Comparison of pass@1 versus pass@5 reveals the performance gap between single-attempt and multi-attempt evaluation. Models sorted by pass@1 descending.}
\label{tab:pass1_comparison}
\small
\begin{tabular}{lcc}
\toprule
\textbf{Model} & \textbf{pass@1 (\%)} & \textbf{pass@5 (\%)} \\
\midrule
Claude Sonnet 4.5 & 87.50 $\pm$ 7.25 & 88.75 $\pm$ 6.92 \\
GPT 5.2 & 85.00 $\pm$ 7.82 & 95.00 $\pm$ 4.78 \\
Claude Opus 4.5 & 83.75 $\pm$ 8.08 & 86.25 $\pm$ 7.55 \\
Claude Haiku 4.5 & 78.75 $\pm$ 8.96 & 87.50 $\pm$ 7.25 \\
GPT 5.1 Codex Max & 73.75 $\pm$ 9.64 & 85.00 $\pm$ 7.82 \\
Qwen3 Max & 65.00 $\pm$ 10.45 & 76.25 $\pm$ 9.33 \\
Qwen3 Coder & 57.50 $\pm$ 10.83 & 75.00 $\pm$ 9.49 \\
Gemini 3 Pro Preview & 55.00 $\pm$ 10.90 & 80.00 $\pm$ 8.77 \\
Grok 4.1 Fast & 35.00 $\pm$ 10.45 & 67.50 $\pm$ 10.26 \\
DeepSeek V3.2 & 31.25 $\pm$ 10.16 & 71.25 $\pm$ 9.92 \\
DeepSeek R1 0528 & 20.00 $\pm$ 8.77 & 46.25 $\pm$ 10.93 \\
Grok Code Fast 1 & 11.25 $\pm$ 6.92 & 32.50 $\pm$ 10.26 \\
Llama 4 Scout & 2.50 $\pm$ 3.42 & 6.25 $\pm$ 5.30 \\
Llama 4 Maverick & 2.50 $\pm$ 3.42 & 8.75 $\pm$ 6.19 \\
Command-R+ 08 2024 & 0.00 $\pm$ 0.00 & 7.50 $\pm$ 5.77 \\
\bottomrule
\end{tabular}
\end{table}

\begin{figure} [H]
    \centering
    \includegraphics[width=0.9\linewidth]{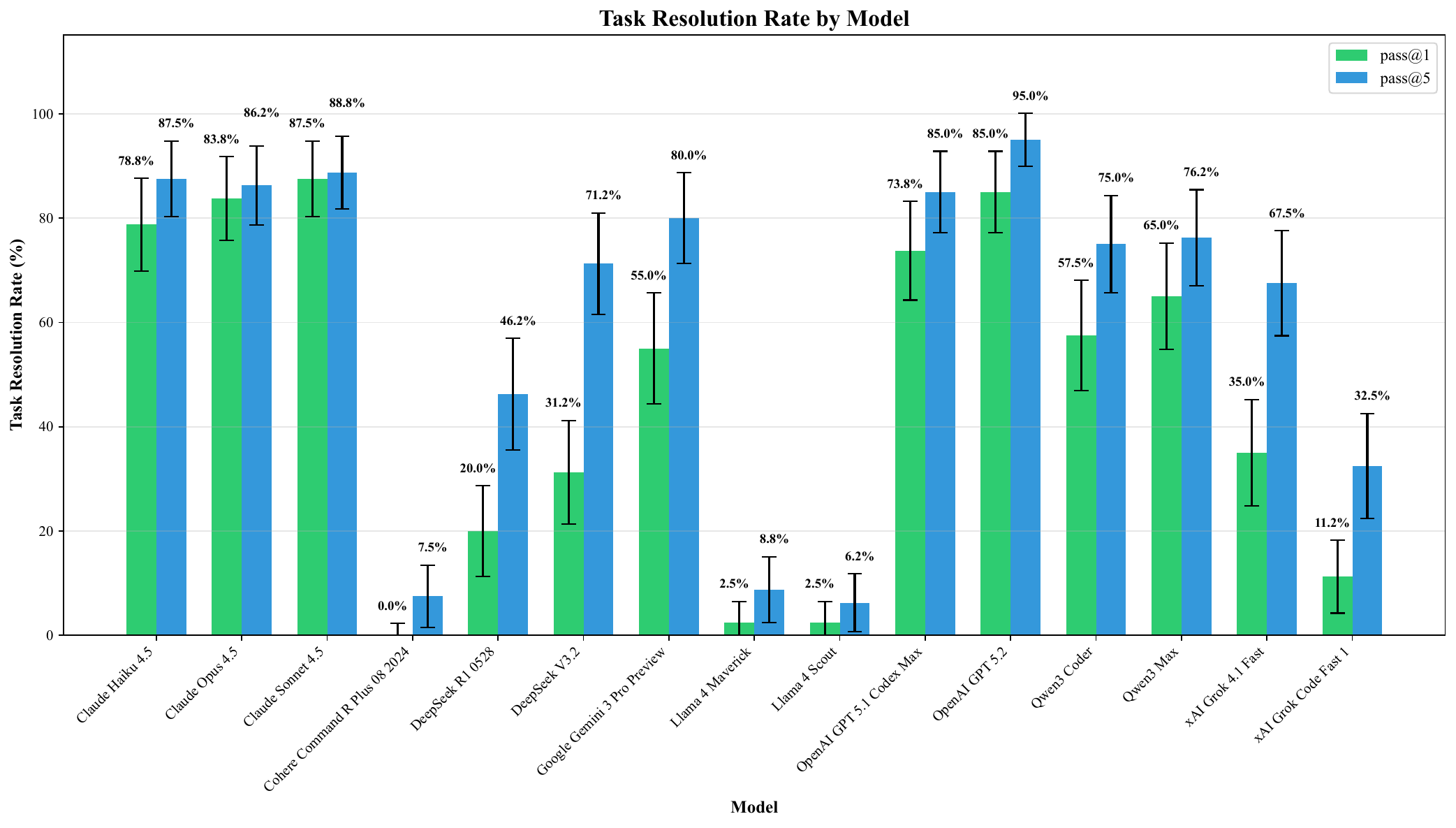}
    \caption{\textbf{Model Performance Comparison: pass@1 vs pass@5} Direct comparison of first-attempt success rates (pass@1) versus best-of-five success rates (pass@5) across all evaluated models. Error bars represent 95\% confidence intervals. This visualization highlights the performance gap between single-attempt and multi-attempt evaluation, demonstrating the non-deterministic nature of LLM agents.}
    \label{fig:pass1_vs_pass5}
\end{figure}

\begin{figure} [H]
\centering
    \includegraphics[width=0.9\linewidth]{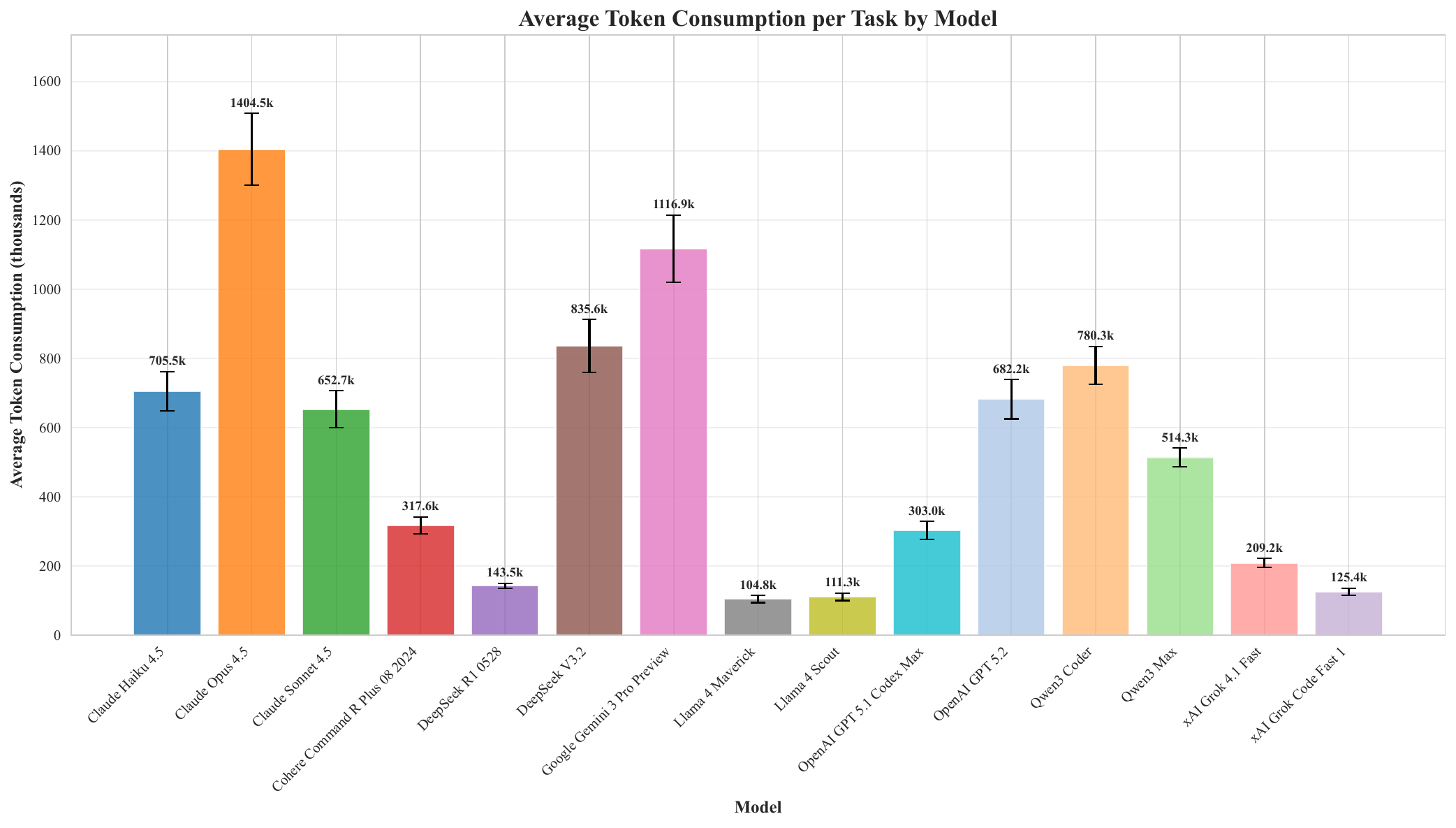}
    \caption{\textbf{Token Consumption by Model} Average token consumption per successful task completion across all evaluated models. Lower values indicate more efficient agents that accomplish tasks with fewer API calls and shorter conversations.}
    \label{fig:token_consumption}
\end{figure}

\begin{figure} [H]
    \centering
    \includegraphics[width=0.9\linewidth]{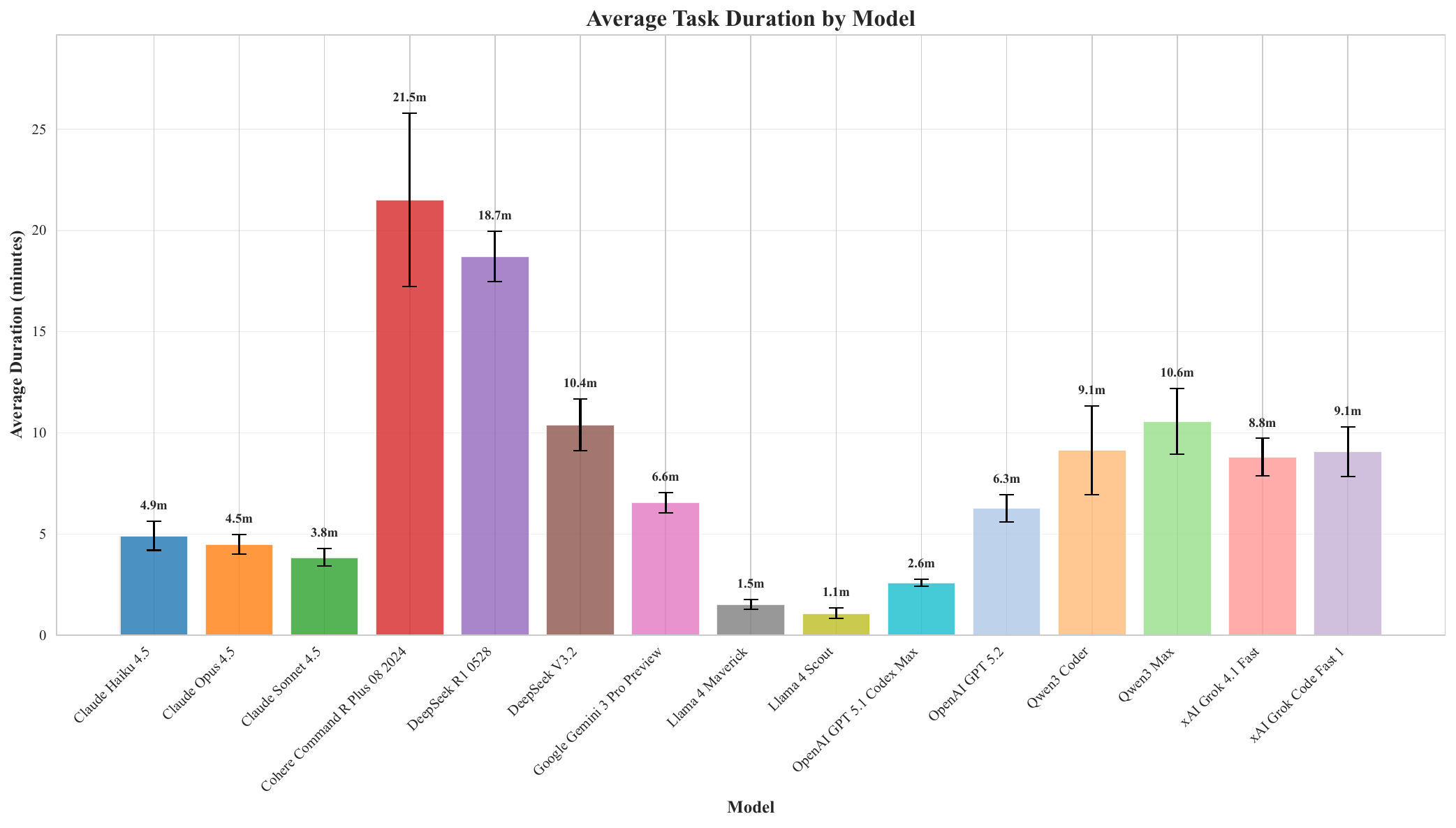}
    \caption{\textbf{Task Duration by Model} Average wall-clock time (in minutes) required for task completion across models. Duration includes agent reasoning time, tool execution latency, and API response time.}
    \label{fig:duration_by_model}
\end{figure}

\begin{figure} [H]
\centering
    \includegraphics[width=0.9\linewidth]{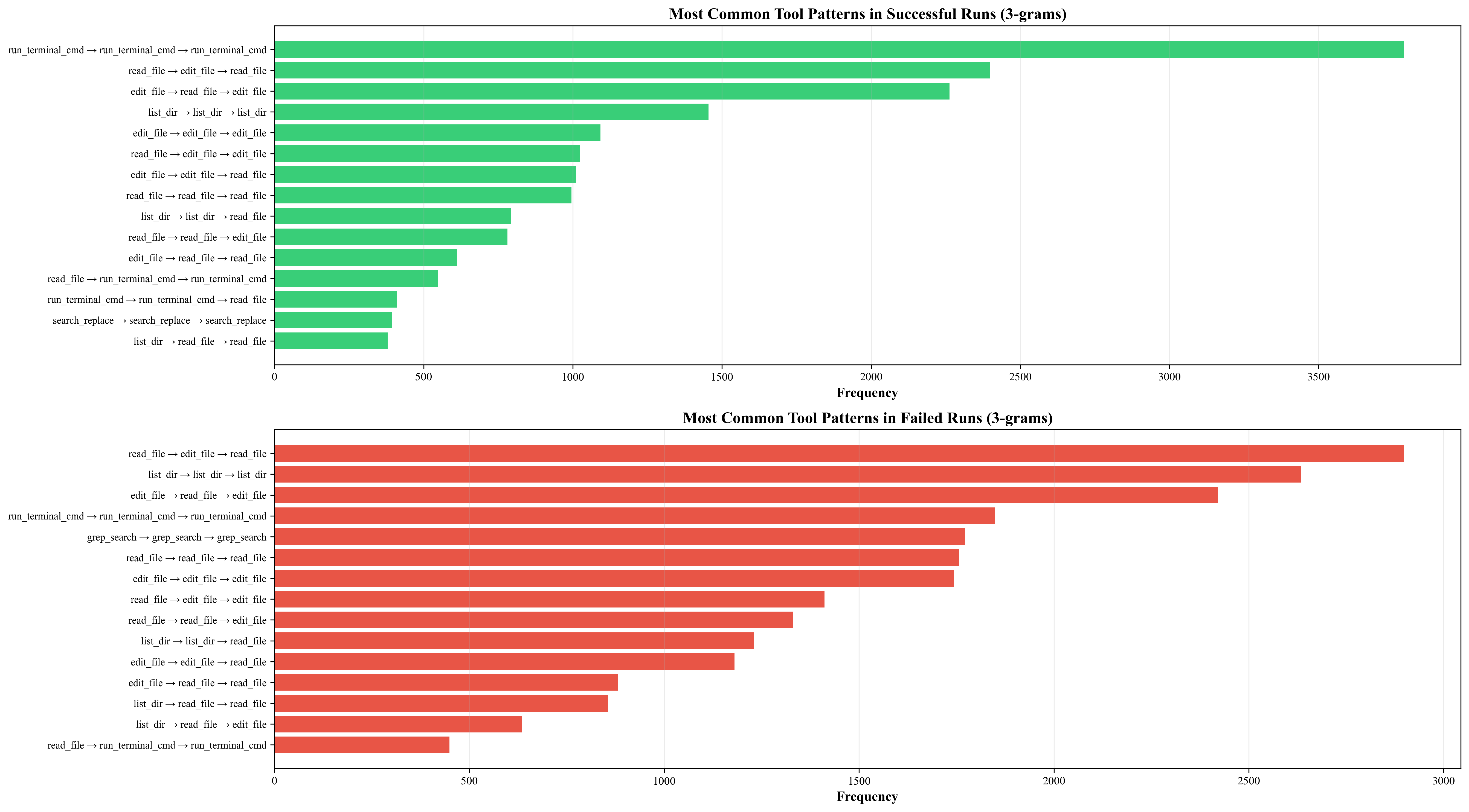}
    \caption{\textbf{Tool usage overview (all models)} Aggregated radar plot summarizing relative tool-call frequency across models. Spikes indicate tool categories that dominate an agent's workflow; flatter profiles indicate more balanced tool use.}
    \label{fig:tool_usage_patterns_overview}
\end{figure}

\begin{figure}[H]
    \centering
    \includegraphics[width=0.95\linewidth]{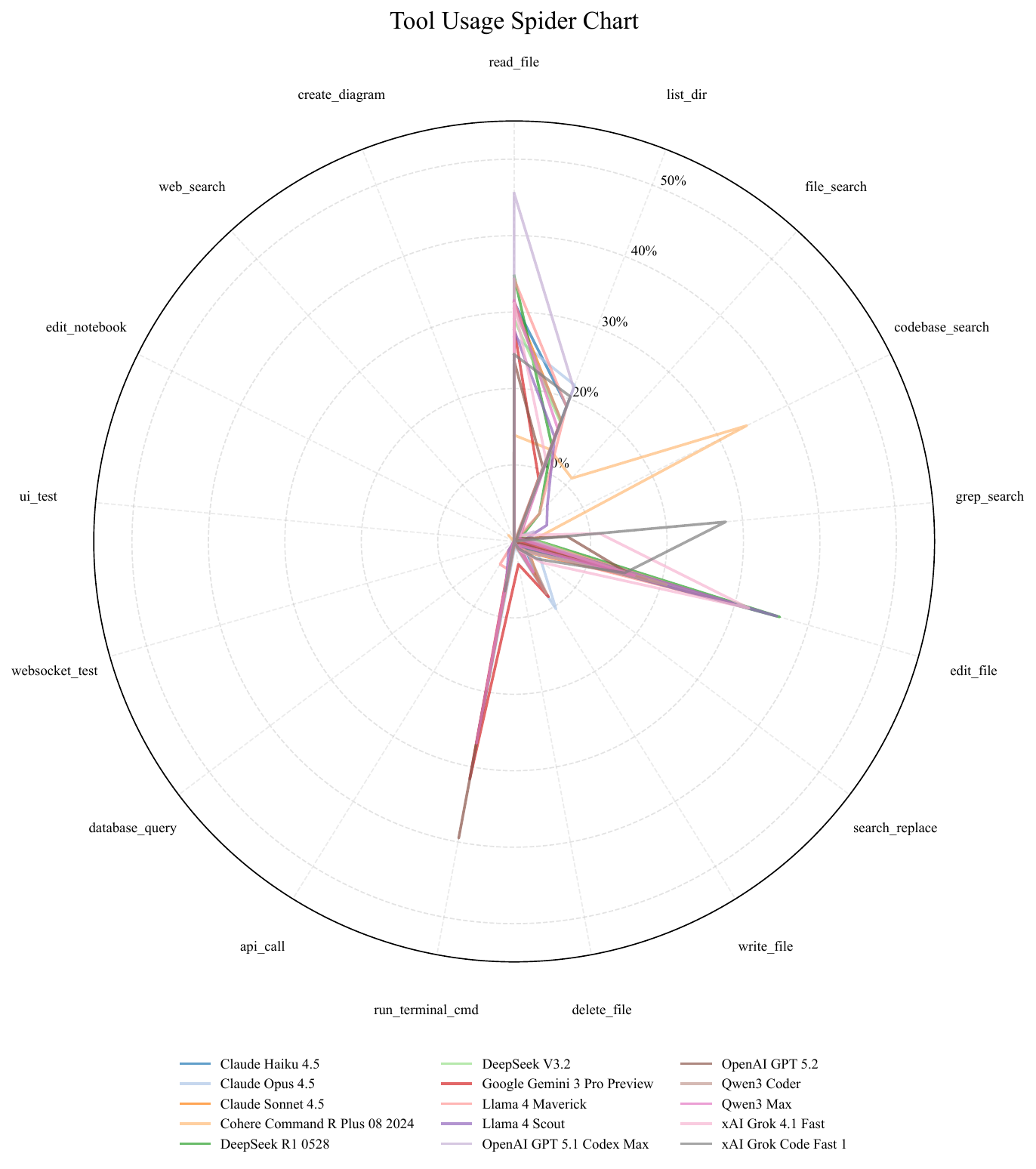}
    \caption{\textbf{Tool Usage Spider Chart} All 15 models overlaid on a single radar plot (one line per model), showing the relative frequency of each tool among that model's tool calls. This makes shared “default workflows” (common spikes) and model-specific tool biases (divergent spikes) visible at a glance.}
    \label{fig:tool_usage_spider}
\end{figure}

\begin{figure} [H]
\centering
    \includegraphics[width=0.9\linewidth]{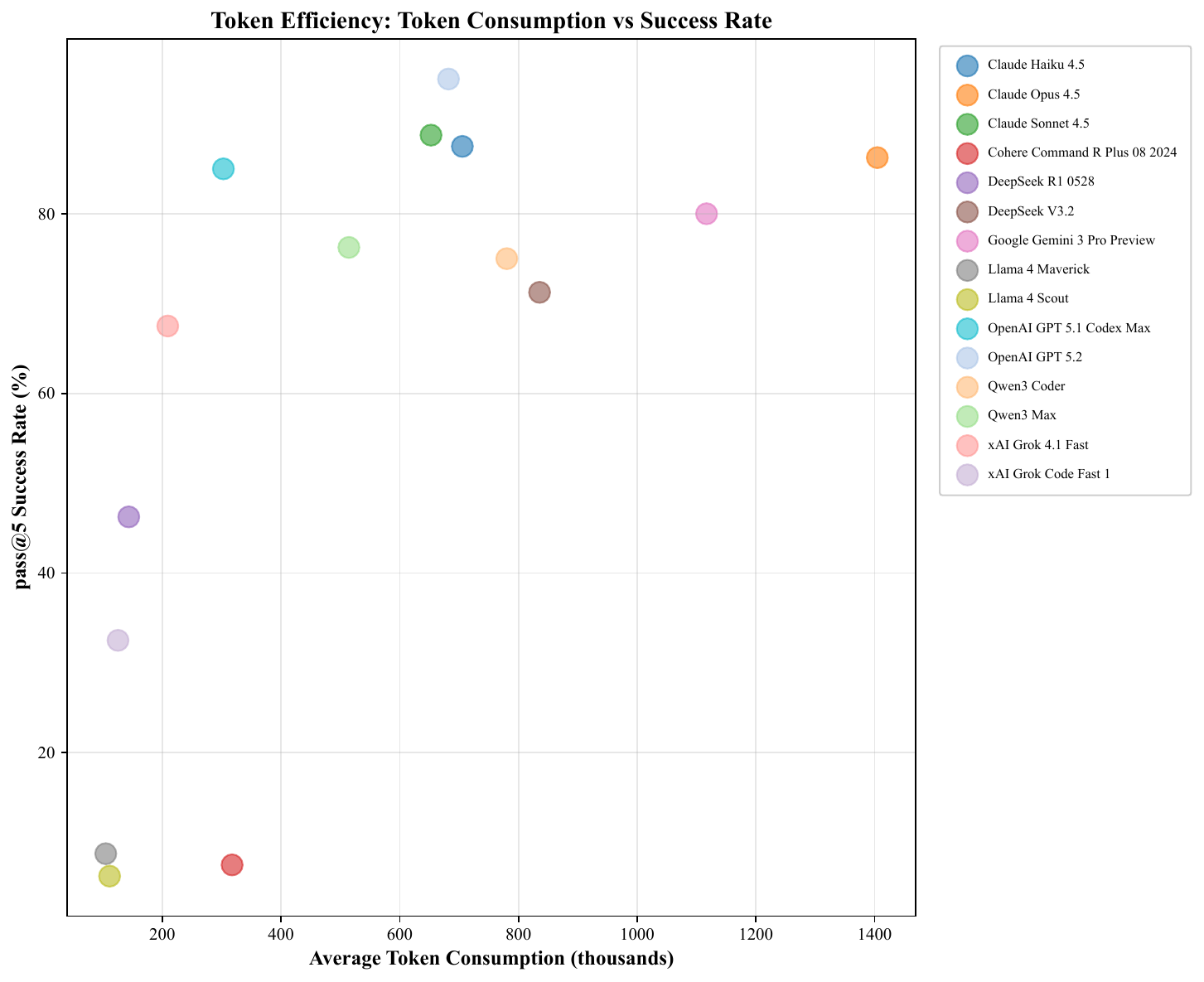}
    \caption{\textbf{Token Efficiency Analysis} Scatter plot of task success rate (pass@5) versus average token consumption. Models in the upper-left quadrant achieve high success with lower token usage, representing optimal efficiency. The plot identifies GPT-5.2 and Claude Haiku as efficiency leaders.}
    \label{fig:token_efficiency}
\end{figure}

\begin{figure} [H]
\centering
    \includegraphics[width=0.9\linewidth]{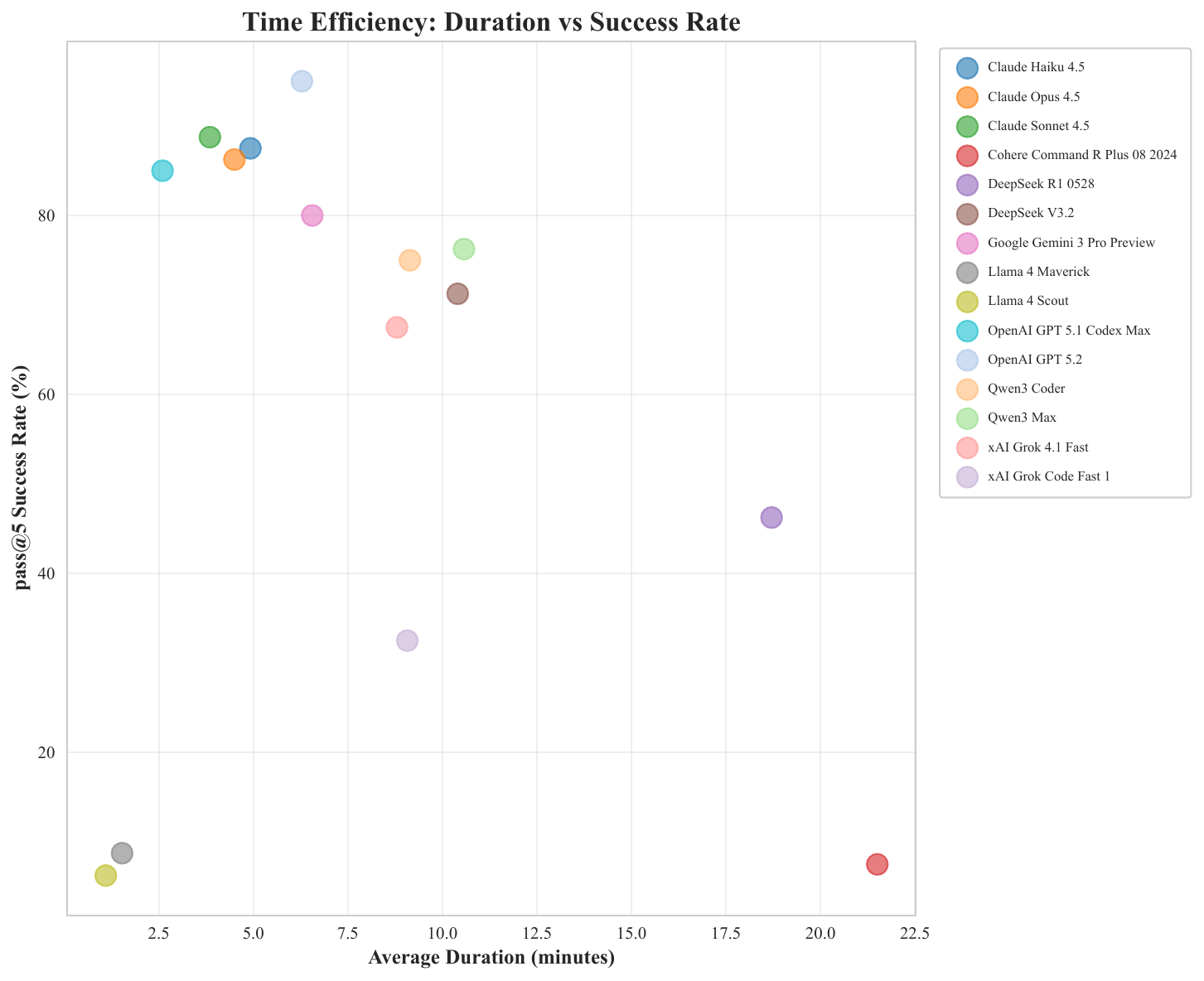}
    \caption{\textbf{Duration Efficiency Analysis} Scatter plot of task success rate (pass@5) versus average task duration. Faster completion times with high success rates indicate efficient problem-solving strategies. This metric captures both model inference speed and strategic efficiency.}
    \label{fig:duration_efficiency}
\end{figure}

\begin{figure} [H]
\centering
    \includegraphics[width=0.9\linewidth]{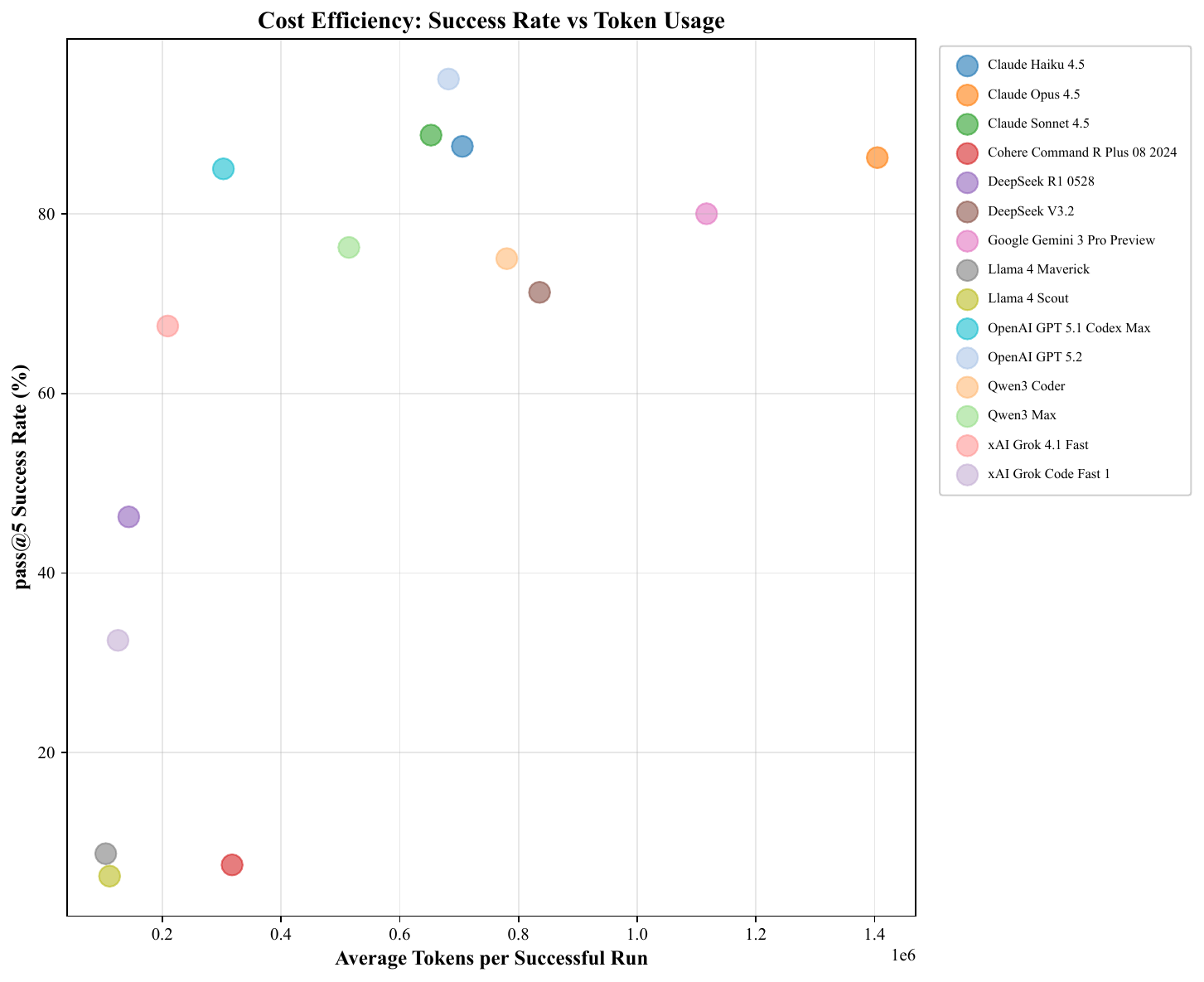}
    \caption{\textbf{Economic Efficiency: Performance vs Cost} Analysis of model performance relative to API costs. This visualization is able to help practitioners choose models based on budget constraints while maintaining acceptable success rates.}
    \label{fig:performance_cost}
\end{figure}

\clearpage
\subsection{Per-Repository Performance Heatmaps}

The following figures show detailed task-level performance breakdowns for each repository in IDE-Bench. Each heatmap displays the success rate for every task (1-10) across all evaluated models, revealing task-specific challenges and model specializations.
\vspace{-6pt}

\begingroup
\captionsetup{skip=2pt}

\begin{figure}[H]
\centering
    \includegraphics[width=\linewidth,height=0.34\textheight,keepaspectratio]{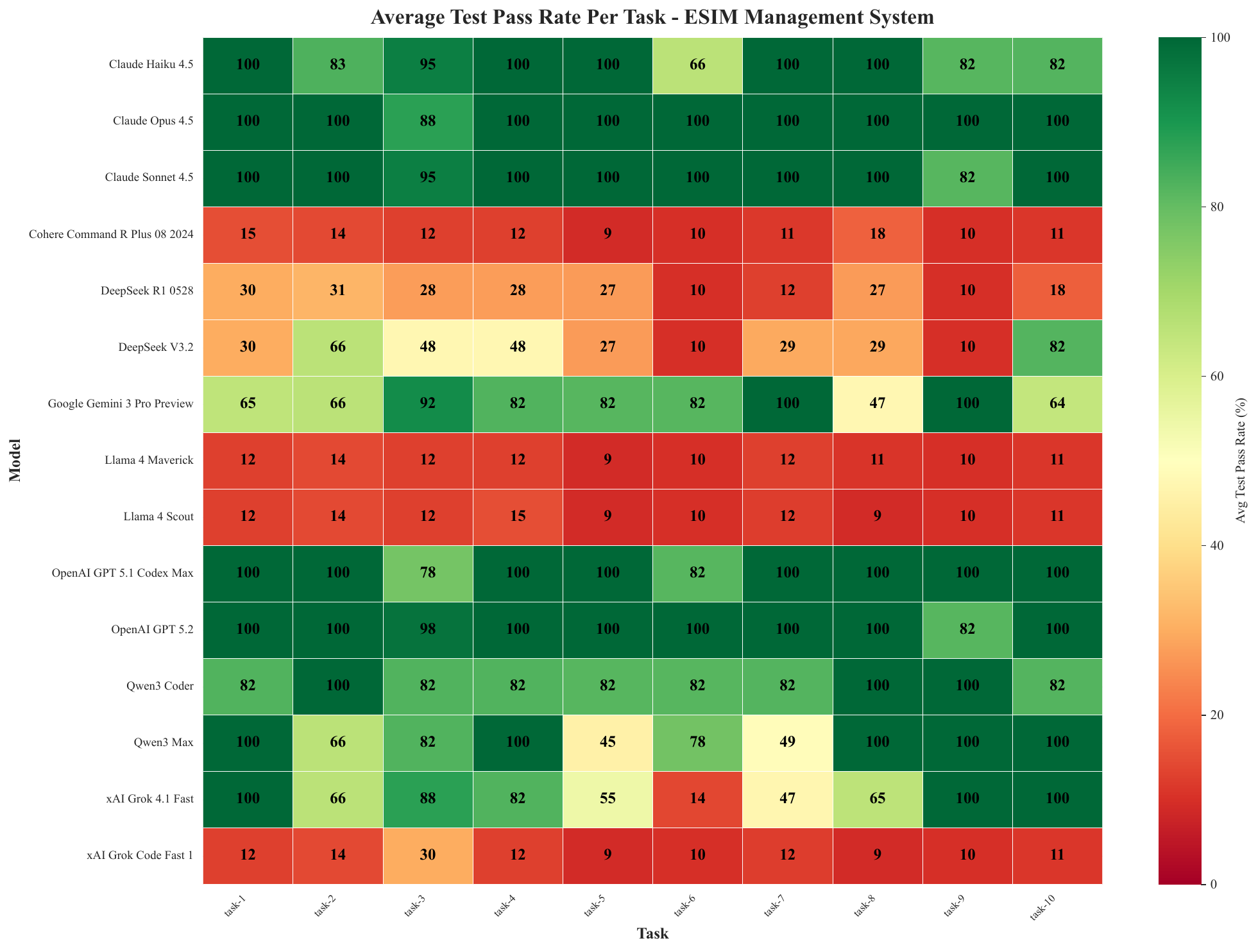}
    \caption{\textbf{ESIM Management System Performance Heatmap} Task-level success rates for the C-based ESIM management repository, emphasizing systems programming and memory management.}
    \label{fig:heatmap_esim}
    \vspace{2pt}
    \includegraphics[width=\linewidth,height=0.34\textheight,keepaspectratio]{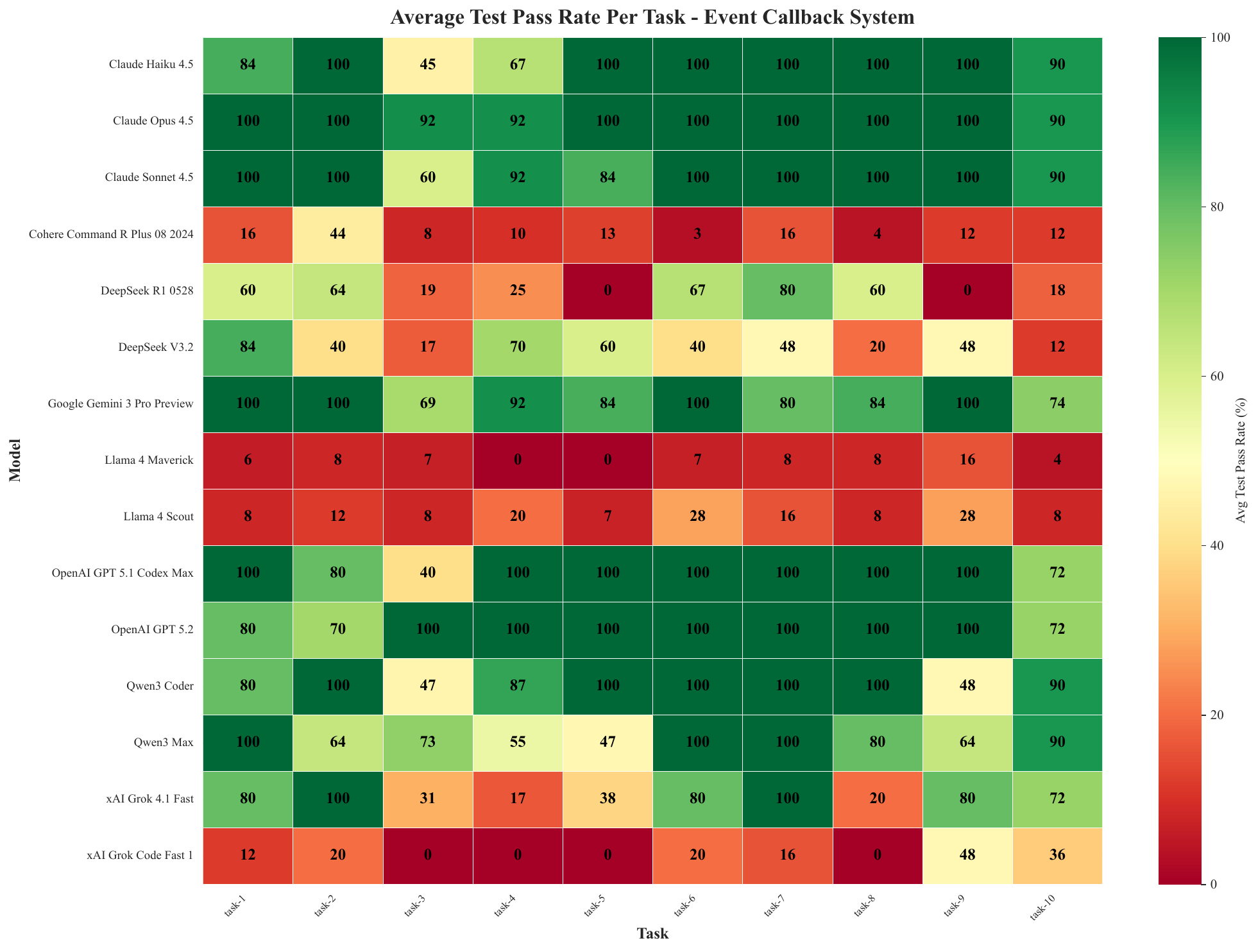}
    \caption{\textbf{Event Callback System Performance Heatmap} Task-level success rates for the TypeScript/Node.js event notification platform, testing asynchronous programming and webhook handling.}
    \label{fig:heatmap_event}
\end{figure}

\begin{figure}[p]
    \centering
    \includegraphics[width=\linewidth,height=0.41\textheight,keepaspectratio]{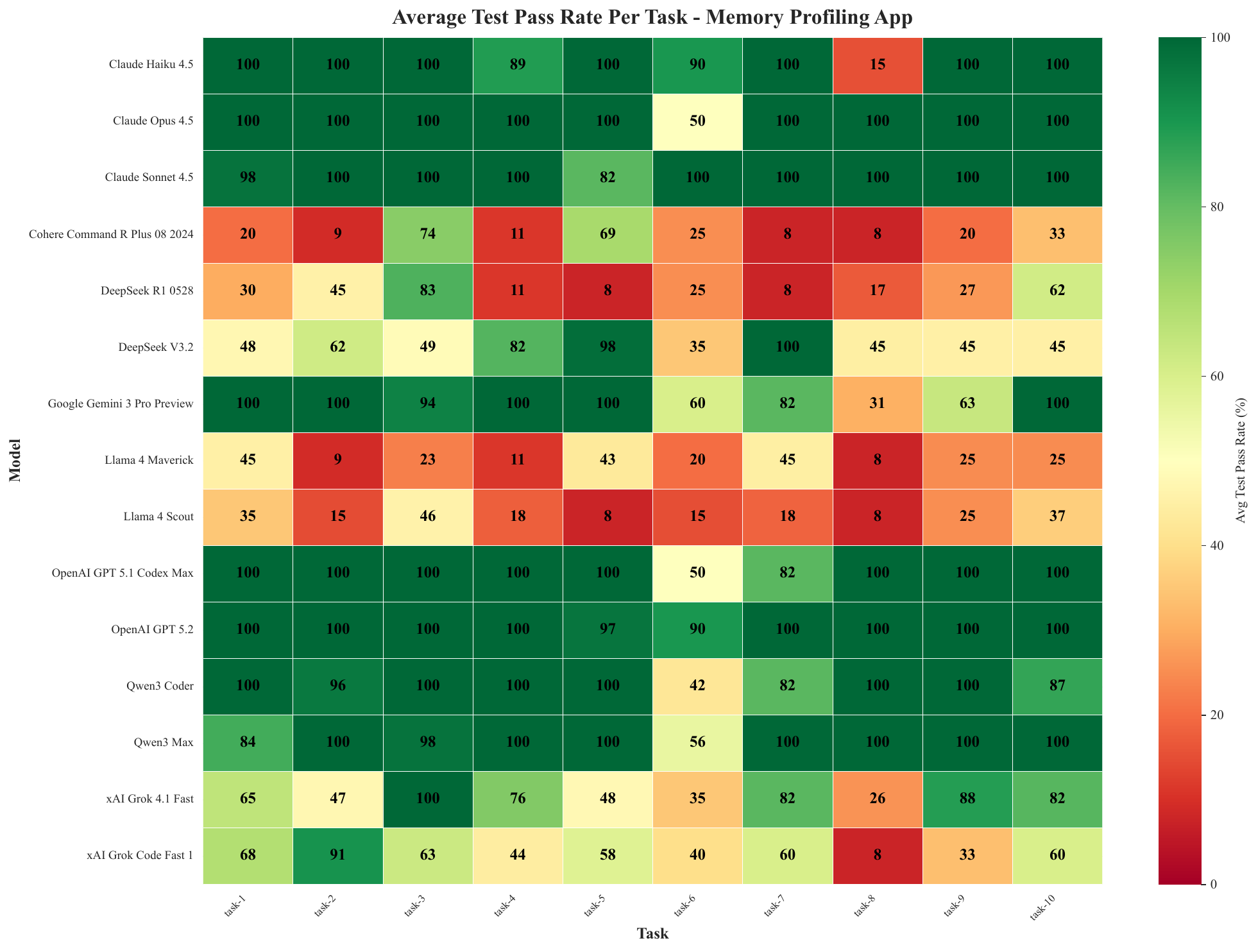}
    \caption{\textbf{Memory Profiling App Performance Heatmap} Task-level success rates for the C++ memory profiler, testing low-level debugging and diagnostic tool development.}
    \label{fig:heatmap_memory}
    \vspace{6pt}
    \includegraphics[width=\linewidth,height=0.41\textheight,keepaspectratio]{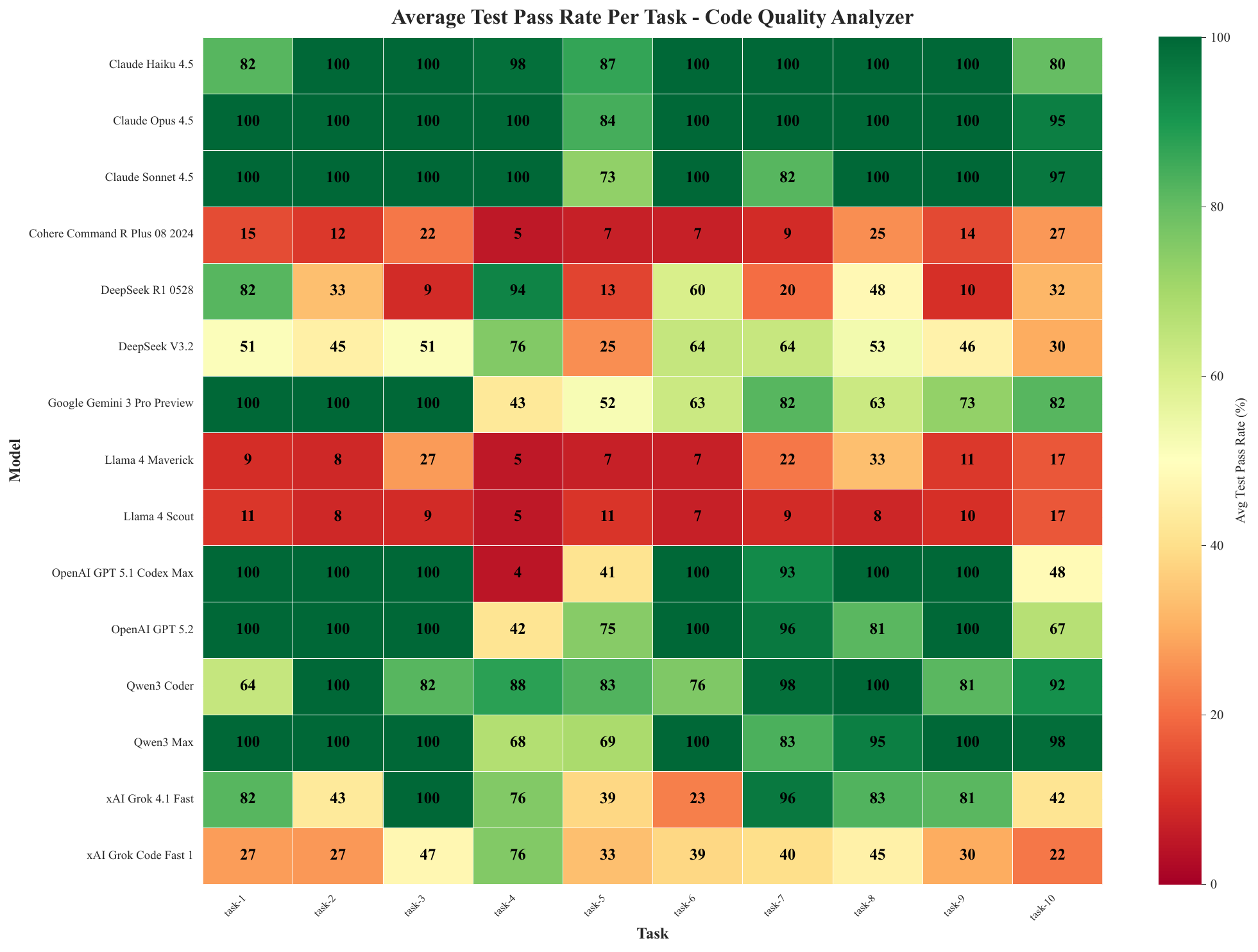}
    \caption{\textbf{Code Quality Analyzer Performance Heatmap} Task-level success rates for the Python code analysis toolkit, testing text processing and static analysis capabilities.}
    \label{fig:heatmap_codeq}
\end{figure}

\begin{figure}[p]
    \centering
    \includegraphics[width=\linewidth,height=0.41\textheight,keepaspectratio]{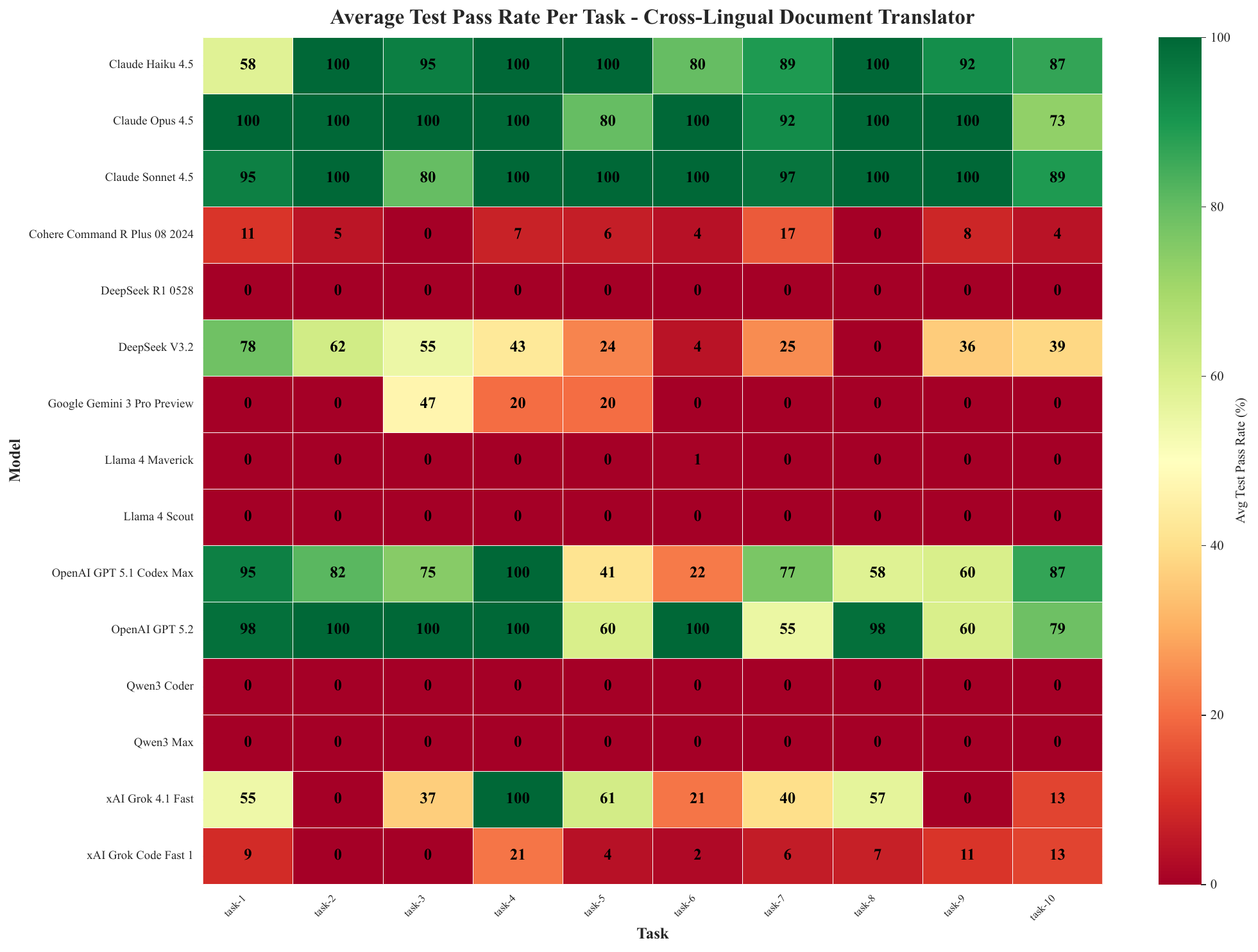}
    \caption{\textbf{Cross-Lingual Document Translator Performance Heatmap} Task-level success rates for the MERN stack translation application, testing full-stack development with MongoDB, Express, React, and Node.js.}
    \label{fig:heatmap_translator}
    \vspace{6pt}
    \includegraphics[width=\linewidth,height=0.41\textheight,keepaspectratio]{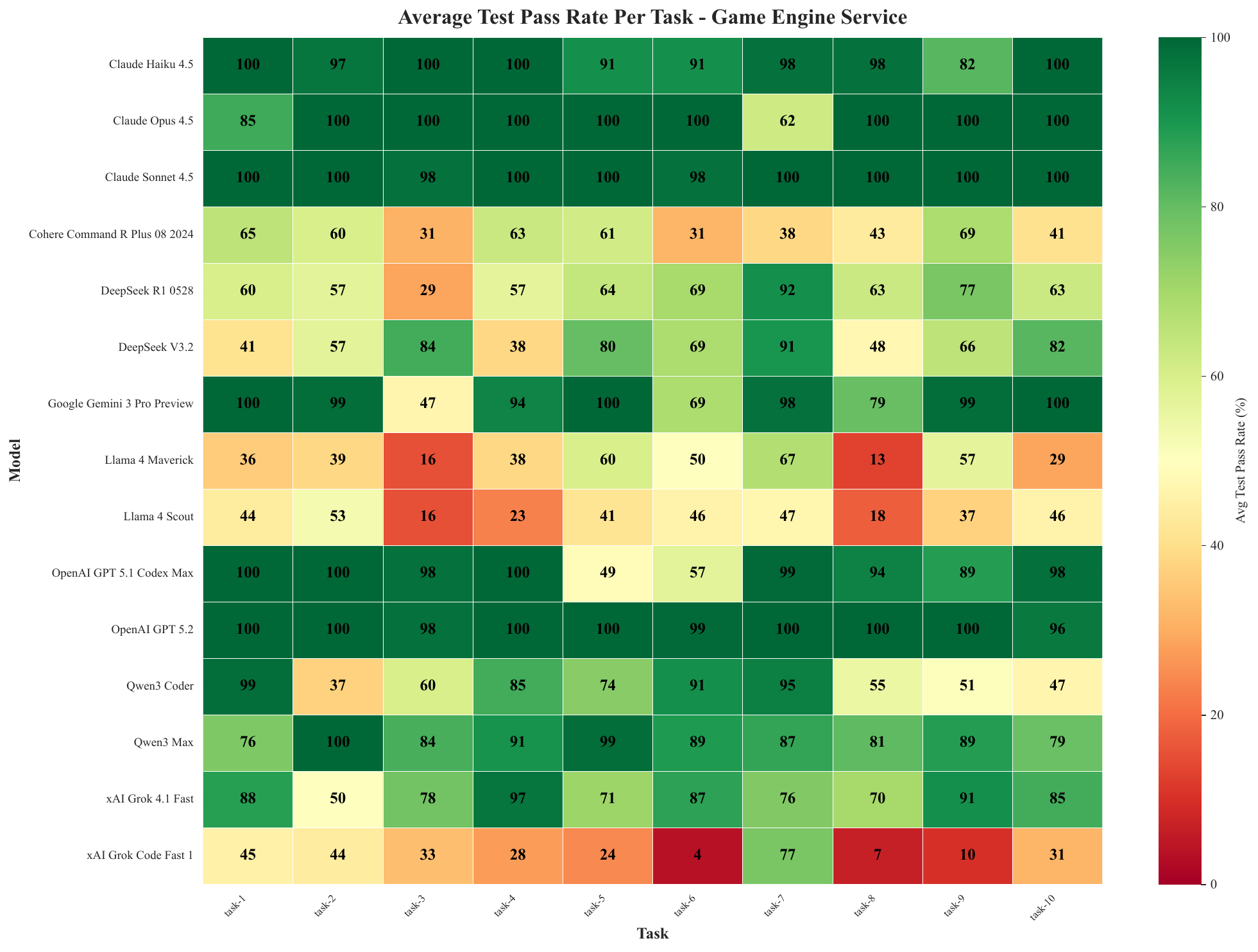}
    \caption{\textbf{Game Engine Service Performance Heatmap} Task-level success rates for the C++ game engine, testing mathematical algorithms, physics calculations, and rendering optimization.}
    \label{fig:heatmap_game}
\end{figure}

\begin{figure}[p]
\centering
    \includegraphics[width=\linewidth,height=0.41\textheight,keepaspectratio]{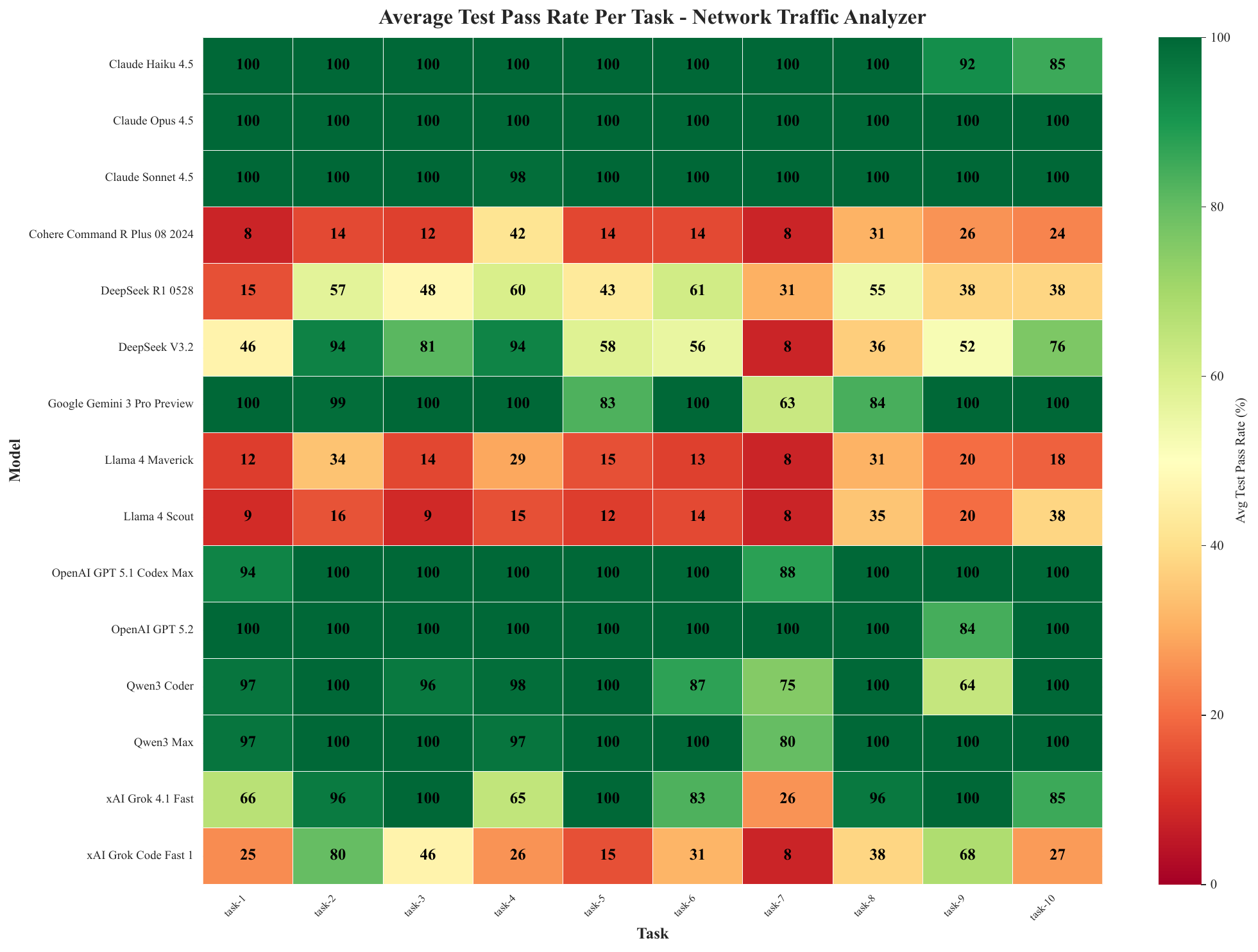}
    \caption{\textbf{Network Traffic Analyzer Performance Heatmap} Task-level success rates for the Python network analysis toolkit, testing protocol understanding and log parsing.}
    \label{fig:heatmap_network}
    \vspace{6pt}
    \includegraphics[width=\linewidth,height=0.41\textheight,keepaspectratio]{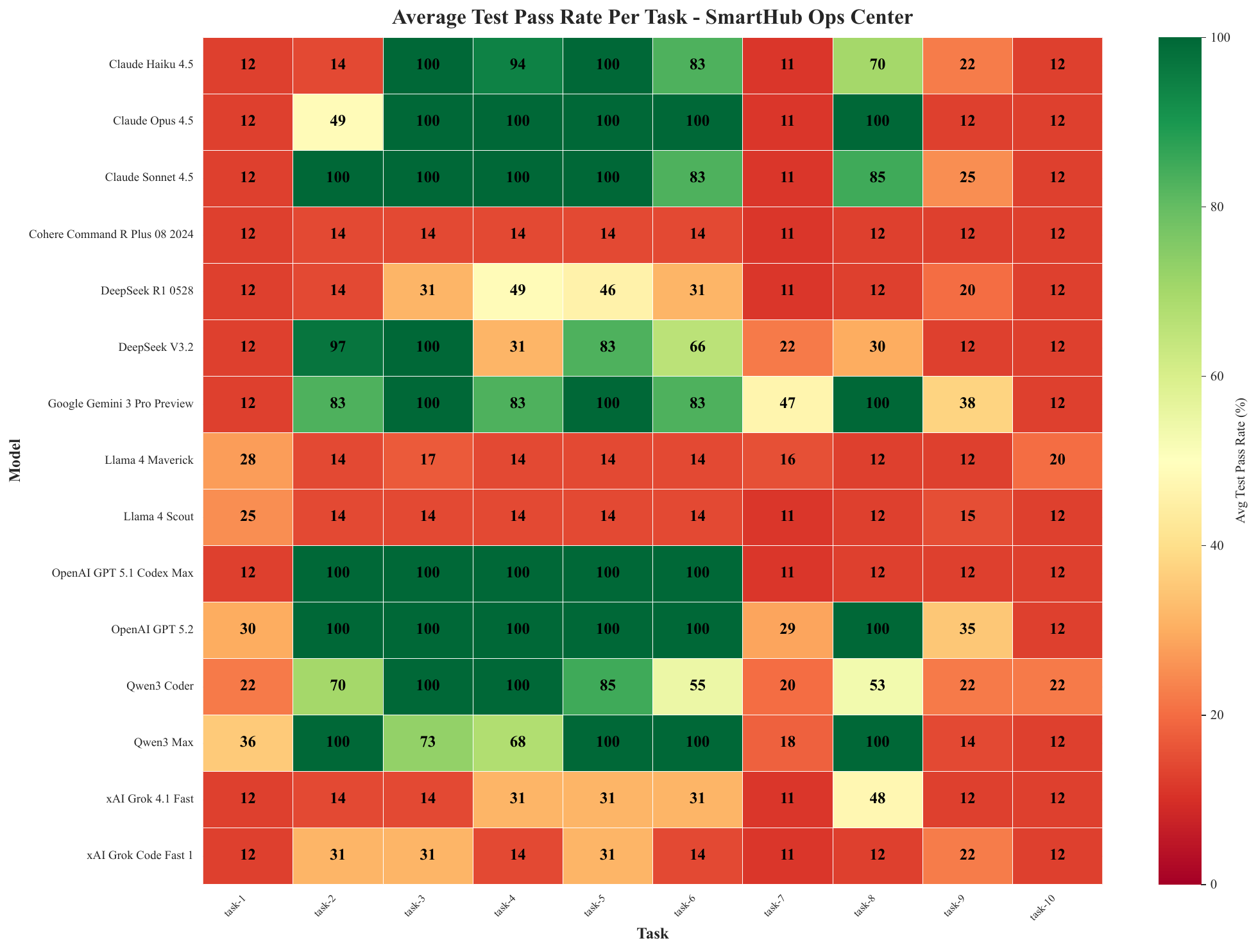}
    \caption{\textbf{SmartHub Operations Center Performance Heatmap} Task-level success rates for the Java web application (Javalin + Thymeleaf), testing MVC-style layering and server-rendered views.}
    \label{fig:heatmap_smarthub}
\end{figure}

\endgroup

\clearpage
\section{Reproducibility and Access}

\subsection{Repository Descriptions}
\label{app:repository_descriptions}

This section provides detailed descriptions of the eight repositories in the IDE-Bench evaluation suite.

\subsubsection{ESIM Management System} The ESIM Management System simulates a C-based command-line platform used for managing eSIM devices, with features including device registration and management, log analysis, network configuration and management. We use this repository to evaluate agents' capabilities in understanding the specifications of a scalable device management. Tasks involve fixing bugs in activation code generation, implementing device management commands, handling certificate parsing, encryption/decryption operations, and configuration file processing. This repository emphasizes correctness in C programming, including memory management challenges, string manipulation, file I/O, and command-line argument parsing.

\subsubsection{Event Callback System} This repository hosts a TypeScript/Node.js–based platform designed for managing event notification endpoints, including subscription handling, event dispatch, automated retries, signature validation, throughput control, and delivery auditing. The system enables external services to subscribe to application events and ensures that notifications are delivered reliably, securely, and with full visibility. This repository focuses on building and maintaining event notification infrastructure, a critical component of microservices. Tasks involve implementing webhook delivery systems, retry mechanisms, rate limiting, event filtering, payload transformation, and delivery status analytics. The repository requires understanding event handling, queue management, statistical calculations, and handling asynchronous operations. Agents must demonstrate the ability to work with complex state machines for tracking notification delivery status.

\subsubsection{Memory Profiling App} The Memory Profiler Toolkit provides tools for monitoring and analyzing memory usage in C++ applications. It detects memory leaks, calculates usage statistics, and generates detailed reports. Tasks involve implementing address validation, leak detection, duplicate address tracking, memory statistics calculation, and generating detailed profiling reports. This repository requires understanding memory allocation patterns, pointer validation, and detecting memory corruption (null pointers, duplicate addresses). Agents must work with low-level memory concepts and gather memory usage statistics while gracefully handling edge cases of empty log files and malformed input. The repository tests the agent's ability to build debugging and diagnostic tools essential for systems programming.

\subsubsection{Code Quality Analyzer} The Code Quality Analyzer repository provides a comprehensive code analysis tool for evaluating code quality metrics, detecting issues, and generating reports. Tasks include implementing log parsers that count HTTP errors, building complexity analyzers, tracking variable usage patterns, validating JSON structures, and performing code quality metrics calculations. This repository requires understanding parsing algorithms, data structure manipulation, and handling malformed input gracefully. These tasks test the agent's ability to work with text processing, pattern matching, and statistical aggregation, useful for maintaining consistency across a codebase.

\subsubsection{Cross-Lingual Document Translator} The cross-lingual document translator implements a full-stack web application for translating documents between languages, featuring user authentication, translation history, and various management features. This repository focuses on feature implementation in a production-like web application stack where the backend runs on Node.js and the database is managed with MongoDB, with JWT authentication middleware and a web-based frontend interface. Tasks involve adding new functionality to a document translation service, including favorites/bookmarks systems, user management features, translation history pagination, and authentication flow modifications. The repository requires understanding REST API design, database schema modifications, and authentication schemes used across many web apps. Agents must demonstrate proficiency in full-stack development workflows.

\subsubsection{Game Engine Service} The Game Engine Service repository is a C++ game engine implementation featuring core systems for animation, physics, rendering, and entity management. It evaluates agents' capabilities in game development, focusing on animation systems, physics calculations, and rendering optimization. Tasks involve implementing animation blending with easing functions, rotation interpolation with proper angle wrapping, collision detection, spatial partitioning, and sprite batching. This repository requires deep understanding of mathematical concepts including easing curves, angular interpolation, coordinate transformations, and spatial algorithms. Agents must correctly handle angle calculations, easing functions, and frame processing. The repository tests a thorough understanding of game engine architecture patterns and handling of dynamic interactions in entity component systems.

\subsubsection{Network Traffic Analyzer} The Network Traffic Analyzer repository is a Python-based toolkit for analyzing network traffic logs. It contains various scripts for processing, aggregating, and analyzing network traffic data. We evaluate agents' capabilities in network programming and log analysis. Tasks include implementing traffic summary generators, session trackers, protocol analyzers, top talker identification, anomaly detection, and bandwidth calculations. This repository requires understanding of network protocols (TCP, UDP), log file parsing with specific format requirements, handling multi-column timestamps, and statistical aggregation. Agents must demonstrate fluency in data processing, calculating statistics for unique IP addresses, and maintaining precise output formats.

\subsubsection{SmartHub Operations Center} The SmartHub Operations Center repository focuses on Java web application development using the Javalin framework. In this use case, it simulates an environment of multiple devices connected in home or edge contexts. The system that this work presents shows a design focusing on the following modules: device observation, device connection, tracking energy use, monitoring climate conditions, assessing security, maintaining logs, replicating data, and conducting tests for developing features. Each module contains its own means for processing those requests, implementing logic, and displaying information to users. Tasks involve fixing bugs in service layer calculations, implementing route handlers, and ensuring correct data flow between service, handler, and view layers. This repository requires understanding of MVC design, template engines, and dependency injection patterns.

\subsection{Agent Tool Interface}
\label{app:tool_interface}

This section provides complete specifications for the 17 tools available to models in the IDE-Bench harness. All tools follow OpenAI's function calling specification and are uniformly available to all evaluated models.

\textbf{File System \& Code Navigation} (6 tools):
\begin{itemize}[noitemsep, topsep=0pt]
    \item \texttt{read\_file}: Read file contents with optional line-range specification (offset and limit parameters). Test files in the \texttt{/tasks} directory and \texttt{run\_tests.sh} are blocked to prevent evaluation integrity violations. Returns file contents with line numbers.
    \item \texttt{list\_dir}: List directory contents recursively or non-recursively. Automatically excludes common ignore patterns (node\_modules, .git, \_\_pycache\_\_, .venv, dist, build, etc.). Returns array of file and directory paths with metadata.
    \item \texttt{codebase\_search}: Lexical keyword search using grep/ripgrep for exact text matches across the codebase. Supports case-insensitive search, context lines (before/after), and file type filtering. Returns matching lines with file paths and line numbers.
    \item \texttt{grep\_search}: Advanced regex-based search with full regular expression support, case sensitivity controls, multiline mode, and flexible include/exclude patterns. More powerful than codebase\_search for complex pattern matching.
    \item \texttt{file\_search}: Fuzzy file path search for locating files by partial name matches. Uses glob patterns to find files matching specified naming patterns across the repository.
    \item \texttt{delete\_file}: Delete files at specified paths. Blocks deletion of test files following the same security model as \texttt{read\_file}.
\end{itemize}

\textbf{Code Editing} (3 tools):
\begin{itemize}[noitemsep, topsep=0pt]
    \item \texttt{edit\_file}: Structured line-based editing supporting three operation types: REPLACE (substitute line ranges), INSERT (add lines at position), and DELETE (remove line ranges). Requires precise line targeting with start\_line and end\_line parameters. For Python files, automatically validates syntax using \texttt{ast.parse()} after edits and reports any syntax errors. Security model prevents editing test files.
    \item \texttt{search\_replace}: Simple string-based find-and-replace within files. Supports single replacement or replace-all mode. Does not require line numbers, making it suitable for simple textual substitutions. Also respects test file security restrictions.
    \item \texttt{write\_file}: Create new files or completely overwrite existing files. Accepts file path and complete file contents as parameters. Useful for creating new modules or configuration files from scratch.
\end{itemize}

\textbf{Execution \& Testing} (1 tool):
\begin{itemize}[noitemsep, topsep=0pt]
    \item \texttt{run\_terminal\_cmd}: Execute shell commands within the Docker container environment. Supports both foreground mode (with 120-second timeout, suitable for build commands, test runs, and quick checks) and background mode (for long-running processes like web servers, development servers, and database instances). Returns command output, exit code, and execution metadata. All commands run in the \texttt{/app} working directory by default.
\end{itemize}

\textbf{Full-Stack Testing} (4 tools, task-dependent):
\begin{itemize}[noitemsep, topsep=0pt]
    \item \texttt{api\_call}: Make HTTP requests to test REST API endpoints. Supports all standard methods (GET, POST, PUT, DELETE, PATCH, HEAD, OPTIONS). Accepts headers, request body (JSON or form data), query parameters, and authentication credentials. Returns response status code, headers, body, and timing information. Essential for testing backend API implementations.
    \item \texttt{database\_query}: Execute MongoDB operations for database testing and validation. Supports standard CRUD operations (find, findOne, insert, update, delete) and aggregation pipelines. Accepts database name, collection name, and operation-specific parameters. Returns query results, document counts, and operation status. Required for tasks involving persistent data storage.
    \item \texttt{websocket\_test}: Test Socket.IO and WebSocket functionality for real-time features. Supports connection establishment, event emission, event listening, and message exchange. Validates bidirectional communication patterns essential for chat systems, live notifications, and real-time collaboration features.
    \item \texttt{ui\_test}: Simulate frontend interactions for UI validation. Provides capabilities for clicking elements, typing into input fields, capturing screenshots, navigating between pages, and verifying DOM states. Uses Playwright-like interface for browser automation. Useful for end-to-end testing of web applications.
\end{itemize}

These four full-stack testing tools are available when tasks include MERN stack components (MongoDB, Express.js, React, Node.js), enabling comprehensive end-to-end verification across the full application stack. For tasks focused on algorithmic implementation, systems programming, or backend services without databases, models primarily rely on file system, editing, and terminal execution tools.

\textbf{Specialized Tools} (3 tools):
\begin{itemize}[noitemsep, topsep=0pt]
    \item \texttt{edit\_notebook}: Jupyter notebook cell editing with support for creating, modifying, and deleting cells. Accepts notebook file path, cell index, cell type (code/markdown), and cell contents. Available in the harness but not used in current benchmark tasks, as IDE-Bench focuses on traditional software engineering workflows rather than notebook-based development.
    \item \texttt{web\_search}: Search the web for documentation, API references, and technical information. Available in tool definitions but not implemented in the evaluation environment, as tasks are designed to be solvable from repository context alone without requiring external knowledge retrieval. This ensures consistent evaluation conditions regardless of network availability.
    \item \texttt{create\_diagram}: Generate Mermaid diagram syntax for visualization purposes (flowcharts, sequence diagrams, class diagrams, etc.). Available but not evaluated in current benchmark, as IDE-Bench focuses on functional code implementation rather than documentation artifacts.
\end{itemize}

All tool invocations require an \texttt{explanation} parameter (string) describing the tool's purpose and intended effect. This parameter serves multiple purposes: (1) encourages models to articulate their reasoning before taking actions, (2) enables trajectory analysis of agent decision-making patterns, and (3) provides interpretable logs for debugging agent behavior. The harness logs all tool calls with timestamps, parameters, results, and execution time for comprehensive post-hoc analysis.

\subsection{Benchmark Details}

The harness operates with the following default execution parameters for consistent evaluation:
\begin{itemize} [noitemsep, topsep=0pt]
    \item Maximum iterations: 100 conversation turns per task
    \item Temperature: 0.1 
    \item Maximum output tokens: 4,000 tokens per model response
    \item API call timeout: 600 seconds per LLM API call
    \item Command timeout: 120 seconds for foreground terminal commands; background processes supported for long-running servers
\end{itemize}

\subsubsection{Execution Workflow and Grading Pipeline}
\label{app:harness_grading}

After the agent completes execution (or reaches the 100-iteration limit), the grader runs the test suite (\texttt{./run\_tests.sh}) to assess functional correctness. The grader then extracts agent code changes through a 4-step git diff pipeline: (1) \texttt{git diff HEAD} for unstaged changes, (2) \texttt{git add -A} then \texttt{git diff --cached} for staged changes, (3) commit-based diff via \texttt{git diff HEAD\textasciitilde 1 HEAD}, and (4) file-by-file comparison against \texttt{git show HEAD:<file>} if all git methods fail. The extracted diff is compared to the reference patch (\texttt{task\_diff.txt}) using semantic similarity scoring. The grader automatically detects test frameworks (pytest, jest, maven, junit, go test, cargo, rspec, phpunit, dotnet test, mocha) and outputs comprehensive metrics including test pass rates, diff similarity scores, iteration counts, and tool usage patterns.

\subsubsection{Harness Implementation Details}
\label{app:harness_impl}

\textbf{Model Integration:} The harness uses LiteLLM for unified API access across model providers with OpenAI-format function calling, enabling evaluation of any LiteLLM-compatible model. Additionally, the system implements exponential backoff retry logic (5 retries, 10-second delay) for transient API errors and handles model-specific quirks (such as the JSON string formatting for Cohere).

\textbf{Context Management:} When conversation history exceeds 25 messages, the system truncates to approximately 80\% of the model's context window, preserving the system message, first user message, and most recent turns.

\textbf{Security Model:} To maintain evaluation integrity, the \texttt{read\_file}, \texttt{edit\_file}, \texttt{search\_replace}, and \texttt{delete\_file} tools block access to test files (\texttt{/tasks} directory, \texttt{run\_tests.sh}, \texttt{test\_*.py}, \texttt{*.test.js}, \texttt{*.spec.ts}) and test directories (\texttt{tests/}, \texttt{\_\_tests\_\_/}). Agents must reason from task requirements without observing ground truth implementations.

\subsubsection{Evaluation Metrics}
\label{app:eval_metrics_full}

The grader collects comprehensive metrics across multiple dimensions:

\textbf{Primary Metrics:}
\begin{itemize} [noitemsep, topsep=0pt]
    \item \textbf{Grounded Performance Bounds:} We establish the Floor (Null Baseline) by executing the test suite on the initial, unmodified repository state to identify pre-existing failures or trivial passes. We establish the Ceiling (Oracle Baseline) by applying the reference patch (\texttt{task\_diff.txt}) and verifying that 100\% of tests pass. This framing ensures that model performance is measured within a calibrated success range.
    \item \textbf{Task Resolution Rate (pass@k):} We evaluate using pass@k, defined as the probability that at least one of k independent attempts succeeds. A task is considered solved under pass@k if any of the k attempts passes all functional tests.
    \item \textbf{Test Pass Rate:} Percentage of test cases passed, with detailed per-test status tracking.
\end{itemize}

\textbf{Additional Metrics:}
\begin{itemize} [noitemsep, topsep=0pt]
    \item \textbf{Agent success:} Whether the agent made code changes and achieved a passing test run.
    \item \textbf{Syntax errors:} Number of Python syntax errors blocked and successfully fixed.
    \item \textbf{Total iterations:} Number of conversation turns used.
    \item \textbf{Files visited vs.\ edited:} Exploration-modification ratio.
    \item \textbf{Tool usage patterns:} Frequency and order of tool calls.
    \item \textbf{Stopping reason:} Completion, iteration limit, or error.
    \item \textbf{Full conversation trace:} Complete agent trajectory.
\end{itemize}

\subsubsection{Model Prompting}

All models receive a comprehensive system prompt that provides structured guidance for IDE agent operation. The prompt lists all 17 available tools organized into five categories (Search \& Discovery, File Operations, Execution \& Automation, Specialized Tools, and MERN Stack Tools), creates guidelines for workflows (discovery before modification, incremental implementation, testing when applicable), and provides guidance for full-stack applications (MERN stack components, using specialized testing tools for APIs, databases, and WebSockets). 

Each tool invocation requires an \texttt{explanation} parameter describing its purpose and reasoning, enabling trajectory analysis of agent decision-making. The complete system prompt is provided in Appendix~\ref{app:system_prompt} for full transparency and reproducibility.

\subsection{Complete System Prompt}
\label{app:system_prompt}

All models in the IDE-Bench evaluation receive the following system prompt, which is provided at the beginning of each task:

\begin{verbatim}
You are a powerful coding assistant with comprehensive 
development capabilities. You have access to the following 
tools:

SEARCH & DISCOVERY:
- codebase_search: Semantic search through codebase to find 
  relevant code snippets
- grep_search: Fast regex-based text search with file 
  filtering
- file_search: Fuzzy search for files by name
- list_dir: List directory contents for exploration

FILE OPERATIONS:
- read_file: Read file contents with optional line range 
  support
- edit_file: Propose structured edits to files
- search_replace: Find and replace text in files
- write_file: Create new files or overwrite existing ones
- delete_file: Remove files from the filesystem

EXECUTION & AUTOMATION:
- run_terminal_cmd: Execute shell commands (with background 
  support)
- create_directory: Create directory structures

SPECIALIZED TOOLS:
- edit_notebook: Edit Jupyter notebook cells
- create_diagram: Generate Mermaid diagrams
- web_search: Search the web for information (when 
  available)

MERN STACK TOOLS:
- api_call: Make HTTP requests to test REST API endpoints
- database_query: Execute MongoDB queries (find, insert, 
  update, delete, aggregate)
- websocket_test: Test Socket.IO real-time functionality
- ui_test: Browser automation for React frontend testing 
  (screenshot, click, type, navigate)

WORKFLOW GUIDELINES:
1. Break down complex tasks into steps
2. Use search tools to understand the codebase first
3. Read relevant files before making changes
4. Use appropriate tools for the task (semantic search vs 
   grep vs file search)
5. Provide clear explanations with each tool use
6. Test changes when possible using terminal commands

FOR MERN STACK APPLICATIONS:
1. Identify if you're working with a MERN (MongoDB, 
   Express, React, Node.js) stack
2. Use api_call to test backend endpoints after making 
   changes
3. Use database_query to verify data persistence in MongoDB
4. Use websocket_test for real-time features (chat, 
   notifications, live updates)
5. Use ui_test to verify frontend functionality and user 
   interactions
6. Look for server/ directory (backend), client/ directory 
   (frontend), and package.json files

Always explain your reasoning and approach clearly.
\end{verbatim}

This prompt establishes consistent evaluation conditions across all models by providing identical tool documentation and workflow guidance. By requiring explanation parameters for each tool invocation, our prompt thus enables detailed trajectory analysis while encouraging systematic exploration over random trial-and-error. The prompt represents a deliberate intervention in agent behavior; models that effectively internalize this structured approach demonstrate measurably better task completion rates than those following ad-hoc strategies.

\subsection{Accessing the Repositories and Evaluation Harness}

\textbf{Evaluation Harness:} The IDE-Bench evaluation harness, including Docker configuration, agent utilities, grader subsystem, and all tool implementations, along with the Game Engine Service repository, serving as a sample dataset, are publicly available on GitHub at \url{https://github.com/AfterQuery/IDE-Bench}. The harness supports any LiteLLM-compatible model and can be extended to additional programming languages and technology stacks.

\textbf{Training Data Contamination Prevention:} To maintain the integrity of the evaluation repositories and to prevent training data contamination, the seven of the eight benchmark repositories remain unpublished. Benchmarks often lose effectiveness once their evaluation data appear publicly on platforms like GitHub and Hugging Face, as future model releases may memorize solutions. We balance reproducibility with contamination prevention through a controlled access model:

\begin{itemize}[noitemsep, topsep=0pt]
    \item \textbf{Sample Repository:} The Game Engine Service repository is made available under the datasets folder at \url{https://github.com/AfterQuery/IDE-Bench} to demonstrate task structure, difficulty, and evaluation methodology.
    \item \textbf{Full Evaluation Set:} The complete set of eight repositories with 80 tasks is available upon request; interested parties should contact \texttt{research@afterquery.com} for access. 
\end{itemize}

\fi 
\end{document}